\documentclass[twocolumn,aps,rmp,eqsecnum,amsmath,amssymb]{revtex4}

\usepackage{graphics}
\usepackage{psfrag}

\begin{document}

\title{
  Projector-based renormalization method (PRM) and its application to 
  many-particle systems
}

\author{Arnd H\"ubsch, Steffen Sykora, and Klaus W.~Becker}

\affiliation{
  Institut f\"{u}r Theoretische Physik,
  Technische Universit\"{a}t Dresden, 01062 Dresden, Germany
}

\date{\today}

\begin{abstract}
  Despite the advances in the development of numerical methods 
  analytical approaches play a key role on the way towards a deeper 
  understanding of strongly interacting systems. In this regards, 
  renormalization schemes for Hamiltonians represent an important new 
  direction in the field. Among these renormalization schemes the 
  projector-based renormalization method (PRM) reviewed here might be the 
  approach with the widest range of possible applications: As demonstrated 
  in this review, continuous unitary transformations, perturbation theory, 
  non-perturbative phenomena, and quantum-phase transitions can be 
  understood within the same theoretical framework. This review starts from 
  the definition of an effective Hamiltonian by means of projection operators 
  that allows the evaluation within perturbation theory as well as the 
  formulation of a renormalization scheme. The developed approach is then 
  applied to three different many-particle systems: At first, we study the 
  electron-phonon problem to discuss several modifications of the method and 
  to demonstrate how phase transitions can be described within the PRM. 
  Secondly, to show that non-perturbative phenomena are accessible by the PRM, 
  the periodic Anderson is investigated to describe heavy-fermion behavior. 
  Finally, we discuss the quantum-phase transition in the one-dimensional 
  Holstein model of spinless fermions where both metallic and insulating phase 
  are described within the same theoretical framework.
\end{abstract}

\maketitle
\tableofcontents

\section{Introduction}

During the last three decades the investigation of phenomena related with
strongly interacting electrons has developed to a central field of condensed
matter physics. In this context, high-temperature superconductivity and
heavy-fermion behavior are maybe the most important examples. It has been
clearly turned out that such systems require true many-body approaches that
properly take into account the dominant strong electronic correlations. 

In the past, many powerful numerical methods like exact diagonalization
\cite{ED}, numerical renormalization group \cite{NRG}, Quantum Monte-Carlo
\cite{MC}, the density-matrix renormalization group \cite{DMRG}, or the
dynamical mean-field theory \cite{DMFT} have been developed to study strongly
correlated electronic systems. In contrast, only very few analytical
approaches are available to tackle such systems. In this regard,
renormalization schemes for Hamiltonians developed in the nineties of the last
century \cite{GW_1993, GW_1994, W_1994} represent an important new direction
in the field where renormalization schemes are implemented in the Liouville
space (that is built up by all operators of the Hilbert space). Thus, these
approaches can be considered as further developments of common renormalization
group theory \cite{RG} that is based on a renormalization within the Hilbert 
space. 

In this review we want to discuss the projector-based renormalization method
[PRM, Ref.~\onlinecite{BHS_2002}] that shares some basic concepts with the
renormalization schemes for Hamiltonians mentioned above 
\cite{GW_1993, GW_1994, W_1994}. All these approaches including the PRM 
generate effective Hamiltonians by applying a sequence of unitary
transformations to the initial Hamiltonian of the physical system. However,
there is one distinct difference between these methods: Both similarity
renormalization \cite{GW_1993,GW_1994} and Wegner's flow equation method
\cite{W_1994} start from a continuous formulation of the unitary transformation
by means of a differential form. In contrast, the PRM is based on discrete
transformations so that a direct link to perturbation theory can be provided. 

\bigskip
This review is organized as follows: 

\smallskip
In the next section we discuss the basic concepts of the PRM: We introduce 
projection operators in the Liouville space that allow the definition 
of an effective Hamiltonian. If these ingredients are combined with unitary 
transformations one can derive a new kind of perturbation theory that is not
restricted to the ground-state but also allows to investigate excitations. 
(To illustrate this point we briefly discuss the triplet dispersion relation 
of a dimerized and frustrated spin chain in the Appendix.)
However, this perturbation theory is not the focus of this review and can be 
considered as an interesting side-product of the development of the PRM, a 
renormalization scheme based on the same ingredients. To illustrate the method 
in some detail, the exactly solvable Fano-Anderson model is considered.
 
Improving our previous publications on the PRM, we show here the relation of 
the PRM to Wegner's flow equation method \cite{W_1994} for the first time. It 
turns out the latter method can be understood within the framework of the PRM 
by choosing a complementary unitary transformations to generate the effective 
Hamiltonian. For demonstration, the Fano-Anderson model is solved with this 
approach, too.  

\smallskip
As a more physical example, the electron-phonon interaction is studied in 
Sec.~\ref{EP}. In particular, the PRM is compared in some detail with the 
flow equation method \cite{W_1994} and the similarity transformation
\cite{GW_1993,GW_1994}. Furthermore, we introduce a possible modification of
the PRM that allows to derive block-diagonal Hamiltonians, and we discuss in
some detail the freedom in choosing the generator of the unitary
transformation the PRM is based on. Finally, we show how phase
transitions can be studied within the PRM by adding symmetry breaking fields
to the Hamiltonian.

\smallskip
In Sec.~\ref{PAM} the PRM is applied to the periodic Anderson model to
describe heavy-fermion behavior. Whereas the famous
slave-boson mean-field theory \cite{C_1984,FKZ_1988} obtains an effectively
free system consisting of two non-interacting fermionic quasi-particles, here
the periodic Anderson model is mapped onto an effective model that still takes
into account electronic correlations. Thus, in principle both mixed and
integral valence solution can be found. However, here we restrict ourself to
an analytical solution of the renormalization equations that is limited to the 
mixed valence case. 

\smallskip
As third application of the PRM  the one-dimensional Holstein model
of spinless fermions is discussed. It is well known that the 
system undergoes a quantum phase transition from a metallic 
to a Peierls distorted state if the electron-phonon coupling exceeds a 
critical value. First, for the metallic state we discuss the 
crossover behavior between the adiabatic and 
anti-adiabatic case in Sec.~\ref{Holstein_CROSS}. 
All physical properties are shown to strongly depend on the ratio of  
phonon and hopping energy in the system.  In Sec.~\ref{Holstein_QP},  
a unified description of the quantum-phase transition is given for the 
one-dimensional model in the adiabatic case.

Finally, as a second example for a quantum phase transition, we discuss in 
Sec.~\ref{Holstein_SC} the competition of charge ordering and 
superconductivity in the two-dimensional Holstein model. Based on the PRM both 
charge density wave and superconductivity are studied within one 
theoretical framework. 

\smallskip
We summarize in Sec.~\ref{summary}.

\section{Projector-based renormalization method (PRM)}
\label{PRM1}

In this section we introduce the concepts of the PRM \cite{BHS_2002} where we
particularly pay attention to a general notation that is used throughout 
the review for all applications of the approach. 

We define projection operators of the Liouville space and define an effective
Hamiltonian where, in contrast to common approaches, \textit{excitations} 
instead of \textit{states} are integrated out. In this way, not only a
perturbation theory is derived but also and more important a 
renormalization scheme (that we call PRM in the following) is established 
which allows to diagonalize or at least to quasi-diagonalize many-particle 
Hamiltonians. As an illustrative example, the exactly solvable Fano-Anderson 
model is discussed. 

The PRM is based on a sequence of \textit{finite} unitary transformations 
whereas Wegner's flow equations start from a continuous formulation of 
unitary transformations by means of a differential form. It turns out that 
such a continuous transformation can also be understood in the framework 
of the PRM if a complementary choice for the generator of the unitary 
transformation is used and infinitely small transformation steps are 
considered. To discuss the differences between the two formulations of the 
PRM in more detail, we also solve the Fano-Anderson model using the 
developed continuous approach.

\subsection{Basic concepts}
\label{basic_concepts}

The projector-based renormalization method (PRM)  \cite{BHS_2002} 
starts from the usual decomposition of a given many-particle Hamiltonian, 
\begin{eqnarray*}
\mathcal{H} = \mathcal{H}_{0} + \mathcal{H}_{1}\, , 
\end{eqnarray*}
where the perturbation $\mathcal{H}_{1}$ should not contain any terms that
commute with the unperturbed part $\mathcal{H}_{0}$. Thus, the interaction
$\mathcal{H}_{1}$ consists of the transitions between eigenstates of
$\mathcal{H}_{0}$ with corresponding \textit{non-zero} transition
energies. The presence of $\mathcal{H}_{1}$ usually prevents an exact solution
of the eigenvalue problem of the full Hamiltonian $\mathcal{H}$ so that suited
approximations are necessary.

\bigskip
The aim is to construct an effective 
Hamiltonian $\mathcal{H}_{\lambda}$ with a renormalized 'unperturbed'
part ${\cal H}_{0,\lambda}$ and a remaining 'perturbation' 
${\cal H}_{1,\lambda}$
\begin{eqnarray}
  \label{B1a}
  \mathcal{H}_{\lambda} &=& 
 \mathcal{H}_{0,\lambda} +   \mathcal{H}_{1,\lambda}
\end{eqnarray}
with  the following properties:
\begin{enumerate}
  \item[(i)] 
  The eigenvalue problem of the renormalized 
  Hamiltonian $\mathcal{H}_{0,\lambda}$ is diagonal
  \begin{eqnarray*}
    \mathcal{H}_{0,\lambda} | n^{\lambda} \rangle &=& 
    E_{n}^\lambda | n^\lambda \rangle
  \end{eqnarray*}
  with $\lambda$-dependent eigenvalues $E_{n}^{\lambda}$ and 
  eigenvectors $|n^{\lambda}\rangle$. 
  \item[(ii)]
  The effective Hamiltonian $\mathcal{H}_{\lambda}$ is
  constructed in such a way so that (measured with respect to 
  $\mathcal{H}_{0,\lambda}$) all non-diagonal contributions with transition 
  energies larger than some cutoff energy $\lambda$ vanish.
  \item[(iii)]
  $\mathcal{H}_{\lambda}$ has the same eigenvalues as 
  the original Hamiltonian ${\cal H}$.
\end{enumerate}

The eigenvalue problem of ${\cal H}_{0,\lambda}$ is crucial for the 
construction of ${\cal H}_\lambda$ because it can be used to define 
projection operators, 
\begin{eqnarray}
  \label{B1}
  \mathbf {P}_{\lambda} {\mathcal{A}} &=&
  \sum_{m,n} | n^{\lambda} \rangle \langle m^{\lambda} |
  \langle n^{\lambda} | \mathcal{A} | m^{\lambda}\rangle \\
  && \qquad\times
  \Theta(\lambda -|E_n^\lambda -E_m^\lambda |) \nonumber\\
  \label{B2}
  \mathbf{Q}_{\lambda} &=& \mathbf{1} - \mathbf{P}_{\lambda}.
\end{eqnarray}
Note that neither
$|n^{\lambda}\rangle$ nor $|m^{\lambda}\rangle$ need to be low- or high-energy
eigenstates of $\mathcal{H}_{0,\lambda}$. $\mathbf{P}_{\lambda}$ and
$\mathbf{Q}_{\lambda}$ are super-operators acting on operators 
${\mathcal{A}}$ of the Hilbert space of the system. Thus,
$\mathbf{P}_{\lambda}$ and $\mathbf{Q}_{\lambda}$ can be
interpreted as projection operators of the Liouville space that is built up by
all operators of the Hilbert space. $\mathbf{P}_{\lambda}$ projects on
those parts of an operator ${\mathcal{A}}$ which only consist of transition
operators 
$| n^{\lambda}\rangle \langle m^{\lambda} |$ with energy differences 
$| E_n^\lambda - E_m^\lambda| $ less than a given cutoff $\lambda$, whereas 
$\mathbf{Q}_{\lambda}$ projects onto the high-energy transitions of
${\mathcal{A}}$.

\smallskip
In terms of the projection operators $\mathbf{P}_\lambda$ and 
$\mathbf {Q}_\lambda$ the property of ${\cal H}_\lambda$ to allow no 
transitions between the eigenstates of ${\cal H}_{0,\lambda}$ with energies
larger than $\lambda$ reads
\begin{eqnarray}
  \label{B3}
  \mathcal{H}_{\lambda} &=& \mathbf{P}_{\lambda} 
  \mathcal{H}_{\lambda}
\quad \mbox{or} \quad
  \mathbf{Q}_{\lambda} \mathcal{H}_{\lambda} = 0.
 \end{eqnarray}

For an actual construction of the effective Hamiltonian we now assume that 
the effective Hamiltonian ${\cal H}_\lambda$ can be obtained from the 
original Hamiltonian by a unitary transformation,
\begin{eqnarray}
  \label{B3a}
  \mathcal{H}_{\lambda} &=& 
  e^{X_{\lambda}}\; \mathcal{H}\;  e^{-X_{\lambda}} \;.
\end{eqnarray}
which shall automatically guarantee that condition (iii) above is 
fulfilled.

In the following the evaluation of the effective Hamiltonian \eqref{B3a} is 
done in two ways: At first a perturbative treatment is derived. After that 
we develop a much more sophisticated renormalization where we interprete the 
unitary transformation of Eq.~\eqref{B3a} as a sequence of small 
transformations. The projector-based perturbation theory discussed in the next 
subsection is important for the understanding of the renormalization scheme 
derived later. However, the main focus of this review is the PRM.

\subsection{Perturbation theory}
\label{B_perturbation}

In the following we evaluate the effective Hamiltonian 
$\mathcal{H}_{\lambda}$ in perturbation theory. For this purpose the effective 
Hamiltonian $\mathcal{H}_{\lambda}$ from Eqs. \eqref{B3} and 
\eqref{B3a} is simplified in a crucial point: The projection operators are 
now defined with respect to the eigenvalue problem of the unperturbed part 
of the \textit{original} Hamiltonian ${\cal H}_0$, 
\begin{eqnarray*}
  \mathcal{H}_{0} | n^{} \rangle &=& E_{n}^{} | n^{} \rangle. 
\end{eqnarray*}
Thus, these projection operators differ from the formerly defined projectors 
$\mathbf{P}_{\lambda}$ and $\mathbf{Q}_{\lambda}$ and can be written as 
follows
\begin{eqnarray}
  \label{B1b}
  \mathbf{\bar{P}}_{\lambda}^{} {\mathcal{A}} &=&
  \sum_{m,n} | n^{} \rangle \langle m^{} |
  \langle n^{} | \mathcal{A} | m^{}\rangle \\[-1ex]
  &&\quad\times\Theta(\lambda -|E_n^{}-E_m^{}|), \nonumber \\
  \label{B2b}
  \mathbf{\bar{Q}}_{\lambda} &=& \mathbf{1} - \mathbf{\bar{P}}_{\lambda}.
\end{eqnarray} 
The renormalized Hamiltonian ${\cal H}_\lambda$ is now obtained from the 
unitary transformation \eqref{B3a},
\begin{eqnarray*}
  \mathcal{H}_{\lambda} &=& \mathbf{\bar{P}}_{\lambda}^{}
  \mathcal{H}_{\lambda}
  \, = \, e^{X_{\lambda}}\; \mathcal{H}\;  e^{-X_{\lambda}} \;,
\end{eqnarray*}
where $X_\lambda$ is the generator of this transformation. To find 
$X_{\lambda}$, we employ the modified condition \eqref{B3}:
All matrix elements of  $\mathcal{H}_{\lambda}$ for transitions 
with energies larger than $\lambda$ vanish, \textit{i.e.}
\begin{eqnarray}
  \label{B4}
  \mathbf{\bar{Q}}_{\lambda}^{} \mathcal{H}_{\lambda} &=& 0
\end{eqnarray}
First we expand ${\cal H}_\lambda$ 
with respect to $X_{\lambda}$,
\begin{eqnarray}
  \label{B5}
  \mathcal{H}_{\lambda} &=&
  \mathcal{H} + \left[ X_{\lambda}, \mathcal{H} \right] +
  \frac{1}{2!}
  \left[ X_{\lambda}, \left[ X_{\lambda}, \mathcal{H} \right] \right]
  \\
  && + \,
  \frac{1}{3!}
  \left[ X_{\lambda}, \left[ X_{\lambda}, 
    \left[ X_{\lambda}, \mathcal{H} \right]
  \right] \right] + \dots \; .
  \nonumber
\end{eqnarray}
and assume that 
the generator $X_{\lambda}$ can be written as a power series
in the interaction $\mathcal{H}_{1}$,
\begin{eqnarray}
  \label{B6}
  X_{\lambda} &=&
  X_{\lambda}^{(1)} + X_{\lambda}^{(2)} + X_{\lambda}^{(3)} +
  \dots \,.
\end{eqnarray}
Thus inserting \eqref{B6} in Eq.~\eqref{B5}, the effective Hamiltonian 
$\mathcal{H}_{\lambda}$ can be rewritten as a power series in the
interaction $\mathcal{H}_{1}$
\begin{eqnarray}
  \label{B7}
  \mathcal{H}_{\lambda} &=&
  \mathcal{H}_{0} + \mathcal{H}_{1} + 
  \left[ X_{\lambda}^{(1)}, \mathcal{H}_{0} \right] +
  \left[ X_{\lambda}^{(1)}, \mathcal{H}_{1} \right] \\
  && + \,
  \left[ X_{\lambda}^{(2)}, \mathcal{H}_{0} \right] +
  \frac{1}{2!} \left[
    X_{\lambda}^{(1)}, \left[ X_{\lambda}^{(1)}, \mathcal{H}_{0} \right]
  \right] + 
  {\cal O}(\mathcal{H}_{1}^{3}) .
  \nonumber
\end{eqnarray}

The contributions $X_{\lambda}^{(n)}$ to the generator of the unitary
transformation can successively be determined by employing Eq.~\eqref{B4}.
One finds
\begin{eqnarray}
  \label{B8}
  \mathbf{\bar{Q}}_{\lambda}^{} X_{\lambda}^{(1)} &=&
  \frac{1}{{\bf L}}_{0} 
  \left( \mathbf{\bar{Q}}_{\lambda}^{} \mathcal{H}_{1} \right) , \\
  \label{B9}
  \mathbf{\bar{Q}}_{\lambda}^{} X_{\lambda}^{(2)} &=& -\,
  \frac{1}{2 \mathbf{L}_{0} } \mathbf{\bar{Q}}_{\lambda}^{}
  \left[
    ( \mathbf{\bar{Q}}_{\lambda}^{}\mathcal{H}_{1} ), 
    \frac{1}{\mathbf{L}_{0}}
    ( \mathbf{\bar{Q}}_{\lambda}^{}\mathcal{H}_{1})
  \right] \phantom{aaa}\\
  && -\,
  \frac{1}{{\bf L}_{0}}
  \mathbf{\bar{Q}}_{\lambda}^{}
  \left[
    (\mathbf{\bar{P}}_{\lambda}^{}\mathcal{H}_{1}), 
    \frac{1}{ \mathbf{L}_{0} }
    (\mathbf{\bar{Q}}_{\lambda}^{}\mathcal{H}_{1})
  \right] .
  \nonumber
\end{eqnarray}
Here, ${\bf L}_{0}$ is the Liouville operator of the unperturbed
Hamiltonian $\mathcal{H}_{0}$ which is defined by 
$\mathbf{L}_{0}\mathcal{A} = [\mathcal{H}_{0}, \mathcal{A}]$
for any operator variable $\mathcal{A}$. 

As one can see from \eqref{B8} and \eqref{B9}, no 
information about the low-energy part
$\mathbf{\bar{P}}_{\lambda}^{}X_{\lambda}$ of the generator $X_{\lambda}$
can be deduced from \eqref{B4}. Therefore, we set for simplicity
\begin{eqnarray}
  \label{B10}
  \mathbf{\bar{P}}_{\lambda}^{}X_{\lambda} &=& 
  \mathbf{\bar{P}}_{\lambda}^{}X_{\lambda}^{(1)}
  \,=\, \mathbf{\bar{P}}_{\lambda}^{}X_{\lambda}^{(2)} \, = \cdots = \, 0.
  \phantom{aaa}
\end{eqnarray} 

Inserting Eqs.~\eqref{B8}, \eqref{B9}, and \eqref{B10} into the power series
\eqref{B7} for $\mathcal{H}_{\lambda}$, 
the desired perturbation theory is found,
\begin{eqnarray}
  {\cal H}_{\lambda} &=&
  {\cal H}_{0} + \mathbf{\bar{P}}_{\lambda}^{}{\cal H}_{1} -
  \frac{1}{2} \mathbf{\bar{P}}_{\lambda}^{}
  \left[
    ( \mathbf{\bar{Q}}_{\lambda}^{}{\cal H}_{1}), \frac{1}{\mathbf{L}_{0}}
    ( \mathbf{\bar{Q}}_{\lambda}^{}{\cal H}_{1})
  \right] \nonumber\\
  \label{B11}
  && -\,
  \mathbf{\bar{P}}_{\lambda}^{}
  \left[
    (\mathbf{\bar{P}}_{\lambda}^{}{\cal H}_{1}), \frac{1}{ \mathbf{L}_{0} }
    (\mathbf{\bar{Q}}_{\lambda}^{}{\cal H}_{1})
  \right] + \mathcal{O}(\mathcal{H}_{1}^{3}),
\end{eqnarray}
which can easily be extended to higher order terms. Note that the correct size
dependence of the Hamiltonian is automatically guaranteed 
by the commutators in Eq.~\eqref{B11}. The limit $\lambda \rightarrow 0$ is
of particular interest because in this case the complete interaction
$\mathcal{H}_{1}$ is integrated out.

Usual perturbation theory derives effective Hamiltonians that are only valid
for a certain range of the system's Hilbert space. In contrast, 
${\cal H}_{\lambda}$, as derived above, has no limitations with respect
to the Hilbert space so that it can also be used to study
excited states. To illustrate this important aspect of our projector-based
perturbation theory, we discuss the dimerized and frustrated spin
chain in the appendix.

\bigskip
At this point we would like to note that 
Eq.~\eqref{B11} can also be derived in a different way. 
It turns out that $X_{\lambda}^{(2)}$ is only needed to
fulfill the requirement 
${\cal H}_{\lambda} = \mathbf{\bar{P}}_{\lambda}^{}{\cal H}_{\lambda}$ if
we restrict ourselves to second order perturbation theory. Thus, in this case
$X_{\lambda}^{(2)}$ can be set to $0$ if the projector
$\mathbf{\bar{P}}_{\lambda}^{}$ is applied to the right hand side of
Eq.~\eqref{B7}, 
\begin{eqnarray}
\label{B7a}
  \mathcal{H}_{\lambda} &=&
  \mathcal{H}_{0} + \mathbf{\bar{P}}_{\lambda}^{}\mathcal{H}_{1} + 
  \mathbf{\bar{P}}_{\lambda}^{}
  \left[ X_{\lambda}^{(1)}, \mathcal{H}_{0} \right] \\
  &+&
  \mathbf{\bar{P}}_{\lambda}^{}
  \left[ X_{\lambda}^{(1)}, \mathcal{H}_{1} \right] +
  \frac{1}{2!} \mathbf{\bar{P}}_{\lambda}^{} \left[
    X_{\lambda}^{(1)}, \left[ X_{\lambda}^{(1)}, \mathcal{H}_{0} \right]
  \right]  + \cdots  .
  \nonumber
\end{eqnarray}
It is easy to proof that Eq.~\eqref{B7a} again leads to the result 
Eq.~\eqref{B11} if \eqref{B10}  and \eqref{B8}
is used.

\bigskip
In Appendix \ref{spin_chain}, the developed perturbation theory \eqref{B11} 
is applied to the dimerized and frustrated spin chain where ground-state 
energy and triplet dispersion relation have been calculated.

\bigskip
A perturbation theory based on Wegner's flow equations \cite{W_1994}, that
also allows a description of the complete Hilbert space, has been derived in
Refs.~\onlinecite{S_1997} and \onlinecite{KU_2000}. However, this approach
requires an equidistant spectrum of the unperturbed Hamiltonian 
$\mathcal{H}_{0}$. In contrast, the perturbation theory presented here can be
applied to systems with arbitrary Hilbert space, and has similarities to a
cumulant approach to effective Hamiltonians \cite{HVB_1999}.

\subsection{Stepwise renormalization}

In the previous subsection the effective Hamiltonian $\mathcal{H}_{\lambda}$ 
as defined by Eqs. \eqref{B3} and \eqref{B3a} has been evaluated within a 
new kind of perturbation theory. However, if the unitary transformation 
\eqref{B3a} is interpreted as a sequence of unitary transformation a 
renormalization scheme can be developed based on the same definition of 
the effective Hamiltonian. Because again the projection operators 
$\mathbf{P}_{\lambda}$ and $\mathbf{Q}_{\lambda}$ play a key role 
we call the derived method \cite{BHS_2002} projector-based renormalization 
method (PRM).

\bigskip
Let us start from a renormalized Hamiltonian 
$\mathcal{H}_{\lambda} = \mathcal{H}_{0,\lambda} + \mathcal{H}_{1,\lambda}$
that has been obtained after all transitions with energy differences larger 
than $\lambda$ have already been integrated out. Of course,
 ${\cal H}_{0,\lambda}$ and ${\cal H}_{1,\lambda}$ will differ from the 
original ${\cal H}_0$ and ${\cal H}_1$. Furthermore, we assume 
$\mathcal{H}_{\lambda}$ has the properties (i)-(iii) proposed in subsection
\ref{basic_concepts}. 

Now we want to eliminate all excitations within the
energy range between $\lambda$ and a smaller new energy cutoff 
$\lambda-\Delta\lambda$. Thereby we use a unitary transformation,  
\begin{eqnarray}
  \label{B16}
  \mathcal{H}_{(\lambda-\Delta\lambda)} &=& 
  e^{X_{\lambda, \Delta\lambda}}\; \mathcal{H}_{\lambda}\;  
  e^{-X_{\lambda, \Delta\lambda}} \;,
\end{eqnarray}
so that the effective Hamiltonian $\mathcal{H}_{\lambda - \Delta \lambda}$ has
the same eigenspectrum as the Hamiltonian $\mathcal{H}_\lambda$. Note that 
the generator $X_{\lambda,\Delta \lambda}$ needs to be chosen anti-Hermitian, 
$X_{\lambda,\Delta \lambda} = -X^{\dagger}_{\lambda, \Delta \lambda}$, 
to ensure that $\mathcal{H}_{\lambda - \Delta \lambda}$
is Hermitian when ${\cal H}_\lambda$ was Hermitian before. 
To find an appropriate generator $X_{\lambda,\Delta \lambda}$
of the unitary transformation, we employ the condition that
$\mathcal{H}_{\lambda}$ has (with respect to 
$\mathcal{H}_{0,\lambda}$) only vanishing matrix 
elements for transitions with energies larger than $\lambda$, \textit{i.e.}
$\mathbf{Q}_{\lambda}^{} \mathcal{H}_{\lambda} = 0$.
Similarly, also
\begin{eqnarray}
  \label{Bed4}
   \mathbf{Q}_{(\lambda- \Delta \lambda)} 
\mathcal{H}_{(\lambda- \Delta \lambda)} &=& 0 
\end{eqnarray}
must be fulfilled, where $\mathbf{Q}_{(\lambda -\Delta \lambda)}$ is now 
defined
with respect to the excitations of ${\cal H}_{0,(\lambda -\Delta \lambda)}$.

\bigskip
In principle, there are two strategies to 
evaluate Eqs.~\eqref{B16} and \eqref{Bed4}:
The first uses perturbation theory as derived in subsection
\ref{B_perturbation}. In this case 
$\mathcal{H}_{(\lambda-\Delta\lambda)}$
can be written as
\begin{eqnarray}
  \label{B17a}
  \lefteqn{\mathcal{H}_{(\lambda-\Delta\lambda)} \,=\,} && \\
  &=&
  \mathcal{H}_{0,\lambda} + 
  \mathbf{P}_{(\lambda-\Delta\lambda)}\mathcal{H}_{1,\lambda} + 
  \mathbf{P}_{(\lambda-\Delta\lambda)}
  \left[ X_{\lambda,\Delta\lambda}, \mathcal{H}_{0,\lambda} \right] 
  \nonumber\\
  && +\,
  \mathbf{P}_{(\lambda-\Delta\lambda)}
  \left[ X_{\lambda,\Delta\lambda}, \mathcal{H}_{1,\lambda} \right] 
  \nonumber\\
  && + \,
  \frac{1}{2} \mathbf{P}_{(\lambda-\Delta\lambda)} \left[
    X_{\lambda,\Delta\lambda}, 
    \left[ X_{\lambda,\Delta\lambda}, \mathcal{H}_{0,\lambda} \right]
  \right] + 
  {\cal O}(\mathcal{H}_{1,\lambda}^{3}) .
  \nonumber
\end{eqnarray}
The generator $X_{\lambda,\Delta\lambda}$ has to be chosen corresponding to
Eq.~\eqref{B8},
\begin{eqnarray}
  \label{B17b}
  \mathbf{Q}_{(\lambda-\Delta\lambda)} X_{\lambda,\Delta\lambda} &=& 
  \frac{1}{\mathbf{L}_{0,\lambda}}
  \left[
    \mathbf{Q}_{(\lambda-\Delta\lambda)} \mathcal{H}_{1,\lambda}
  \right] + \cdots \hspace*{0.2cm}.
\end{eqnarray}
For details of the derivation we refer to subsection \ref{B_perturbation}.
This approach has been successfully applied to the electron-phonon interaction
to describe superconductivity \cite{HB_2003}. 

Alternatively, one can also start from an appropriate ansatz for the
generator in order to calculate $\mathcal{H}_{(\lambda-\Delta\lambda)}$ in a
non-perturbative manner \cite{HB_2005}. An ansatz for the generator 
with the same operator structure as Eq.~\eqref{B17b} is often a very good
choice. This approach has been applied to the periodic Anderson
model to describe heavy-fermion behavior \cite{HB_2005,HB_2006}.

\bigskip
It turns out that the second strategy has the great advantage to successfully
prevent diverging renormalization contributions. However, in both cases,
Eqs.~\eqref{B16} and \eqref{Bed4} describe a renormalization step that lowers
the energy cutoff of the effective Hamiltonian from $\lambda$ to
$\lambda-\Delta\lambda$. Consequently, difference equations for the
Hamiltonian $\mathcal{H}_{\lambda}$ can be derived, and the resulting
equations for the $\lambda$ dependence of the parameters of the Hamiltonian
are called renormalization equations. By starting from the original 
model ${\cal H} =:{\cal H}_{\lambda =\Lambda}$ the Hamiltonian is
renormalized by reducing the cutoff $\lambda$ in steps $\Delta \lambda$.  
The limit $\lambda \rightarrow 0$
provides the desired effective Hamiltonian ${\cal H}_{\lambda =0} =:
\tilde{\cal H} $ without any
interaction. Note 
that the results strongly depend on
the parameters of the original Hamiltonian $\mathcal{H}$.

\subsection{
  Generator of the unitary transformation and further approximations
}
\label{GenUT}

It turns out that the generator $X_{\lambda,\Delta\lambda}$ of the unitary
transformation is not yet completely determined by Eqs. \eqref{B16} and
\eqref{Bed4}. Instead, the low-energetic
excitations included in $X_{\lambda,\Delta\lambda}$, namely the part 
$\mathbf{P}_{(\lambda-\Delta\lambda)}X_{\lambda,\Delta\lambda}$, can be chosen
arbitrarily. The result of the renormalization scheme should not depend on
the particular choice of
$\mathbf{P}_{(\lambda-\Delta\lambda)}X_{\lambda,\Delta\lambda}$ as long as all
renormalization steps are performed without approximations. However,
approximations will be necessary for practically all interacting systems of
interest so the choice 
$\mathbf{P}_{(\lambda-\Delta\lambda)}X_{\lambda,\Delta\lambda}$ becomes
relevant. If $\mathbf{P}_{(\lambda-\Delta\lambda)}X_{\lambda,\Delta\lambda}=0$
is chosen the minimal transformation is performed to match the requirement
\eqref{Bed4}. Such an approach of "minimal" transformations avoid errors caused
by approximations necessary for every renormalization step as much as
possible. Note that in order to derive the expression \eqref{B17a} 
this choice of 
$\mathbf{P}_{(\lambda-\Delta\lambda)}X_{\lambda,\Delta\lambda}$
was used.
However, in particular cases a non-zero choice for 
$\mathbf{P}_{(\lambda-\Delta\lambda)}X_{\lambda,\Delta\lambda}$ might help to
circumvent problems in the evaluation of the renormalization equations.

\bigskip
In general, new interaction terms can be generated in every renormalization
step. This might allow the investigation of competing interactions which
naturally emerge within the renormalization procedure. However, actual
calculations require a closed set of renormalization equations. Thus, often a
factorization approximation has to be performed in order to trace back
complicated operators to terms already appearing in the renormalization
ansatz. Consequently, derived effective Hamiltonians might be limited in
their possible applications if important operators have not been appropriately
included in the renormalization scheme.

If a factorization approximation needs to be performed the obtained
renormalization equations will contain expectation values that must be
calculated separately. In principle, these expectation values are defined with
respect to $\mathcal{H}_{\lambda}$ because the factorization approximation was
employed for the renormalization step that transformed $\mathcal{H}_{\lambda}$
to $\mathcal{H}_{(\lambda-\Delta\lambda)}$. However, $\mathcal{H}_{\lambda}$
still contains interactions that prevent a straight evaluation of required
expectation values. The easiest way to circumvent this difficulty is to
neglect the interactions and to use the diagonal unperturbed part
$\mathcal{H}_{0,\lambda}$ instead of $\mathcal{H}_{\lambda}$ for the
calculation of the expectation values. This approach has been successfully
applied to the Holstein model to investigate single-particle excitations and
phonon softening \cite{SHBWF_2005}. However, it turns out that often the
interaction term in $\mathcal{H}_{\lambda}$ is crucial for a proper
calculation of the required expectation values. Thus, usually a more involved
approximation has been used that neglects the $\lambda$ dependence of the
expectation values but includes interaction effects by calculating the
expectation values with respect to the full Hamiltonian $\mathcal{H}$ instead
of $\mathcal{H}_{\lambda}$. In this case, the renormalization equations need
to be solved in a self-consistent manner because they depend on expectation
values defined with respect to the full Hamiltonian $\mathcal{H}$ which are
not known from the very beginning but can be determined from the fully
renormalized (and diagonal) Hamiltonian  
$\tilde{\mathcal{H}}= \lim_{\lambda \rightarrow 0} \mathcal{H}_{\lambda}$.

There exist two ways to calculate expectation values of the full Hamiltonian
from the renormalized Hamiltonian. The first one is based on the free energy
that can be calculated either from the original model $\mathcal{H}$ or the
renormalized Hamiltonian $\tilde{\mathcal{H}}$,
\begin{eqnarray*}
  F &=& 
  - \frac{1}{\beta} \mathrm{ln} \, \mathrm{Tr} \, e^{-\beta\mathcal{H}}
  \,=\,
  - \frac{1}{\beta} \mathrm{ln} \, \mathrm{Tr} \, 
e^{-\beta\tilde{\mathcal{H}}},
\end{eqnarray*}
because $\tilde{\mathcal{H}}$ is obtained from $\mathcal{H}$ by unitary
transformations. The desired expectation values can then be determined from
the free energy by functional derivatives. This approach has advantages as
long as the derivatives can be evaluated analytically as, for example, in
Refs.~\onlinecite{HB_2003} and \onlinecite{HB_2005}. 

The second way to calculate expectation values of the full Hamiltonian
employs unitarity for any operator variable $\mathcal{A}$,
\begin{eqnarray*}
\label{B17c}
  \langle \mathcal{A} \rangle &=&
  \frac{\mathrm{Tr}\left( \mathcal{A} \, e^{-\beta\mathcal{H}} \right)}
  {\mathrm{Tr} \,e^{-\beta\mathcal{H}}} \,=\,
  \frac{
    \mathrm{Tr}\left( \tilde{\mathcal{A}} 
\,e^{-\beta\tilde{\mathcal{H}}} \right)
  }
  { \mathrm{Tr} \,e^{-\beta\tilde{\mathcal{H}}} },
\end{eqnarray*}
where we defined 
$\tilde{\mathcal{A}} = \lim_{\lambda \rightarrow 0}
\mathcal{A}_{\lambda}$. Thus, additional renormalization equations need to be
derived for the required operator variables $\mathcal{A}_{\lambda}$ where 
the same sequence of unitary transformations has to be applied to the 
operator variable $\mathcal{A}$ as to the Hamiltonian $\mathcal{H}$.

\subsection{Example: Fano-Anderson model}
\label{B_Fano-Anderson}

In this subsection we want to illustrate the PRM discussed above by
considering an exactly solvable model, namely the Fano-Anderson model
\cite{A_1961, F_1961},
\begin{eqnarray}
  \label{B18}
  {\cal H} &=& {\cal H}_{0} + {\cal H}_{1},\\
  {\cal H}_{0} &=&
  \sum_{{\bf k},m}
  \left(
    \varepsilon_{f} \, f^{\dagger}_{{\bf k}m} f_{{\bf k}m} +
    \varepsilon_{{\bf k}} \, c^{\dagger}_{{\bf k}m} c_{{\bf k}m}
  \right) , \nonumber \\
  {\cal H}_{1} &=&
  \sum_{{\bf k},m} V_{{\bf k}}
  \left(
    f_{{\bf k}m}^{\dagger}c_{{\bf k}m} + c_{{\bf k}m}^{\dagger}f_{{\bf k}m}
  \right) . \nonumber
\end{eqnarray}
The Hamiltonian \eqref{B18} describes dispersion-less $f$ electrons
interacting with conduction electrons where all correlation effects are
neglected. $\mathbf{k}$ denotes the wave vector, and the one-particle energies
are measured with respect to the chemical potential. Both types of electrons
are assumed to have the same orbital index $m$ with values
$1,\dots,\nu_{f}$. The model \eqref{B18} is easily diagonalized,
\begin{eqnarray}
  \label{B19}
  {\cal H} &=&
  \sum_{{\bf k},m} \omega_{{\bf k}}^{(\alpha)}
  \alpha_{{\bf k}m}^{\dagger} \alpha_{{\bf k}m} +
  \sum_{{\bf k},m} \omega_{{\bf k}}^{(\beta)}
  \beta_{{\bf k}m}^{\dagger} \beta_{{\bf k}m},
\end{eqnarray}
where $\alpha_{{\bf k}m}^{\dagger}$ and $\beta_{{\bf k}m}^{\dagger}$ are given
by linear combinations of the original fermionic operators 
$c_{{\bf k}m}^{\dagger}$ and $f_{{\bf k}m}^{\dagger}$,
\begin{eqnarray}
  \label{B20}
  \alpha_{{\bf k}m}^{\dagger} & = &
  u_{{\bf k}} \, f_{{\bf k}m}^{\dagger} + v_{{\bf k}} \,
  c_{{\bf k}m}^{\dagger}, \\
  \label{B21}
  \beta_{{\bf k}m}^{\dagger} & = &
  -v_{{\bf k}} \, f_{{\bf k}m}^{\dagger} + u_{{\bf k}} \,
  c_{{\bf k}m}^{\dagger},
\end{eqnarray}
\begin{eqnarray*}
  |u_{{\bf k}}|^{2} & = &
  \frac{1}{2}
  \left( 1 - \frac{\varepsilon_{{\bf k}}-\varepsilon_{f}}
  {W_{\mathbf{k}}} \right),
  \\
  |v_{{\bf k}}|^{2} & = &
  \frac{1}{2}
  \left( 1 + \frac{\varepsilon_{{\bf k}}-\varepsilon_{f}}
{W_{\mathbf{k}}} \right).
\end{eqnarray*}
Here, we defined 
$
  W_{\mathbf{k}} = 
  \sqrt{
    \left( \varepsilon_{{\bf k}}-\varepsilon_{f} \right)^{2} +
    4 |V_{{\bf k}}|^{2}
  }
$, and the eigenvalues of $\mathcal{H}$ are given by
\begin{eqnarray}
  \label{B22}
  \omega_{{\bf k}}^{(\alpha,\beta)} &=&
  \frac{\varepsilon_{{\bf k}} + \varepsilon_f}{2}
  \pm \frac{1}{2} W_{\mathbf{k}}.
\end{eqnarray}

\bigskip
In the following, we want to apply the PRM as introduced above to the
Fano-Anderson model \eqref{B18} where we mainly use the formulation of
Ref.~\cite{HB_2005}. The goal is to integrate out the hybridization term 
$\mathcal{H}_{1}$ so that we finally obtain an effectively free
model. Therefore, having in mind the exact solution of the model, we make the
following renormalization ansatz:
\begin{eqnarray}
  \label{B23}
  {\cal H}_{\lambda} &=& 
  {\cal H}_{0,\lambda} + {\cal H}_{1,\lambda}, \\
  {\cal H}_{0, \lambda} &=&
   \sum_{{\bf k},m}
  \left(
    \varepsilon_{{\bf k},\lambda}^{f} \, f^{\dagger}_{{\bf k}m} f_{{\bf k}m} +
    \varepsilon_{{\bf k},\lambda}^{c} \, c^{\dagger}_{{\bf k}m} c_{{\bf k}m}
  \right) ,
  \nonumber \\
 {\cal H}_{1,\lambda}
  &=&
  \sum_{{\bf k},m}
  V_{{\bf k},\lambda}
  \left(
    f_{{\bf k}m}^{\dagger}c_{{\bf k}m} + c_{{\bf k}m}^{\dagger}f_{{\bf k}m}
  \right),
 \nonumber
\end{eqnarray}
Note that $V_{\mathbf{k},\lambda}$ includes a cutoff function in
order to ensure that the requirement
$\mathbf{Q}_{\lambda}\mathcal{H}_{\lambda}=0$
is fulfilled.

In the next step we want to eliminate excitations with energies within the
energy shell between $\lambda$ and $\lambda - \Delta\lambda$ by means of an
unitary transformation similar to \eqref{B16}. By inspecting the perturbation
expansion corresponding to subsection \ref{B_perturbation}, the generator of
the unitary transformation must have the following form:
\begin{eqnarray}
  X_{\lambda,\Delta\lambda} &=&
  \sum_{{\bf k},m}
  A_{\mathbf{k}}(\lambda,\Delta\lambda)
  \left(
    f_{\mathbf{k}m}^{\dagger} c_{\mathbf{k}m} - 
    c_{\mathbf{k}m}^{\dagger} f_{\mathbf{k}m}
  \right), \nonumber\\[-1ex]
  \label{B24}
\end{eqnarray}
where the parameters $A_{\mathbf{k}}(\lambda,\Delta\lambda)$ need to be
properly determined so that Eq.~\eqref{Bed4} is fulfilled. To evaluate the
transformation \eqref{B16}, we now consider the transformations of the
operators appearing in the renormalization ansatz \eqref{B23}. For example, we
obtain
\begin{eqnarray*}
  \lefteqn{
    e^{X_{\lambda,\Delta\lambda}} \, 
    c^{\dagger}_{\mathbf{k}m}c_{\mathbf{k}m} \,
    e^{-X_{\lambda,\Delta\lambda}} - c^{\dagger}_{\mathbf{k}m}c_{\mathbf{k}m}
    \,=\,
  }&& \\
  &=&
  \frac{1}{2}
  \left\{\cos\left[ 2 A_{\mathbf{k}}(\lambda,\Delta\lambda)\right] - 1 \right\}
  \left( 
    c^{\dagger}_{\mathbf{k}m}c_{\mathbf{k}m} - 
    f^{\dagger}_{\mathbf{k}m}f_{\mathbf{k}m} 
  \right) \\
  && +\,
  \sin\left[ 2 A_{\mathbf{k}}(\lambda,\Delta\lambda)\right]
  \left(
    f^{\dagger}_{\mathbf{k}m}c_{\mathbf{k}m} + 
    c^{\dagger}_{\mathbf{k}m}f_{\mathbf{k}m} 
  \right).
\end{eqnarray*}
Here it is important to notice that due to the fermionic anti-commutator
relations the different $\mathbf{k}$ are not coupled with each other. Very
similar transformations can also be found for
$f^{\dagger}_{\mathbf{k}m}f_{\mathbf{k}m}$ and 
$\left(
  f^{\dagger}_{\mathbf{k}m}c_{\mathbf{k}m} + 
  c^{\dagger}_{\mathbf{k}m}f_{\mathbf{k}m} 
\right)$. Inserting these transformations into \eqref{B16} leads to the
following renormalization equations:
\begin{eqnarray}
  \label{B25}
  \lefteqn{
    \varepsilon^{f}_{\mathbf{k},(\lambda-\Delta\lambda)} - 
    \varepsilon^{f}_{\mathbf{k},\lambda} \,=\,
  }&&\\
  &=&
  -\frac{1}{2}
  \left\{
    \cos\left[ 2 A_{\mathbf{k}}(\lambda,\Delta\lambda)\right] - 1
  \right\}
  \left( 
    \varepsilon^{c}_{\mathbf{k},\lambda} -
    \varepsilon^{f}_{\mathbf{k},\lambda}
  \right) \nonumber \\
  && + \,
  V_{\mathbf{k},\lambda} \sin\left[2 A_{\mathbf{k}}
(\lambda,\Delta\lambda)\right],
  \nonumber
\end{eqnarray}
\begin{eqnarray}
  \label{B26}
  \varepsilon^{c}_{\mathbf{k},(\lambda-\Delta\lambda)} - 
  \varepsilon^{c}_{\mathbf{k},\lambda} 
  & =&
  -\, \left(
    \varepsilon^{f}_{\mathbf{k},(\lambda-\Delta\lambda)} - 
    \varepsilon^{f}_{\mathbf{k},\lambda}
  \right).
\end{eqnarray}

Now we need to determine the parameters
$A_{\mathbf{k}}(\lambda,\Delta\lambda)$. For this purpose we employ the
condition \eqref{Bed4}: First, from 
$\mathbf{Q}_{\lambda}\mathcal{H}_{\lambda} = 0$
we conclude $V_{\mathbf{k},\lambda} = 
\Theta_{\mathbf{k},\lambda} V_{\mathbf{k}}$, where we have defined 
$
  \Theta_{\mathbf{k}\lambda} = 
  \Theta\left(
    \lambda - 
    |\varepsilon^{f}_{\mathbf{k},\lambda} -
    \varepsilon^{c}_{\mathbf{k},\lambda}| 
  \right)
$. Moreover, from  
$\mathbf{Q}_{(\lambda-\Delta\lambda)}\mathcal{H}_{(\lambda-\Delta\lambda)}=0$
we find
\begin{eqnarray}
  \label{B27}
  \lefteqn{
    \tan\left[ 2 A_{\mathbf{k}}(\lambda,\Delta\lambda) \right] \,=\,
  }&&\\
  &=&
    \left[1 - \Theta_{\mathbf{k}(\lambda- \Delta \lambda )} \right] 
 \Theta_{\mathbf{k}\lambda} \,
  \frac{2V_{\mathbf{k},\lambda}}
  {\varepsilon^{f}_{\mathbf{k},\lambda} -
    \varepsilon^{c}_{\mathbf{k},\lambda}}
  \nonumber
\end{eqnarray}
which shows that also $A_{\bf k}(\lambda, \Delta \lambda)$ contains the 
cutoff factor $\Theta_{{\bf k},\lambda}$.
Note that in the expression \eqref{B27} the low excitation-energy part
of the generator was chosen to be zero 
$\mathbf{P}_{(\lambda-\Delta\lambda)}X_{\lambda,\Delta\lambda}=0$.
As one can see from Eqs.~\eqref{B25}-\eqref{B27}, the renormalization of the
parameters of a given $\mathbf{k}$ is \textit{not} affected by other
$\mathbf{k}$ values. Furthermore, it is important to notice that  
$
  |\varepsilon^{f}_{\mathbf{k},\lambda} -
  \varepsilon^{c}_{\mathbf{k},\lambda}| \le
  |\varepsilon^{f}_{\mathbf{k},(\lambda-\Delta\lambda)} -
  \varepsilon^{c}_{\mathbf{k},\lambda-\Delta\lambda}|
$.
Consequently, each $\mathbf{k}$ value is renormalized only once during the
renormalization procedure eliminating excitations from large to small
$\lambda$ values. Such a steplike renormalization allows an easy solution of
the renormalization equations \eqref{B25}-\eqref{B27} where $\lambda$ is
replaced by the cutoff $\Lambda$ of the original model and we set
$\lambda-\Delta\lambda=0$. 
Here, one needs to consider that the parameter 
$A_{\mathbf{k}}$ changes its sign if the difference 
$\varepsilon_{f} - \varepsilon_{\mathbf{k}}$ changes its sign. Thus, we find
the following renormalized Hamiltonian
\begin{eqnarray}
  \tilde{\mathcal{H}} &:=& 
  \lim_{\lambda\rightarrow 0} \mathcal{H}_{\lambda} \,=\,
  \sum_{\mathbf{k},m} \left(
    \tilde{\varepsilon}_{\mathbf{k}}^{f}
    f^{\dagger}_{\mathbf{k}m} f_{\mathbf{k}m} + 
    \tilde{\varepsilon}_{\mathbf{k}}^{c}
    c^{\dagger}_{\mathbf{k}m} c_{\mathbf{k}m}
  \right), \nonumber\\[-1ex]
  \label{B28}
\end{eqnarray}
where the renormalized energies are given by
\begin{eqnarray}
  \label{B29}
  \tilde{\varepsilon}_{\mathbf{k}}^{f} &=&
  \frac{\varepsilon_{f} +\varepsilon_{\mathbf{k}}}{2} +
  \frac{
    \mathrm{sgn}( \varepsilon_{f} -\varepsilon_{\mathbf{k}} )
  }{2}
  W_{\mathbf{k}} , \\[2ex]
  \label{B30}
  \tilde{\varepsilon}_{\mathbf{k}}^{c} &=&
  \frac{\varepsilon_{f} +\varepsilon_{\mathbf{k}}}{2} -
  \frac{
    \mathrm{sgn}( \varepsilon_{f} -\varepsilon_{\mathbf{k}} )
  }{2}
  W_{\mathbf{k}}.
\end{eqnarray}

\bigskip
The results of the renormalization and the diagonalization are completely
comparable for physical accessible quantities like quasiparticle energies
[compare \eqref{B22} with Eqs. \eqref{B29} and \eqref{B30}] or expectation
values. However, there is also an important difference between the two
approaches: Whereas the eigenmodes $\alpha_{\mathbf{k}m}^{\dagger}$ and
$\beta_{\mathbf{k}m}^{\dagger}$ of the diagonalized Hamiltonian \eqref{B19}
change there character as function of the wave vector $\mathbf{k}$ [compare
\eqref{B20} and \eqref{B21}], the operators $f^{\dagger}_{\mathbf{k}m}$ and
$c^{\dagger}_{\mathbf{k}m}$ of $\tilde{\mathcal{H}}$ remain
$f$-like and $c$-like for all $\mathbf{k}$ values. In return, the
quasi-particle energies $\tilde{\varepsilon}_{\mathbf{k}}^{f}$ and
$\tilde{\varepsilon}_{\mathbf{k}}^{c}$ show a steplike behavior as function of
$\mathbf{k}$ at $\varepsilon_{f} - \varepsilon_{\mathbf{k}} = 0$ so that 
the deviations from the original one-particle energies $\varepsilon_{f}$ and
$\varepsilon_{\mathbf{k}}$ remain relatively small for all $\mathbf{k}$ values.

\subsection{Generalized generator of the unitary transformation}
\label{B_generalize_G}

As already mentioned in subsection \ref{GenUT}, the low-energetic excitations 
included in the generator $X_{\lambda, \Delta \lambda}$ of the unitary 
transformation \eqref{B16} can be chosen arbitrarily, i.e. 
${\mathbf P}_{(\lambda - \Delta \lambda)} X_{\lambda, \Delta \lambda}$ is not 
determined by the condition \eqref{Bed4}. 

In the previous subsection an approach of ``minimal'' transformations has 
been applied to the Fano-Anderson model where  
${\mathbf P}_{(\lambda - \Delta \lambda)} X_{\lambda, \Delta \lambda}$ is 
set to zero. However, in the following we want to demonstrate that it is also 
possible to take advantage of this freedom to choose the generator  
$X_{\lambda, \Delta \lambda}$ and to derive a continuous version of the PRM. 
As it will turn out in Sec.~\ref{EP_flow} the PRM can also be connected to 
Wegner's flow equation method \cite{W_1994}.

\bigskip 
By allowing a nonzero part 
${\mathbf P}_{(\lambda - \Delta \lambda)} X_{\lambda, \Delta \lambda} \neq 0$ 
the generator $X_{\lambda, \Delta \lambda}$ of the unitary transformation 
\eqref{B16} can be written as follows 
\begin{eqnarray}
  \label{B30.aa}
  X_{\lambda, \Delta \lambda} &=&
  {\mathbf P}_{(\lambda - \Delta \lambda)} X_{\lambda, \Delta \lambda} + 
  {\mathbf Q}_{(\lambda - \Delta \lambda)} X_{\lambda, \Delta \lambda} \ \ 
\end{eqnarray} 
Here the part 
${\mathbf Q}_{(\lambda - \Delta \lambda)} X_{\lambda, \Delta \lambda}$
ensures that Eq. \eqref{Bed4}, 
$
  \mathbf{Q}_{(\lambda-\Delta\lambda)} 
  \mathcal{H}_{(\lambda-\Delta\lambda)} = 0
$,
is fulfilled. Note however, one may also choose the remaining part 
${\mathbf P}_{(\lambda - \Delta \lambda)} X_{\lambda, \Delta \lambda}$
in such a way that it almost completely 
integrates out all the interactions \textit{before} the cutoff 
energy $\lambda$ approaches their corresponding transition energies. 

\bigskip
As it will be discussed in Sec.~\ref{EP} in more detail, the flow equation 
method \cite{W_1994} and the PRM (in its minimal form) take advantage of the 
freedom to chose the generator of the unitary transformation in a very 
different way. In the PRM, the low transition-energy projection part of the 
generator, ${\bf P}_{\lambda} X_\lambda$, is set to zero for convenience. The 
flow equation approach instead uses exactly this part to eliminate the 
interaction. 

Even though the PRM resembles the similarity transformation 
\cite{GW_1993, GW_1994} and Wegner's flow equation method \cite{W_1994} in 
some aspects there is an important difference: The latter two methods 
start from \textit{continuous} transformations in differential form. This has 
the advantage that one can use available computer subroutines to solve the 
differential flow equations. In contrast,  the PRM is based on 
\textit{discrete} transformations which lead to coupled difference equations. 
The advantage of the PRM is to provides a direct link to 
perturbation theory (as already discussed in subsection \ref{B_perturbation}). 
Moreover, the stepwise renormalization of the PRM allows a unified treatment 
on both sides of a quantum phase transition (see for example 
Sec.~\ref{Holstein_QP}) which seems not to be possible in the flow equation 
method. However, as we show in the following the idea of continuous unitary 
transformations can also be implemented in the framework of the PRM.

\subsection{Fano-Anderson model revisited}
\label{FA_rev}

Now we want to demonstrate that the freedom in choosing the generator of the 
unitary transformation can be employed in order to derive a continuous 
renormalization scheme within the framework of the PRM. As an example we again 
discuss the Fano-Anderson model. 

\bigskip 
As already discussed, the part 
$\mathbf{P}_{(\lambda-\Delta\lambda)} X_{\lambda,\Delta\lambda}$
of the generator $X_{\lambda, \Delta \lambda}$ 
of the unitary transformation is not fixed by the
PRM. In the former treatment of the Fano-Anderson model in subsection 
\ref{B_Fano-Anderson}  we had chosen 
$\mathbf{P}_{(\lambda-\Delta\lambda)} X_{\lambda,\Delta\lambda}=0$
for simplicity. In the following we want to take advantage of this
freedom in a different way. 

According to Eq. \eqref{B24}, the generator of the Fano-Anderson model 
is given by  
\begin{eqnarray*}
  X_{\lambda,\Delta\lambda} &=&
  \sum_{{\bf k},m}
  A_{\mathbf{k}}(\lambda,\Delta\lambda)
  \left(
    f_{\mathbf{k}m}^{\dagger} c_{\mathbf{k}m} - 
    c_{\mathbf{k}m}^{\dagger} f_{\mathbf{k}m}
  \right) 
\end{eqnarray*}
where the most general form of 
$A_{\mathbf{k}}(\lambda,\Delta\lambda)$ can be written as
\begin{eqnarray}
\label{B30a}
  A_{\mathbf{k}}(\lambda,\Delta\lambda) &=&
  A'_{\mathbf{k}}(\lambda,\Delta\lambda) \,
  \Theta_{\mathbf{k}, \lambda}
  \left[ 1 - \Theta_{\mathbf{k}, \lambda - \Delta\lambda} \right]
  \nonumber \\
  && +
  A''_{\mathbf{k}}(\lambda,\Delta\lambda) \,
  \Theta_{\mathbf{k}, \lambda} \Theta_{\mathbf{k}, \lambda - \Delta\lambda}.
\end{eqnarray} 
Here, the renormalization contributions related with 
$\mathbf{P}_{(\lambda-\Delta\lambda)} X_{\lambda,\Delta\lambda}$ and 
$\mathbf{Q}_{(\lambda-\Delta\lambda)} X_{\lambda,\Delta\lambda}$ are described
by the parameters $A''_{\mathbf{k}}(\lambda,\Delta\lambda)$ and 
$A'_{\mathbf{k}}(\lambda,\Delta\lambda)$, respectively.

\bigskip
A possible choice for $A''_{\mathbf{k}}(\lambda,\Delta\lambda)$ is
\begin{eqnarray}
\label{B30ba}
A''_{\mathbf{k}}(\lambda,\Delta\lambda) &=&  
  \frac{
    \left(
        \varepsilon_{{\mathbf k},\lambda}^f  - 
        \varepsilon_{\mathbf{k},\lambda}^c
    \right)
    V_{\mathbf{k},\lambda}
  }{
    \kappa
    \left[
      \lambda - \left|
        \varepsilon_{{\mathbf k},\lambda}^f - 
        \varepsilon_{\mathbf{k},\lambda}^c
      \right|
    \right]^{2}
  }
  \, \Delta\lambda . 
\end{eqnarray}
Of course, there is no derivation for Eq.~\eqref{B30ba} but it will turn out 
that this is indeed a reasonable choice. In particular we will show that 
in the limit of small $\Delta\lambda$ a rapid decay for the
hybridization $V_{\mathbf{k},\lambda}$ is obtained in this way. Thus, the part 
$A'_{\mathbf k}(\lambda, \Delta \lambda)$ of the generator is not important 
anymore for the renormalization procedure and can be neglected in the 
following. In Eq.~\eqref{B30ba}, $\kappa$ denotes an energy constant to 
ensure a dimensionless $A''_{\mathbf{k}}(\lambda,\Delta\lambda)$. Note that
$A''_{\mathbf{k}}(\lambda,\Delta\lambda)$ is chosen proportional to
$\Delta\lambda$ to reduce the impact of the actual value of 
$\Delta\lambda$ on the final results of the renormalization.

\bigskip
In order to derive continuous renormalization equations 
note that the parameter $A_{\mathbf k}(\lambda, \Delta \lambda)$ is
approximately proportional to $\Delta \lambda$. By neglecting the part 
$A'_{\mathbf k}(\lambda, \Delta \lambda)$ of the generator 
one can rewrite Eqs. \eqref{B25} and \eqref{B26} in the limit 
$\Delta \lambda \rightarrow 0$   
\begin{eqnarray}
\label{B30b}
\frac{d\varepsilon_{{\mathbf k},\lambda}^f}{d\lambda} &=& 
- 2 V_{{\mathbf k},\lambda} \alpha_{\mathbf k}(\lambda) \\
\frac{\varepsilon_{{\mathbf k}, \lambda}^c}{d\lambda} &=& 
+ 2 V_{{\mathbf k},\lambda} \alpha_{\mathbf k}(\lambda) \label{B30c}
\end{eqnarray} 
where higher order terms have been neglected. Furthermore, we defined 
\begin{eqnarray}
  \label{B30d}
  \alpha_{\mathbf k}(\lambda) &=& \lim_{\Delta \lambda \rightarrow 0}
  \frac{A_{\mathbf k}''(\lambda, \Delta \lambda)}{\Delta \lambda}, \\
  &=&
  \frac{
    \left(
        \varepsilon_{{\mathbf k},\lambda}^f  - 
        \varepsilon_{\mathbf{k},\lambda}^c
    \right)
    V_{\mathbf{k},\lambda}
  }{
    \kappa
    \left[
      \lambda - \left|
        \varepsilon_{{\mathbf k},\lambda}^f - 
        \varepsilon_{\mathbf{k},\lambda}^c
      \right|
    \right]^{2}
  } \nonumber .
\end{eqnarray} 
A similar equation can also be derived for $V_{{\mathbf k}, \lambda}$,
\begin{eqnarray}
  \label{B30e}
  \frac{d V_{{\mathbf k},\lambda}}{d\lambda}
  &=& (\varepsilon_{{\mathbf k},\lambda}^f - 
  \varepsilon_{{\mathbf k},\lambda}^c)\, 
  \alpha_{{\mathbf k}, \lambda}.
\end{eqnarray} 
To solve these equations we rewrite \eqref{B30e}, 
\begin{eqnarray}
  \label{B30f}
  \alpha_{{\mathbf k}, \lambda} &=& 
  \frac{1} {\varepsilon_{{\mathbf k},\lambda}^f - 
  \varepsilon_{{\mathbf k},\lambda}^c} 
  \frac{d V_{{\mathbf k},\lambda}}{d\lambda} ,
\end{eqnarray} 
and insert into \eqref{B30d}. Using 
$
  \varepsilon_{{\mathbf k}, \lambda}^f
  + \varepsilon_{{\mathbf k}, \lambda}^c = \varepsilon_{{\mathbf k}}^f
  + \varepsilon_{{\mathbf k}}^c
$ we obtain
\begin{eqnarray}
  \label{B30g}
  0 &=& \frac{d}{d\lambda} 
  \left\{
    (\varepsilon_{{\mathbf k}, \lambda}^c)^2  -
    (\varepsilon_{{\mathbf k}}^f +\varepsilon_{{\mathbf k}}^c)
    \varepsilon_{{\mathbf k}, \lambda}^c + V_{{\mathbf k}, \lambda}^2
  \right\}.
\end{eqnarray} 
Eq.~\eqref{B30g} is easily integrated and leads to 
a quadratic equation for 
$\tilde{\varepsilon}_{{\mathbf k}}^c = \lim_{\lambda \rightarrow 0}
\varepsilon_{{\mathbf k}, \lambda}^c$
which corresponds to the former result \eqref{B30}. Moreover, 
$\tilde{\varepsilon}_{{\mathbf k}}^f$ is found from 
$\varepsilon_{{\mathbf k}, \lambda}^f
+ \varepsilon_{{\mathbf k}, \lambda}^c = \varepsilon_{{\mathbf k}}^f
+ \varepsilon_{{\mathbf k}}^c$. According to \eqref{B30e} and 
\eqref{B30d} the $\lambda$-dependence of 
 $V_{{\mathbf k}, \lambda}$ is governed by
\begin{eqnarray}
  \frac{d \ln V_{{\mathbf k},\lambda}}{d\lambda}
  &=& \frac{(\varepsilon_{{\mathbf k},\lambda}^f - 
  \varepsilon_{{\mathbf k},\lambda}^c)^2}
  {\kappa [\lambda - |\varepsilon_{{\mathbf k}, \lambda}^f -
  \varepsilon_{{\mathbf k}, \lambda}^c|]^2   } \,
  \Theta(\lambda -|\varepsilon^f_{{\mathbf k},\lambda }
  -\varepsilon^c_{{\mathbf k},\lambda }| ) \nonumber \\
  \label{B30e2}
  &&
\end{eqnarray} 
As one can easily see from Eq. \eqref{B30e2}, 
\begin{enumerate}
  \item[(i)]
  the interaction $V_{{\mathbf k},\lambda}$ is always renormalized  
  to smaller values when the cutoff energy $\lambda$ is lowered,
  \item[(ii)]
  and at 
  $
    \lambda = 
    \left|
      \varepsilon_{{\mathbf k}, \lambda}^f - 
      \varepsilon_{{\mathbf k}, \lambda}^c
    \right|
  $
  the renormalized coupling $V_{{\mathbf k},\lambda}$ vanishes, i.e.~it 
  has completely integrated out by the present choice of the generator 
  ${\bf P}_{\lambda -\Delta \lambda}X_{\lambda - \Delta \lambda}$.
\end{enumerate}


\section{
  Renormalization of the electron-phonon interaction
}
\label{EP}

The classical BCS-theory \cite{BCS} is essentially based on attractive
electron-electron interactions \cite{C_1956}. It is well-known that such an
interaction can be mediated via phonons coupled to the electronic system
\cite{F_1952}. In this section we want to revisit this problem because it has
been studied \cite{LW_1996, M_1997, HB_2003} by Wegner's flow equation method
\cite{W_1994}, by a similarity transformation proposed by G{\l}azek and Wilson
\cite{GW_1993, GW_1994}, and by the PRM \cite{BHS_2002}. Therefore, the
electron-phonon interaction is a perfectly suited test case to discuss
differences and similarities of the three methods. In this section we consider
the following Hamiltonian
\begin{eqnarray}
  \label{EP1}
  {\cal H} &=&
  \sum_{{\bf k},\sigma} \varepsilon_{\bf k} \,
    c_{{\bf k}\sigma}^{\dagger}c_{{\bf k}\sigma}  +
  \sum_{\bf q} \omega_{\bf q} \,
    b_{\bf q}^{\dagger}b_{\bf q} \\
  && + \,
  \sum_{{\bf k},{\bf q},\sigma}
  g_{\bf q}\left[
    b_{\bf q}^{\dagger} c_{{\bf k}\sigma}^{\dagger} c_{({\bf k}+{\bf q})\sigma}
     +
    b_{\bf q} c_{({\bf k}+{\bf q})\sigma}^{\dagger} c_{{\bf k}\sigma}
  \right]
  \nonumber
\end{eqnarray}
which describes electrons $c_{{\bf k},\sigma}^{\dagger}$ and phonons 
$b_{\bf q}^{\dagger}$ that interact with each other.

\bigskip
In the following we apply a slightly modified version of the PRM to the
electron-phonon problem \eqref{EP1} in order to derive an effective
electron-electron interaction. It turns out that Fr\"{o}hlich's transformation 
\cite{F_1952} is re-examined in this way. 

In \ref{EP_flow} the approach is modified in the spirit of the ideas 
developed in  subsections \ref{B_generalize_G} and \ref{FA_rev}. Thus, 
allowing a more continuous renormalization of the electron-phonon 
interaction we derive the result of Ref.~\onlinecite{LW_1996}
obtained by the flow equation method. 

In subsection \ref{EP_BCS} a much more sophisticated scheme is introduced by 
adding a symmetry breaking field to the Hamiltonian so that a gap equation 
can be derived. The effective electron-electron interaction is then obtained 
by comparing with the famous BCS-gap equation. The strategy to introduce 
symmetry breaking fields turns out to be of general importance for the 
investigation of phase transitions within the PRM. 

Finally, the different results for the electron-phonon interaction \eqref{EP1} 
are discussed in subsection \ref{EP_disc}.

\subsection{Fr\"{o}hlich's transformation}
\label{EP_Froehlich}

In this subsection we want to apply the PRM to the electron-phonon problem
\eqref{EP1} in order to derive an effective electron-electron
interaction. Here, we start from the renormalization ansatz,
\begin{eqnarray}
  \label{EP2}
  \mathcal{H}_{\lambda} &=& 
  \mathcal{H}_{0} + \mathcal{H}_{1,\lambda}, \\[1ex]
  \mathcal{H}_{0} &=&
  \sum_{\mathbf{k},\sigma} \varepsilon_{\mathbf{k}} \,
  c_{\mathbf{k}\sigma}^{\dagger}c_{\mathbf{k}\sigma} +   
  \sum_{\mathbf{q}} \omega_{\mathbf{q}} \,
  b_{\mathbf{q}}^{\dagger}b_{\mathbf{q}}
  , \nonumber
\end{eqnarray}
\begin{eqnarray}
  \mathcal{H}_{1,\lambda}&=&
  \mathcal{H}_{1,\lambda}^{\mathrm{el,ph}} + 
  \mathcal{H}_{1,\lambda}^{\mathrm{el,el}} , \nonumber \\[1ex]
  \mathcal{H}_{1,\lambda}^{\mathrm{el,ph}} &=&
  \sum_{\mathbf{k}, \mathbf{q}, \sigma}\left[
    g_{\mathbf{k}, \mathbf{q},\lambda} \,
    b_{-\mathbf{q}}^{\dagger} 
  \right. \nonumber\\
  && \qquad
  \left. \phantom{b_{-\mathbf{q}}^{\dagger}}+\,
    g_{\mathbf{k}+\mathbf{q}, -\mathbf{q}, \lambda} \, b_{\mathbf{q}}
  \right] 
  c_{(\mathbf{k}+\mathbf{q})\sigma}^{\dagger}c_{\mathbf{k}\sigma},
  \nonumber \\[1ex]
  \mathcal{H}_{1,\lambda}^{\mathrm{el,el}} &=&
  \sum_{\mathbf{k}, \sigma, \mathbf{k'}, \sigma', \mathbf{q}}
  V_{\mathbf{k}, \mathbf{k'}, \mathbf{q}, \lambda} \,
  c_{(\mathbf{k}+\mathbf{q})\sigma}^{\dagger}
  c_{(\mathbf{k'}-\mathbf{q})\sigma'}^{\dagger}
  c_{\mathbf{k'}\sigma'}c_{\mathbf{k}\sigma},
  \nonumber
\end{eqnarray}
that was also used in Ref.~\onlinecite{LW_1996} where the flow equation
method was applied to the same system. Note that the parameters 
of $\mathcal{H}_{1,\lambda}$ contain a cutoff function in
order to ensure that only transitions with energies smaller
than $\lambda$ are included. The parameters of
$\mathcal{H}_{\lambda}$ depend on the energy cutoff $\lambda$ because all
transitions with energies larger than $\lambda$ have already been integrated
out. However, we shall restrict ourselves to the second order
renormalization contributions to $\mathcal{H}_{1,\lambda}$. Therefore, 
$\mathcal{H}_{0}$ is assumed to be $\lambda$ independent.

\bigskip
In the following we want to integrate out all transitions which create or
annihilate phonons, however keeping all electronic transitions. 
Therefore, the present calculation differs from
the previous ones where all parts of the 
'unperturbed Hamiltonian' ${\cal H}_{0, \lambda}$ were subject to the
renormalization procedure. As it turns out, the electron-phonon 
coupling will be replaced by an effective elec\-tron-electron interaction. 
However, the
final Hamiltonian containing the electron-electron interaction is not diagonal
any more as required for the standard PRM. 
Instead, we want to derive a block-diagonal
Hamiltonian so that the renormalization approach has to be modified. For this
purpose, we define projection operators $\mathbf{P}_{\lambda}^{\mathrm{ph}}$
and $\mathbf{Q}_{\lambda}^{\mathrm{ph}}$ that are defined with respect to the
phonon part of the unperturbed Hamiltonian $\mathcal{H}_{0}$. These new
projectors now replace those of the full unperturbed Hamiltonian. 

Thus, from
$\mathbf{Q}_\lambda^{\mathrm{ph}}{\cal H}_{1,\lambda} =0$
we conclude $g_{{\bf k}, {\bf q},\lambda}= \Theta_{\mathbf{q},\lambda}\,
g_{{\bf k}, {\bf q},\lambda}$, where we have defined
$\Theta_{\mathbf{q},\lambda} = \Theta(\lambda - \omega_{\bf q})$. Moreover, 
following Ref.~\onlinecite{LW_1996}, the generated electron-electron
interaction $\mathcal{H}_{1,\lambda}^{\mathrm{el,el}}$ is not considered
in determining the generator of the unitary transformation
\eqref{B16}. Thus, the generator can be written as
\begin{eqnarray}
  \label{EP3b}
  \lefteqn{X_{\lambda, \Delta\lambda} \,=\,}&&\\
  &=&
  \sum_{\mathbf{k},\mathbf{q},\sigma}
    A_{\mathbf{k},\mathbf{q}}(\lambda, \Delta\lambda) 
\left[
    \, b_{\mathbf{q}}^{\dagger}   
       c_{\mathbf{k}\sigma}^\dag c_{(\mathbf{k}+\mathbf{q})\sigma}
- 
      b_{\mathbf{q}} c_{(\mathbf{k}+\mathbf{q})\sigma}^\dag
      c_{\mathbf{k}\sigma}
\, \right]
  \nonumber
\end{eqnarray}
where the parameter $A_{\mathbf{k},\mathbf{q}}(\lambda, \Delta\lambda)$
needs to be
properly determined in the following: Corresponding to \eqref{Bed4}, 
\begin{eqnarray}
  \label{EP4}
  \mathbf{Q}_{(\lambda-\Delta\lambda)}^{\mathrm{ph}} 
  \mathcal{H}_{(\lambda-\Delta\lambda)} &=& 0
\end{eqnarray}
must be fulfilled. 

As already discussed, the part 
$\mathbf{P}_{(\lambda-\Delta\lambda)}^{\mathrm{ph}}X_{\lambda,\Delta\lambda}$
of the generator \eqref{EP3b} of the unitary transformation is not fixed by the
PRM. Thus, the parameters $A_{\mathbf{k},\mathbf{q}}(\lambda, \Delta \lambda)$
have the following general form
\begin{eqnarray}
  \label{EP3a}
  A_{\mathbf{k},\mathbf{q}}(\lambda,\Delta\lambda) &=&
  A'_{\mathbf{k},\mathbf{q}}(\lambda,\Delta\lambda) \,
  \Theta_{\mathbf{q}, \lambda}
  \left[ 1 - \Theta_{\mathbf{q}, \lambda - \Delta\lambda} \right]
  \nonumber \\
  && + \,
  A''_{\mathbf{k},\mathbf{q}}(\lambda,\Delta\lambda) \,
  \Theta_{\mathbf{q}, \lambda}
  \Theta_{\mathbf{q}, \lambda - \Delta\lambda}.
\end{eqnarray} 
Note that both parts of $A_{\mathbf{k},\mathbf{q}}(\lambda,\Delta\lambda)$ 
include the factor $\Theta_{\mathbf{q}, \lambda}$. However, in the following 
${\mathbf P}_{(\lambda - \Delta \lambda)} X_{\lambda, \Delta \lambda}$  and 
$ A''_{\mathbf{k},\mathbf{q}}(\lambda,\Delta\lambda)$ are set to zero for 
simplicity. Note that a different choices for 
$A''_{\mathbf{k},\mathbf{q}}(\lambda,\Delta\lambda)$ will be used in the 
subsequent subsection.

We restrict ourselves to second order renormalization contributions so that 
the unitary transformation \eqref{B16} can easily be evaluated where operator
terms are only kept if they are included in the ansatz \eqref{EP2}. Thus, we
directly obtain difference equation for the electron-phonon coupling,
\begin{eqnarray}
  \label{EP5}
    \lefteqn{g_{\mathbf{k},\mathbf{q},\lambda-\Delta\lambda} -
    g_{\mathbf{k},\mathbf{q},\lambda} \,=\,} \\
   &=&
  -\left[
    \varepsilon_{\mathbf{k}+\mathbf{q}} - \varepsilon_{\mathbf{k}} +
    \omega_{\mathbf{q}}
  \right] \, A_{\mathbf{k}+\mathbf{q},-\mathbf{q}}(\lambda, \Delta\lambda), 
   \nonumber
 \end{eqnarray}
and for the effective electron-electron interaction,
\begin{eqnarray}
  \label{EP7}
  \lefteqn{
    V_{\mathbf{k}, \mathbf{k'}, \mathbf{q}, \lambda-\Delta\lambda} - 
    V_{\mathbf{k}, \mathbf{k'}, \mathbf{q}, \lambda}
    \,=\,
  } && \\
  &&=\,-\,
  A_{\mathbf{k'}-\mathbf{q},\mathbf{q}}(\lambda, \Delta\lambda)
  g_{\mathbf{k}+\mathbf{q},-\mathbf{q},\lambda} \nonumber\\
  && -\,
  A_{\mathbf{k'} ,-\mathbf{q}}(\lambda, \Delta\lambda)
  g_{\mathbf{k},\mathbf{q},\lambda} \nonumber \\
  && -\,
  \frac{1}{2} 
  \left(
    \varepsilon_{\mathbf{k}+\mathbf{q}} - \varepsilon_{\mathbf{k}} -
    \omega_{\mathbf{q}}
  \right)
  A_{\mathbf{k'}-\mathbf{q},\mathbf{q}}(\lambda, \Delta\lambda) \,
  A_{\mathbf{k},\mathbf{q}}(\lambda, \Delta\lambda) \nonumber\\
  && + \,
  \frac{1}{2} 
  \left(
    \varepsilon_{\mathbf{k}+\mathbf{q}} - \varepsilon_{\mathbf{k}} +
    \omega_{\mathbf{q}}
  \right)
  A_{\mathbf{k}+\mathbf{q},-\mathbf{q}}(\lambda, \Delta\lambda) \,
 A_{\mathbf{k'},-\mathbf{q}}(\lambda, \Delta\lambda).
\nonumber
\end{eqnarray}
Because we have set 
${\mathbf P}_{(\lambda - \Delta \lambda)} X_{\lambda, \Delta \lambda} = 0$, 
renormalization contributions only appear if the phonon energy
$\omega_{\mathbf{q}}$ is in the energy shell between $(\lambda-\Delta\lambda)$ 
and $\lambda$. Consequently, we find a step-like renormalization of the
electron-phonon coupling $g_{\mathbf{k},\mathbf{q},\lambda}$ and the generated
electron-electron interaction
$V_{\mathbf{k},\mathbf{k'},\mathbf{q},\lambda}$. The parameter 
$A_{\mathbf{k},\mathbf{q}}(\lambda, \Delta\lambda)$ 
defined in \eqref{EP3a} has to be chosen in such a
way that $g_{\mathbf{k},\mathbf{q},\lambda - \Delta \lambda} = 
\Theta_{{\bf q},\lambda - \Delta \lambda}
g_{\mathbf{k},\mathbf{q},\lambda - \Delta \lambda}$. 
From
equation \eqref{EP5} we obtain
\begin{eqnarray}
  \label{EP8}
  A_{\mathbf{k},\mathbf{q}}(\lambda, \Delta\lambda) &=&
  \frac{g_{\mathbf{q}}}{
    \varepsilon_{\mathbf{k}} - \varepsilon_{\mathbf{k}+\mathbf{q}} +
    \omega_{\mathbf{q}} } \, \Theta_{{\bf q},\lambda} \,
  [1-  \Theta_{{\bf q},\lambda -\Delta \lambda}]. \nonumber \\
  && 
  \end{eqnarray}
As one can see by inserting Eq.~\eqref{EP8} into \eqref{EP5},  the
electron-phonon coupling has no $k$-dependence in the present approximation, 
i.e. $g_{\mathbf{k},\mathbf{q},\lambda} = g_{\mathbf{q},\lambda}$.

Now we insert Eq.~\eqref{EP8} into the renormalization
equation \eqref{EP7} and consider the limit $\lambda\rightarrow 0$,
\begin{eqnarray}
  \tilde{V}_{\mathbf{k}, \mathbf{k'}, \mathbf{q}} &=&
  \lim_{\lambda\rightarrow 0}
  V_{\mathbf{k}, \mathbf{k'}, \mathbf{q}, \lambda} \,=\,
  \frac{\omega_{\mathbf{q}} \left| g_{\mathbf{q}} \right|^{2}}
  {
    \left( 
      \varepsilon_{\mathbf{k}+\mathbf{q}} - \varepsilon_{\mathbf{k}}
    \right)^{2} - 
    \omega_{\mathbf{q}}^{2}
  },\nonumber\\
  \label{EP10}
  &&
\end{eqnarray}
where we exactly find Fr\"{o}hlich's result \cite{F_1952}.

\subsection{Continuous transformation}
\label{EP_flow}

Wegner's flow equation method \cite{W_1994} was applied to the
electron-phonon system \eqref{EP1} in Ref.~\onlinecite{LW_1996} where a
renormalization ansatz similar to \eqref{EP2} was used. However, a
less singular expression for the effective electron-electron interaction could
be derived in this way. In the following we want to analyze how this different
result can be understood in the framework of the PRM. 

In order to derive continuous renormalization equations the part 
$\mathbf{P}_{(\lambda-\Delta\lambda)}^{\mathrm{ph}}X_{\lambda,\Delta\lambda}$
of the generator of the unitary transformation is chosen to be non-zero so 
that now $A''_{\mathbf{k},\mathbf{q}}(\lambda,\Delta\lambda)$ needs to be 
considered in Eq.~\eqref{EP3a}. Furthermore, 
$A'_{\mathbf{k},\mathbf{q}}(\lambda,\Delta\lambda)$ can be neglected if 
$A''_{\mathbf{k},\mathbf{q}}(\lambda,\Delta\lambda)$ leads to a rapid decay 
of the interaction terms. Thus, neglecting 
$A'_{\mathbf{k},\mathbf{q}}(\lambda,\Delta\lambda)$ and employing the limit 
$\Delta\lambda\rightarrow 0$ we obtain from Eqs.~\eqref{EP5} and \eqref{EP7}
\begin{eqnarray}
  \label{EP11}
  \frac{\mathrm{d}}{\mathrm{d}\lambda} g_{\mathbf{k},\mathbf{q},\lambda} &=&
  \left[
    \varepsilon_{\mathbf{k}+\mathbf{q}} - \varepsilon_{\mathbf{k}} +
    \omega_{\mathbf{q}}
  \right] \,
  {\alpha}_{\mathbf{k},\mathbf{q},\lambda},\\
  \label{EP13}
  \frac{\mathrm{d}}{\mathrm{d}\lambda}
  V_{\mathbf{k}, \mathbf{k'}, \mathbf{q}, \lambda} &=&
  g_{\mathbf{k}+\mathbf{q},-\mathbf{q},\lambda} \,
  {\alpha}_{\mathbf{k'},-\mathbf{q},\lambda} \\
  && +\,
  g_{\mathbf{k},\mathbf{q},\lambda} \,
  {\alpha}_{\mathbf{k'} + \mathbf{q},\mathbf{q},\lambda}.
  \nonumber
\end{eqnarray}
Here, we introduced
$
{\alpha}_{\mathbf{k},\mathbf{q},\lambda} = 
\lim_{\Delta\lambda\rightarrow 0} 
A''_{\mathbf{k},\mathbf{q}}(\lambda, \Delta\lambda) / \Delta\lambda
$. Again the parameter $A''_{\mathbf{k},\mathbf{q}}(\lambda, \Delta\lambda)$ 
is  chosen proportional to
$\Delta\lambda$ so that the third and the fourth term on the right side of
Eq.~\eqref{EP7} can be neglected in the limit $\Delta\lambda\rightarrow 0$. 

\bigskip
The commonly used generator of the flow equation method is chosen in such a
way that the matrix elements of the interaction, which shall be integrated
out, show an exponential decay with respect to the flow
parameter. Consequently, \textit{all} matrix elements change continuously
during the renormalization procedure. We adapt the
idea of such a continuous renormalization and assume an exponential decay for
the electron-phonon 
interaction,
\begin{eqnarray}
  g_{\mathbf{k},\mathbf{q},\lambda} &=& 
  g_{\mathbf{q}}
  \exp\left\{ 
    -\frac{
      \left(
        \varepsilon_{\mathbf{k}+\mathbf{q}} - \varepsilon_{\mathbf{k}} +
        \omega_{\mathbf{q}}
      \right)^{2}
    }{\kappa\left(\lambda-\omega_{\mathbf{q}}\right)}
  \right\} \,
  \Theta(\lambda - \omega_{\mathbf{q}}), \nonumber \\
  \label{EP14}
  &&
\end{eqnarray}
where $\kappa$ is just a constant to ensure a dimensionless exponent. 
Note that ansatz \eqref{EP14} is inspired by the results of
Ref.~\onlinecite{LW_1996}. Of course, Eq.~\eqref{EP14} is only useful as long
as the considered renormalization contributions are restricted to second order
in the original electron-phonon interaction. Note also that ansatz
\eqref{EP14} meets the basic requirement \eqref{Bed4} of the
PRM, 
$
  \mathbf{Q}_{(\lambda-\Delta\lambda)} \mathcal{H}_{(\lambda-\Delta\lambda)} 
  = 0
$.

\bigskip
Now we need to determine the parameter 
${\alpha}_{\mathbf{k},\mathbf{q},\lambda}$ 
of the unitary transformation. For this purpose, Eq.~\eqref{EP11} is 
divided by $g_{\mathbf{k},\mathbf{q},\lambda}$ and 
integrated between the cutoff
$\lambda>\omega_{\mathbf{q}}$ and $\infty$ by using Eq.~\eqref{EP14}.
We find
\begin{eqnarray}
  \label{EP15}
  {\alpha}_{\mathbf{k},\mathbf{q},\lambda} &=&
  \frac{
    g_{\mathbf{k},\mathbf{q},\lambda} \,
    \left[
      \varepsilon_{\mathbf{k}+\mathbf{q}} - \varepsilon_{\mathbf{k}} +
      \omega_{\mathbf{q}}
    \right]
  }
  {\kappa \left( \lambda - \omega_{\mathbf{q}}\right)^{2}}.
 \end{eqnarray}
Note that this result is equivalent to the choice for 
$A''_{\mathbf{k},\mathbf{q}}(\lambda,\Delta\lambda)$ used for the Fano-Anderson
model in \ref{FA_rev} [compare with equations \eqref{B30ba} and \eqref{B30d}].

Using this solution and the ansatz \eqref{EP14} for the electron-phonon
coupling $g_{\mathbf{k},\mathbf{q},\lambda}$, Eq.~\eqref{EP13} is easily
integrated where the constant $\kappa$ is canceled. Thus, the renormalized
values 
$
  \tilde{V}_{\mathbf{k}, \mathbf{k'}, \mathbf{q}} = 
  \lim_{\lambda\rightarrow 0} V_{\mathbf{k}, \mathbf{k'}, \mathbf{q}, \lambda}
$
can be obtained and reads
\begin{eqnarray}
  \label{EP17}
  \lefteqn{\tilde{V}_{\mathbf{k}, \mathbf{k'}, \mathbf{q}} \,=\,}&& \\
  &=& 
  \frac{
    \left| g_{\mathbf{q}} \right|^{2}
    \left(
      \varepsilon_{\mathbf{k'}-\mathbf{q}} - \varepsilon_{\mathbf{k'}} -
      \omega_{\mathbf{q}}
    \right)
  }{
    \left(
      \varepsilon_{\mathbf{k}+\mathbf{q}} - \varepsilon_{\mathbf{k'}} +
      \omega_{\mathbf{q}}
    \right)^{2} +
    \left(
      \varepsilon_{\mathbf{k'}-\mathbf{q}} - \varepsilon_{\mathbf{k'}} -
      \omega_{\mathbf{q}}
    \right)^{2} 
  }
  \nonumber\\
  && -\,
  \frac{
    \left| g_{\mathbf{q}} \right|^{2}
    \left(
      \varepsilon_{\mathbf{k'}-\mathbf{q}} - \varepsilon_{\mathbf{k'}} +
      \omega_{\mathbf{q}}
    \right)
  }{
    \left(
      \varepsilon_{\mathbf{k}+\mathbf{q}} - \varepsilon_{\mathbf{k'}} -
      \omega_{\mathbf{q}}
    \right)^{2} +
    \left(
      \varepsilon_{\mathbf{k'}-\mathbf{q}} - \varepsilon_{\mathbf{k'}} +
      \omega_{\mathbf{q}}
    \right)^{2} 
  }.
  \nonumber
\end{eqnarray}
This is the final version of the effective electron-electron interaction after
eliminating the electron-phonon interaction. Obviously, \eqref{EP17} differs
from Fr\"{o}hlich's result \cite{F_1952} that had been derived above
\eqref{EP10}. However, Eq.~\eqref{EP17} coincides with the result of
Ref.~\onlinecite{LW_1996} that had been obtained by Wegner's flow equation
method \cite{W_1994}.

At this point it is important to notice that the approaches of
\ref{EP_Froehlich} and \ref{EP_flow} are based on the same renormalization
ansatz \eqref{EP2}. Therefore, the different results are only caused by 
different choices for the generator. Due to the continuous renormalization, the
electron-phonon coupling becomes dependent on the electronic one-particle
energies $\varepsilon_{\mathbf{k}}$ so that the approach of \ref{EP_flow}
involves more degrees of freedom.

The main goal of this subsection was to demonstrate that Wegner's flow equation
method \cite{W_1994} can be understood within the PRM\cite{BHS_2002}, as 
already for the case of the Fano-Anderson model in the previous section. 
However, the idea of a continuous renormalization, as implemented here, can 
also be very useful for other applications. In this regards, the 
discussion line needs to be changed: One starts from an ansatz for the 
generator $X_{\lambda,\Delta\lambda}$ of the unitary transformation similar to
Eqs.~\eqref{EP3b}, \eqref{EP15}, and demonstrates \textit{afterwards} that 
the interaction decays as function of $\lambda$ as required.

\subsection{Improved renormalization scheme and BCS-gap equation}
\label{EP_BCS}

So far the discussion of the electron-phonon problem was focused on the
phonon-induced electron-electron interaction. Thus, we derived block-diagonal
Hamiltonians with constant phonon occupation numbers within each
block. However, in the following we want to tackle the electron-phonon problem
\eqref{EP1} in a different way because an effective phonon mediated
electron-electron interaction is mainly discussed with respect to
superconductivity. The idea is to obtain the superconducting properties
directly from the electron-phonon system.

The goal is again to decouple the electron and the phonon system but now we
want to derive a truly diagonal renormalized Hamiltonian. For this purpose the
PRM shall be applied to the electron-phonon system \eqref{EP1} in conjunction
with a Bogoliubov transformation \cite{B_1958} as it was done in
Ref.~\onlinecite{HB_2003}. 

Whereas the Hamiltonian \eqref{EP1} is gauge invariant, a BCS-like Hamiltonian
breaks this symmetry \cite{BCS}. Therefore, in order to describe
superconducting properties, the renormalized Hamiltonian should contain a
symmetry breaking field as well so that the renormalization ansatz reads
\begin{eqnarray}
  \label{EP18}
  \mathcal{H}_{\lambda} &=& 
  \mathcal{H}_{0,\lambda} + \mathcal{H}_{1,\lambda}, \\[1ex]
  \mathcal{H}_{0,\lambda} &=&
  \sum_{\mathbf{k},\sigma} \varepsilon_{\mathbf{k}} \,
  c_{\mathbf{k}\sigma}^{\dagger}c_{\mathbf{k}\sigma} +   
  \sum_{\mathbf{q}} \omega_{\mathbf{q}} \,
  b_{\mathbf{q}}^{\dagger}b_{\mathbf{q}} \nonumber\\
  && -\,
  \sum_{\bf k}
  \left(
    \Delta_{{\bf k},\lambda} \,
    c_{{\bf k}\uparrow}^{\dagger} c_{-{\bf k}\downarrow}^{\dagger} +
    \Delta_{{\bf k},\lambda}^{*} \,
    c_{-{\bf k}\downarrow} c_{{\bf k}\uparrow}
  \right) +
  C_{\lambda}, \nonumber\\[1ex]
  {\cal H}_{1,\lambda} &=&
  {\bf P}_{\lambda}
  \sum_{{\bf k},{\bf q},\sigma}
  g_{\bf q} \left[
    c_{{\bf k}\sigma}^{\dagger} c_{({\bf k}+{\bf q})\sigma}
    b_{\bf q}^{\dagger} +
    c_{({\bf k}+{\bf q})\sigma}^{\dagger} c_{{\bf k}\sigma}
    b_{\bf q}
  \right]. \nonumber
\end{eqnarray}
Here, the 'fields' $\Delta_{{\bf k},\lambda}$ and 
$\Delta_{{\bf k},\lambda}^{*}$  break the gauge invariance and can be
interpreted as the superconducting gap function. The initial values for
$\Delta_{{\bf k},\lambda}$ and the energy shift $C_{\lambda}$ are given by
those of the original model, $\Delta_{\mathbf{k},\Lambda}=0$,
$C_{\Lambda}=0$. Note that in the following the projectors 
$\mathbf{P}_{\lambda}$ and $\mathbf{Q}_{\lambda}$ are defined as usual with
respect to $\mathcal{H}_{0,\lambda}$ and not only to the phonon
part. Furthermore, renormalization contributions to electronic and phononic
one-particle energies and to the electron-phonon coupling will be neglected
for simplicity.

At this point it is important to realize that the introduction of symmetry
breaking fields is a general concept to study phase transitions within the
PRM. The same approach has also been successfully applied to the Holstein
model and its quantum phase transition \cite{SHB_2006_1, SHB_2006_2}; this 
model will be discussed in Sec.~\ref{Holstein_QP}.

\bigskip
To perform our renormalization scheme as introduced in section \ref{PRM1} we 
need to solve the eigenvalue problem of $\mathcal{H}_{0,\lambda}$. For this 
purpose we utilize the well-known Bogoliubov transformation \cite{B_1958} and 
introduce new $\lambda$ dependent fermionic operators,
\begin{eqnarray}
  \label{EP19}
  \alpha_{{\bf k}\lambda}^{\dagger} &=&
  u_{{\bf k},\lambda}^{*} c_{{\bf k}\uparrow}^{\dagger} -
  v_{{\bf k},\lambda}^{*} c_{-{\bf k}\downarrow} ,\\
  \beta_{{\bf k}\lambda}^{\dagger} &=&
  u_{{\bf k},\lambda}^{*} c_{-{\bf k}\downarrow}^{\dagger} +
  v_{{\bf k},\lambda}^{*} c_{{\bf k}\uparrow}, \nonumber
\end{eqnarray}
where the coefficients read
\begin{eqnarray}
  \label{EP20}
  \left|
    u_{{\bf k},\lambda}
  \right|^{2}
  &=&
  \frac{1}{2}
  \left(
    1 +
    \frac{
      \varepsilon_{\bf k}
    }{
      \sqrt{
        \varepsilon_{\bf k}^{2} +
        \left| \Delta_{{\bf k},\lambda}  \right|^{2}
      }
    }
  \right), \\
  \left| v_{{\bf k},\lambda} \right|^{2}
  &=&
  \frac{1}{2}
  \left(
    1 -
    \frac{
      \varepsilon_{\bf k}
    }{
      \sqrt{
        \varepsilon_{\bf k}^{2} +
        \left| \Delta_{{\bf k},\lambda}  \right|^{2}
      }
    }
  \right). \nonumber
\end{eqnarray}
Hence, $\mathcal{H}_{0,\lambda}$ can be rewritten in diagonal form,
\begin{eqnarray}
  \label{EP21}
  {\cal H}_{0,\lambda} &=&
  \sum_{\bf k} E_{{\bf k},\lambda}
  \left(
    \alpha_{{\bf k}\lambda}^{\dagger} \alpha_{{\bf k}\lambda} +
    \beta_{{\bf k}\lambda}^{\dagger} \beta_{{\bf k}\lambda}
  \right) \\
  && +\,
  \sum_{\bf k}
  \left(
    \varepsilon_{\bf k} - E_{{\bf k},\lambda}
  \right) +
  \sum_{\bf q} \omega_{\bf q} \, b_{\bf q}^{\dagger}b_{\bf q} + C_{\lambda}
  \nonumber
\end{eqnarray}
where the fermionic excitation energies are given by
$
  E_{{\bf k},\lambda} =
  \sqrt{
    \varepsilon_{\bf k}^{2} + \left| \Delta_{{\bf k},\lambda}  \right|^{2}
  }
$.

In the following, we restrict ourselves to second order renormalization
contributions so that the first order of the generator 
$X_{\lambda,\Delta\lambda}$  of the unitary transformation is sufficient [see
Eq.~\eqref{B8} and the discussion in \ref{B_perturbation}]. Thus, 
$X_{\lambda,\Delta\lambda}$ can be written as
\eqref{B16},
\begin{eqnarray}
  \label{EP3}
  \lefteqn{X_{\lambda, \Delta\lambda} \,=\,}&&\\
  &=&
  \sum_{\mathbf{k},\mathbf{q},\sigma}
    A_{\mathbf{k},\mathbf{q}}(\lambda, \Delta\lambda) 
\left[
    \, b_{\mathbf{q}}^{\dagger}   
       c_{\mathbf{k}\sigma}^\dag c_{(\mathbf{k}+\mathbf{q})\sigma} 
- 
      b_{\mathbf{q}} 
       c_{(\mathbf{k}+\mathbf{q})\sigma}^\dag  c
_{\mathbf{k}\sigma}
\, \right]
  \nonumber
\end{eqnarray}
where
\begin{eqnarray}
  A_{\mathbf{k},\mathbf{q}}(\lambda, \Delta\lambda) &=&
  \frac{g_{\mathbf{q}}}{
    \varepsilon_{\mathbf{k}} - \varepsilon_{\mathbf{k}+\mathbf{q}} +
    \omega_{\mathbf{q}}
  } \,
  \Theta_{\mathbf{k},\mathbf{q}}(\lambda, \Delta\lambda), 
  \nonumber\\
  \label{EP23}
  && \\[2ex]
 \Theta_{\mathbf{k},\mathbf{q}}(\lambda, \Delta\lambda) &=&
  \left[ 1 -
    \Theta\left(
      \lambda - \Delta\lambda - 
      \left|
        \varepsilon_{\mathbf{k}} - \varepsilon_{\mathbf{k}+\mathbf{q}} +
        \omega_{\mathbf{q}}
      \right|
    \right)
  \right]
  \nonumber \\
  && \times
  \Theta\left(
    \lambda - 
    \left|
      \varepsilon_{\mathbf{k}} - \varepsilon_{\mathbf{k}+\mathbf{q}} +
      \omega_{\mathbf{q}}
    \right|
  \right) .
  \nonumber
\end{eqnarray}
Note that the generator $X_{\lambda, \Delta\lambda}$ as defined in
Eq.~\eqref{EP3} almost completely agrees with the one used to re-examine
Fr\"{o}hlich's transformation in subsection \ref{EP_Froehlich} [see
Eqs. \eqref{EP3b} and \eqref{EP8}]. However, now the $\Theta$
functions do not only refer to the phonon energies $\omega_{\mathbf{q}}$ but
also to the electronic one-particle energies $\varepsilon_{\mathbf{k}}$ because
of the different definitions of the $\mathbf{P}_{\lambda}$ projection
operators. 

To perform the renormalization step reducing the cutoff from $\lambda$ to
$\lambda-\Delta\lambda$, one would need to express the electronic creation and
annihilation operators by the quasi-particle operators \eqref{EP19}. After
considering the renormalization contributions, the quasi-particle operators
have to be transformed back to the original electron operators. However, this
involved procedure is only necessary if we are interested in renormalization
contributions beyond second order perturbation theory. Therefore, here the
symmetry breaking fields $\Delta_{\mathbf{k},\lambda}$ and
$\Delta_{\mathbf{k},\lambda}^{*}$ are only generated by the renormalization
scheme but not considered in the evaluation of energy denominators or
projection operators. 

Taking into account all simplifications related with second order perturbation
theory, the unitary transformation \eqref{B16} is easily evaluated where
generated operator terms are only kept if their mean-field approximations
renormalize the symmetry breaking fields, $\Delta_{\mathbf{k},\lambda}$ and
$\Delta_{\mathbf{k},\lambda}^{*}$, or the energy shift, $C_{\lambda}$. Thus,
for sufficiently small steps $\Delta\lambda$ we obtain the following
renormalization equations 
\begin{eqnarray}
  \label{EP25}
  \lefteqn{
    \Delta_{{\bf k},\lambda-\Delta\lambda} - \Delta_{{\bf k},\lambda} \,=\,
  } && \\
  &=&
  2 \sum_{\bf q}
  \Theta\left[
    \lambda - 
    \left|
      \varepsilon_{\bf k} - \varepsilon_{({\bf k}+{\bf q})}
    \right|
    + \omega_{\bf q}
  \right] \nonumber\\
  && \times \,
  \left\{
    1 - 
    \Theta\left[
      \lambda - \Delta\lambda - 
      \left|
        \varepsilon_{\bf k} - \varepsilon_{({\bf k}+{\bf q})}
      \right|
      + \omega_{\bf q}
    \right]
  \right\} \nonumber \\
  &&\times \,
  \frac{
    \left| g_{\bf q} \right|^{2}
    \Theta\left[
      \omega_{\bf q} -
      \left| \varepsilon_{\bf k} - \varepsilon_{({\bf k}+{\bf q})} \right|
    \right]
  }{
    \left|
      \varepsilon_{\bf k} - \varepsilon_{({\bf k}+{\bf q})}
    \right| + \omega_{\bf q}
  }
  \left\langle
    c_{-({\bf k}+{\bf q}),\downarrow} c_{({\bf k}+{\bf q}),\uparrow}
  \right\rangle,\nonumber
\end{eqnarray}
\begin{eqnarray}
  C_{(\lambda-\Delta\lambda)} - C_{\lambda} &=&
  \sum_{\bf k}
  \left\langle
    c_{{\bf k},\uparrow}^{\dagger} c_{-{\bf k},\downarrow}^{\dagger}
  \right\rangle
  \left[\Delta_{{\bf k},\lambda-\Delta\lambda}-\Delta_{{\bf k},\lambda}\right].
  \nonumber \\[-1ex]
  \label{EP26}
  &&
\end{eqnarray}
By summing up all difference equations between the cutoff $\Lambda$ of the
original model and the lower cutoff
$\lambda\rightarrow 0$, one easily finds
\begin{eqnarray}
  \tilde\Delta_{\bf k} &=&
  \Delta_{{\bf k},\Lambda} +
  2\sum_{\bf q}
  \frac{
    \left| g_{\bf q} \right|^{2}
    \Theta\left[
      \omega_{\bf q} -
      \left| \varepsilon_{\bf k} - \varepsilon_{({\bf k}+{\bf q})} \right|
    \right]
  }{
    \left|
      \varepsilon_{\bf k} - \varepsilon_{({\bf k}+{\bf q})}
    \right| + \omega_{\bf q}
  } \nonumber \\
  \label{EP27}
  &&\qquad\times\,
  \left\langle
    c_{-({\bf k}+{\bf q}),\downarrow} c_{({\bf k}+{\bf q}),\uparrow}
  \right\rangle, \\[2ex]
  \label{EP28}
  \tilde C &=&
  C_{\Lambda} +
  \sum_{\bf k}
  \left\langle
    c_{{\bf k},\uparrow}^{\dagger} c_{-{\bf k},\downarrow}^{\dagger}
  \right\rangle
  \left( \tilde\Delta_{\bf k} - \Delta_{{\bf k},\Lambda} \right).
\end{eqnarray}
Here we defined 
$\tilde\Delta_{\bf k}=\lim_{\lambda\rightarrow 0}\Delta_{{\bf k},\lambda}$, 
$\tilde C=\lim_{\lambda\rightarrow 0} C_{\lambda}$. 

The final Hamiltonian 
$\tilde{\mathcal{H}} = \lim_{\lambda\rightarrow 0}\mathcal{H}_{\lambda}$ can
easily be diagonalized by a Bogoliubov transformation and reads according
\eqref{EP21}
\begin{eqnarray}
  \label{EP29}
  \tilde{\cal H} &=&
  \sum_{\bf k} \tilde{E}_{\bf k}
  \left(
    \tilde{\alpha}_{\bf k}^{\dagger} \tilde{\alpha}_{\bf k} +
    \tilde{\beta}_{\bf k}^{\dagger}
    \tilde{\beta}_{\bf k}
  \right) \\
  && +\,
  \sum_{\bf k}
  \left(
    \varepsilon_{\bf k} - \tilde{E}_{\bf k}
  \right) +
  \sum_{\bf q} \omega_{\bf q} \, b_{\bf q}^{\dagger}b_{\bf q} + \tilde{C}
  \nonumber
\end{eqnarray}
where $\tilde{E}_{\bf k} = \lim_{\lambda\rightarrow 0}E_{{\bf k},\lambda}$,
$\tilde\alpha_{\bf k} = \lim_{\lambda\rightarrow 0}\alpha_{{\bf k},\lambda}$, 
and $\tilde\beta_{\bf k} = \lim_{\lambda\rightarrow 0}\beta_{{\bf k},\lambda}$.
Its parameters depend on the original system \eqref{EP1}, on the initial
conditions, $\Delta_{\mathbf{k},\Lambda}=0$, $C_{\Lambda}=0$, and on 
expectation values
$
  \left\langle
    c_{{\bf k},\uparrow}^{\dagger} c_{-{\bf k},\downarrow}^{\dagger}
  \right\rangle
$
that need to be determined self-consistently. Following the approach of
Ref.~\onlinecite{HB_2003}, we consider the free energy which can be calculated
either from $\mathcal{H}$ or from the renormalized Hamiltonian 
$\tilde{\mathcal{H}}$. Thus, the required expectation values are easily found
by functional derivatives,
$
  \left\langle
    c_{{\bf k},\uparrow}^{\dagger} c_{-{\bf k},\downarrow}^{\dagger}
  \right\rangle
  =
  -\frac{\displaystyle\partial F}
  {\displaystyle\partial \Delta_{\mathbf{k},\Lambda}}
$,
so that Eq.~\eqref{EP27} can be rewritten as
\begin{eqnarray}
  \label{EP30}
  \tilde\Delta_{\bf k} &=&
  \sum_{\bf q}
  \left\{
    \frac{
      2 \left| g_{\bf q} \right|^{2}
      \Theta\left[
        \omega_{\bf q} -
        \left| \varepsilon_{\bf k} - \varepsilon_{({\bf k}+{\bf q})} \right|
      \right]
    }{
      \left|
        \varepsilon_{\bf k} - \varepsilon_{({\bf k}+{\bf q})}
      \right| + \omega_{\bf q}
    }
  \right\} \\
  && \qquad \times\,
  \frac{
    \tilde{\Delta}_{{\bf k}+{\bf q}}^{*}
    \left[ 1 - 2f( \tilde{E}_{{\bf k}+{\bf q}} ) \right]
  }{
    2\sqrt{
      \varepsilon_{{\bf k}+{\bf q}}^{2} +
      \left| \tilde \Delta_{{\bf k}+{\bf q}} \right|^{2}
    }
  }
  \nonumber
\end{eqnarray}
where the initial condition $\Delta_{\mathbf{k},\Lambda}=0$ has been
used. Eq.~\eqref{EP30} has the form of the famous BCS-gap equation so
that the term inside the braces $\{\cdots\}$ can be interpreted as
parameter of the effective phonon induced electron-electron interaction,
\begin{eqnarray}
  \label{EP31}
  V_{\mathbf{k},\mathbf{-k},\mathbf{q}} &=&
  - \frac{
    \left| g_{\bf q} \right|^{2}
    \Theta\left[
      \omega_{\bf q} -
      \left| \varepsilon_{\bf k} - \varepsilon_{({\bf k}+{\bf q})} \right|
    \right]
  }{
    \left|
      \varepsilon_{\bf k} - \varepsilon_{({\bf k}+{\bf q})}
    \right| + \omega_{\bf q}
  }
\end{eqnarray}
which is responsible for the formation of Cooper pairs. Even though we have
here derived an effective electron-electron interaction as well there is a
significant difference to the approaches of \ref{EP_Froehlich} and
\ref{EP_flow}: In the present formalism both the attractive electron-electron
interaction and the superconducting gap function were derived in 
\textit{one step} by applying the PRM to the electron-phonon system
\eqref{EP1} with additional symmetry breaking fields.

\subsection{Discussion}
\label{EP_disc}

In the following we want to discuss the different approaches to the
phonon-induced effective electron-electron interaction in more detail. At
first we summarize the results derived above where we focus on the interaction
between electrons of a Cooper pair. Fr\"{o}hlich's classical result [see
Ref.~\onlinecite{F_1952} and Eq.~\eqref{EP10}] reads
\begin{eqnarray}
  \label{EP32}
  V_{\mathbf{k}, -\mathbf{k}, \mathbf{q}}^{\mbox{\tiny Fr\"{o}hlich}} &=&
  \frac{\left| g_{\mathbf{q}} \right|^{2} \omega_{\mathbf{q}}}
  {
    \left( 
      \varepsilon_{\mathbf{k}+\mathbf{q}} - \varepsilon_{\mathbf{k}}
    \right)^{2} - 
    \omega_{\mathbf{q}}^{2}
  }.
\end{eqnarray}
However, there is an important problem related with Eq.~\eqref{EP32}: It
diverges at 
$|\varepsilon_{\bf k} - \varepsilon_{({\bf k}+{\bf q})}|=\omega_{\mathbf{q}}$.
Thus, a cutoff function is introduced by hand in the classical BCS-theory to
suppress repulsive contributions to the effective electron-electron
interaction. 

In contrast to the Fr\"ohlich interaction \eqref{EP32}, the results obtained
by Wegner's flow equation method \cite{LW_1996}, by similarity transformation 
\cite{M_1997}, and by the PRM \cite{HB_2003} are less singular,
\begin{eqnarray}
  \label{EP33}
  V_{\mathbf{k}, -\mathbf{k}, \mathbf{q}}^{\mbox{\tiny Lenz/Wegner}} &=& -\,
  \frac{\left| g_{\mathbf{q}} \right|^{2} \omega_{\mathbf{q}}}
  {
    \left( 
      \varepsilon_{\mathbf{k}+\mathbf{q}} - \varepsilon_{\mathbf{k}}
    \right)^{2} + 
    \omega_{\mathbf{q}}^{2}
  }, \\[1ex]
  V_{\mathbf{k}, -\mathbf{k}, \mathbf{q}, \lambda}^{\mbox{\tiny Mielke}} 
  &=& -\,
  \frac{
    \left| g_{\mathbf{q}} \right|^{2}
    \Theta\left(
      \left|
        \varepsilon_{\mathbf{k}+\mathbf{q}} - \varepsilon_{\mathbf{k}}
      \right| +
      \omega_{\mathbf{q}} - \lambda
    \right)
  }
  {
    \left|
      \varepsilon_{\mathbf{k}+\mathbf{q}} - \varepsilon_{\mathbf{k}}
    \right| + 
    \omega_{\mathbf{q}}
  }, \nonumber\\[-1ex]
  \label{EP34}
  && \\[1ex]
  \label{EP35}
  V_{\mathbf{k},-\mathbf{k},\mathbf{q},\lambda}^{
    \mbox{\tiny H\"{u}bsch/Becker}
  } 
  &=& -\,
  \frac{
    \left| g_{\mathbf{q}} \right|^{2}
    \Theta\left(
      \omega_{\mathbf{q}} - 
      \left|
        \varepsilon_{\mathbf{k}+\mathbf{q}} - \varepsilon_{\mathbf{k}}
      \right|
    \right)
  }
  {
    \left|
      \varepsilon_{\mathbf{k}+\mathbf{q}} - \varepsilon_{\mathbf{k}}
    \right| + 
    \omega_{\mathbf{q}}
  }.
\end{eqnarray}
(Note that Eqs.~\eqref{EP33} and \eqref{EP35} have already been derived above,
compare with \eqref{EP17} and \eqref{EP31}. The $\lambda$ dependence of the
electronic and phononic one-particle energies are suppressed in \eqref{EP34}
for simplicity.) All three results for the effective phonon-mediated
electron-electron interaction are never repulsive as long as
$\omega_{\mathbf{q}}>0$ is fulfilled. 

\bigskip
At first we want to discuss Mielke's result \cite{M_1997}, an effective
electron-electron interaction \eqref{EP34} that depends on the energy cutoff
$\lambda$. As Wegner's flow equation method \cite{W_1994}, the used similarity
transformation \cite{GW_1993,GW_1994} is based on continuous unitary
transformations and leads to differential equations for the parameters of the
Hamiltonian. However, like the PRM, the similarity transformation leads to a
band-diagonal structure of the renormalized Hamiltonian with respect to the
eigenenergies of the unperturbed Hamiltonian whereas the flow equation method
generates block-diagonal Hamiltonians.

Mielke derived the phonon-mediated electron-electron interaction \eqref{EP34}
by eliminating excitations with energies larger than $\lambda$ where
excitation energies are measured with respect to the unperturbed Hamiltonian
consisting of both electronic and bosonic degrees of freedom. The obtained
effective interaction becomes $\lambda$ independent for the Einstein model (of
dispersion-less phonons) if $\lambda$ is chosen smaller than the phonon
frequency $\omega_{0}$. For this case Mielke's result \eqref{EP34} is very
similar to ours \eqref{EP35} obtained by the PRM with symmetry-breaking
fields. However, in contrast to our result \eqref{EP35}, the cutoff function
$
  \Theta\left(
    \omega_{\mathbf{q}} - 
    \left| \varepsilon_{\mathbf{k}+\mathbf{q}}-\varepsilon_{\mathbf{k}}\right|
  \right)
$
is absent in \eqref{EP34}. This difference might be related with different
choices for the generator of the unitary transformation in the two methods
but could also be caused by a systematic problem in Mielke's approach: Setting
$\lambda=0$, the final renormalized Hamiltonian contains non-diagonal terms
with respect to the used unperturbed Hamiltonian. This seems to contradict a
basic premise of the similarity transformation.

\begin{figure}
  \begin{center}
    \scalebox{0.64}{
      \psfrag{Ref. HB}[1][0]{\large Ref. \onlinecite{HB_2003}}
      \psfrag{Ref. LW}[1][0]{\large Ref. \onlinecite{LW_1996}}
      \includegraphics*[0,15][375,275]{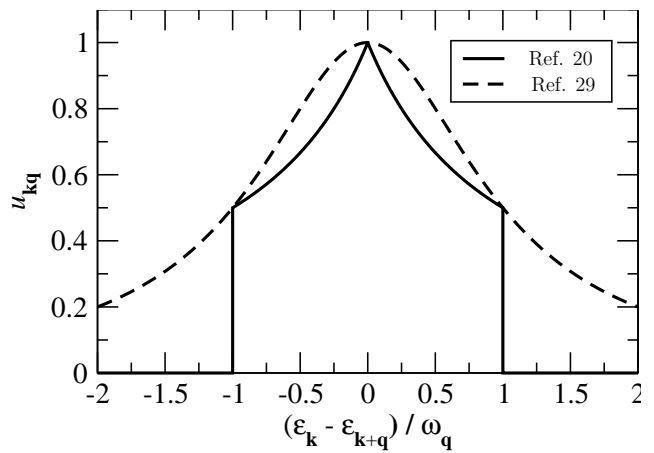}
    }
  \end{center}
  \caption{
    Comparison of the effective electron-electron interaction obtained by
    the PRM (full line) and by Wegner's flow equations (dashed line). Here,
    the dimensionless quantity 
    $
      u_{\mathbf{kq}} = 
      - \frac{\omega_{\mathbf{q}} V_{\mathbf{kq}}}
      {|g_{\mathbf{q}}|^{2}}
    $
    has been introduced.
  }
  \label{Fig_BCS}
\end{figure}

\bigskip
Lenz and Wegner \cite{LW_1996} applied the flow equation method to the
electron-phonon problem as discussed here and obtained an effective
electron-electron interaction as shown in Eq.~\eqref{EP33}. As one can see in
Fig.~\ref{Fig_BCS}, their result is quite similar to ours \eqref{EP35} derived
using the PRM as long as 
$
  \omega_{\mathbf{q}}\ge
  \left| \varepsilon_{\mathbf{k}+\mathbf{q}} - \varepsilon_{\mathbf{k}} \right|
$
is fulfilled. However, in contrast to our result \eqref{EP35}, the interaction
\eqref{EP33} remains finite even for 
$
  \omega_{\mathbf{q}} < 
  \left| \varepsilon_{\mathbf{k}+\mathbf{q}} - \varepsilon_{\mathbf{k}} \right|
$.
Probably, this difference is caused by the different choices for the
generator of the unitary transformation that also require different
approximations in order to obtain closed sets of renormalization equations.

\section{
  Heavy-fermion behavior in the periodic Anderson 
  model 
}
\label{PAM}

The periodic Anderson model (PAM) is considered to be the basic microscopic
model for the theoretical investigation of heavy-fermion (HF) systems
\cite{LRSSW_1986}. It describes localized, strongly correlated $f$ electrons
interacting with itinerant conduction electrons. Here we focus on the limit of
infinitely large Coulomb repulsion on $f$ sites so that the Hamiltonian of the
PAM can be written as
\begin{eqnarray}
  \label{PAM1}
  \mathcal{H} &=& \mathcal{H}_{0} + \mathcal{H}_{1} ,
\end{eqnarray}
\begin{eqnarray}
  \mathcal{H}_0 & = & \varepsilon_{f} \sum _{i,m} \hat{f}^{\dagger} _{im}
  \hat{f} _{im}
  + \sum _{{\bf k},m} \varepsilon_{{\bf k}} \ c^{\dagger}_{{\bf k}m}
  c_{{\bf k}m} , \nonumber\\
  \mathcal{H}_1 & = & \frac{1}{\sqrt{N}} \sum _{{\bf k},i,m} V_{{\bf k}}
  \left(
    \hat{f}^{\dagger} _{im} c_{{\bf k}m} \,
    e^{{\rm i}{\bf k}{\bf R}_i} + {\rm h.c.}
  \right ) .
  \nonumber
\end{eqnarray}
The one-particle energies $\varepsilon_{f}$ and $\varepsilon_{{\bf k}}$, and,
as a simplification, both types of electrons have the same angular momentum
index $m=1\dots\nu_{f}$. The Hubbard operators,
\begin{eqnarray*}
  \hat{f}^{\dagger} _{im} &=& f^{\dagger} _{im} \prod_{\tilde m (\ne m)}
  (1- f_{i\tilde{m}}^\dagger f_{i\tilde{m}}),
\end{eqnarray*}
take into account the infinitely large local Coulomb repulsion and only allow
either empty or singly occupied $f$ sites.

\bigskip
The PRM has already been applied to the PAM in
Ref.~\onlinecite{HB_2005,HB_2006} where approximations have been employed that
allow to map the renormalization equations of the PAM onto those of the
uncorrelated Fano-Anderson model (see subsection \ref{B_Fano-Anderson}). Thus, 
HF behavior and a possible valence transition between mixed and integral
valent states could be studied. However, the approach of
Refs.~\onlinecite{HB_2005,HB_2006} has a significant disadvantage: the
renormalization of the one-particle energies show as function the cutoff
$\lambda$ a steplike behavior that leads to serious problems in the
(numerical) evaluation. Therefore, a constant renormalized $f$ energy had to
be chosen for all values of the energy cutoff $\lambda$ to ensure a continuous
behavior of the one-particle energies as required for physical reasons.

In the following we modify the approach of Refs.~\onlinecite{HB_2005,HB_2006} 
to ensure a more continuous renormalization of all parameters of the 
Hamiltonian. For this purpose, the ideas of \ref{FA_rev} and \ref{EP_flow} 
are transferred to the PAM. However, to explore all features of this continuous 
approach is beyond the scope of this review, we re-derive the analytical 
solution of Ref.~\onlinecite{HB_2005} instead.

\subsection{Renormalization ansatz}

Much of the physics of the PAM \eqref{PAM1} can be understood in terms of
an effective uncorrelated model that consists of two non-interacting fermionic
quasi-particle bands. Various theoretical approaches have been used to generate
such effective Hamiltonians; the most popular among them is the slave-boson
mean-field (SB) theory \cite{C_1984,FKZ_1988}. However, as discussed in
Ref.~\onlinecite{HB_2006}, such approaches \textit{do not prevent} from
unphysical multiple occupation of $f$ sites and are therefore restricted to
heavy-fermion like solutions. [The SB solutions break down if the original $f$
level $\varepsilon_{f}$ is located too far below the Fermi level or if the
hybridization between $f$ and conduction electrons becomes too weak
\cite{FFF_2002}.]

To reliably prevent the system from unphysical states with multiple
occupations of $f$ sites we here follow Ref.~\onlinecite{HB_2006} and start 
from a renormalization ansatz that keeps the Hubbard operators during the whole
renormalization procedure,
\begin{eqnarray}
  \label{PAM2}
  \mathcal{H}_{\lambda} & = &
  \mathcal{H}_{0,\lambda} + \mathcal{H}_{1,\lambda} ,\\[1ex]
  \mathcal{H}_{0,\lambda} & = &
  e_{f,\lambda} \sum_{\mathbf{k},m}
  \hat{f}^{\dagger}_{\mathbf{k}m} \hat{f}_{\mathbf{k}m} +
  \sum_{\mathbf{k},m} \Delta_{\mathbf{k},\lambda}
  \left(
    \hat{f}^{\dagger} _{\mathbf{k}m} \hat{f}_{\mathbf{k}m}
  \right)_{\mathrm{NL}}
  \nonumber \\
  &&
  + \sum _{{\bf k},m}
  \varepsilon_{{\bf k},\lambda} \ c^{\dagger}_{{\bf k}m} c_{{\bf k}m}  +
  E_{\lambda} , \nonumber \\[1ex]
  \mathcal{H}_{1,\lambda} & = &
  \mathbf{P}_{\lambda} \mathcal{H}_{1,\lambda} \,=\,
  \sum_{\mathbf{k},m} V_{\mathbf{k},\lambda} \
  \left(
    \hat{f}^{\dagger}_{\mathbf{k}m} c_{{\bf k}m} + \mathrm{h.c.}
  \right).
  \nonumber
\end{eqnarray}
Eq.~\eqref{PAM2} is obtained after all excitations between eigenstates of 
${\cal H}_{0,\lambda}$ with transition energies
larger than the cutoff $\lambda$ have been eliminated, i.e. 
$\mathbf{Q}_\lambda {\cal H}_\lambda =0$ holds. Furthermore, we introduced 
Fourier transformed Hubbard operators,
\begin{eqnarray*}
  \hat{f}^{\dagger}_{\mathbf{k}m} &=&
  \frac{1}{\sqrt{N}}
  \sum_{i} \hat{f}^{\dagger}_{im} e^{i\mathbf{k}\cdot\mathbf{R}_{i}}.
\end{eqnarray*}
The $\lambda$ dependencies of the parameters are caused by the renormalization
procedure. Note that $V_{\mathbf{k},\lambda}$ includes a cutoff function in
order to ensure that the requirement
$\mathbf{Q}_{\lambda}\mathcal{H}_{\lambda}=0$
is fulfilled. Furthermore, an additional energy shift $E_{\lambda}$ and direct
hopping between $f$ sites,
\begin{eqnarray*}
  \left(
    \hat{f}^{\dagger} _{\mathbf{k}m} \hat{f}_{\mathbf{k}m}
  \right)_{\mathrm{NL}}
  & = &
  \frac{1}{N}\sum_{i, j(\not{=}i)}
  \hat{f}_{im}^{\dagger} \hat{f}_{jm}
  e^{i\mathbf{k}(\mathbf{R}_{i}-\mathbf{R}_{j})},
\end{eqnarray*}
have been generated. Finally, we need the initial parameter values of the
original model (with cutoff $\Lambda$) to fully determine the renormalization,
\begin{eqnarray}
  \label{PAM3}
  e_{f,\Lambda} &=& \varepsilon_{f}, \;
  \Delta_{\mathbf{k},\Lambda} \,=\, 0, \;
  \varepsilon_{{\bf k},\Lambda} \,=\, \varepsilon_{{\bf k}}, \;
  E_{\Lambda} \,=\, 0,\\
  V_{\mathbf{k},\Lambda} &=& V_{\mathbf{k}}.\nonumber
\end{eqnarray}

To implement our PRM scheme we also need the commutator of the unperturbed part
$\mathcal{H}_{0,\lambda}$ of the $\lambda$ dependent Hamiltonian
$\mathcal{H}_{\lambda}$ with the interaction $\mathcal{H}_{1,\lambda}$ (in the
present case the hybridization between $f$ and conduction electrons). To
shorten the notation we here introduce the (unperturbed) Liouville operator
$\mathbf{L}_{0,\lambda}$ that is defined as 
$\mathbf{L}_{0,\lambda} \mathcal{A} = [\mathcal{H}_{0,\lambda}, \mathcal{A}]$
for any operator $\mathcal{A}$. Because of the correlations included in the
Hubbard operators $\hat{f}^{\dagger}_{\mathbf{k}m}$, the required commutator
relation can not be calculated exactly and additional approximations are
necessary. Here, the one-particle operators $\hat{f}^{\dagger}_{\mathbf{k}m}$
and $c^{\dagger}_{\mathbf{k}m}$ are considered as approximative eigenoperators
of $\mathbf{L}_{0,\lambda}$ so that we obtain
\begin{eqnarray}
  \mathbf{L}_{0,\lambda} \ \hat{f}^{\dagger} _{\mathbf{k}m} c_{\mathbf{k}m}
  &\approx&
  \left(
    \varepsilon_{f,\lambda} + D\Delta_{\mathbf{k},\lambda} -
    \varepsilon_{\mathbf{k},\lambda}
  \right)
  \hat{f}^{\dagger}_{\mathbf{k}m} c_{\mathbf{k}m}. \nonumber\\[-1ex]
  \label{PAM4}
  &&
\end{eqnarray}
Here we introduced the local $f$ energy,
\begin{eqnarray}
  \label{PAM5}
  \varepsilon_{f,\lambda}  & = &
  e_{f,\lambda} - D \bar{\Delta}_{\lambda} ,
\end{eqnarray}
the averaged $f$ dispersion,
$
  \bar{\Delta}_{\lambda} =
  \frac{1}{N}\sum_{\mathbf{k}}\Delta_{\mathbf{k},\lambda}
$,
and defined
$
  D = 1 - \langle \hat{n}_{i}^{f} \rangle +
  \langle \hat{n}_{i}^{f} \rangle / \nu_{f}
$.
Note that the factors $D$ in Eqs.~\eqref{PAM4} and \eqref{PAM5} are caused by
the Hubbard operators $\hat{f}^{\dagger}_{\mathbf{k}m}$ where a factorization
approximation has been employed. 

To ensure that  
$
  \mathbf{Q}_{\lambda} \mathcal{H}_{\lambda} = 0
$
is fulfilled by \eqref{PAM2}, the hybridization matrix elements 
must include an additional
$\Theta$-function, 
$V_{\mathbf{k},\lambda} = \Theta(\mathbf{k},\lambda)V_{\mathbf{k},\lambda}$,
where we have defined
\begin{eqnarray}
  \nonumber
  \Theta(\mathbf{k},\lambda) &=& 
  \Theta\left( 
    \lambda - |\varepsilon_{f,\lambda} + D\Delta_{\mathbf{k},\lambda} -
    \varepsilon_{\mathbf{k},\lambda}| 
  \right).
\end{eqnarray}

\subsection{Generator of the unitary transformation}

In order to derive the renormalization equations for the parameters of
$\mathcal{H}_{\lambda}$ we have to consider the unitary transformation to
eliminate excitations within the energy shell between $\lambda-\Delta\lambda$
and $\lambda$. Corresponding to Eq.~\eqref{B16}, such a unitary transformation
is determined by its generator $X_{\lambda,\Delta\lambda}$. As in
Ref.~\onlinecite{HB_2006} we use an ansatz that is motivated by 
perturbation theory [see Eq.~\eqref{B8}],
\begin{eqnarray}
  X_{\lambda,\Delta\lambda} &=&
  \sum_{\mathbf{k},m} A_{\mathbf{k}}(\lambda,\Delta\lambda) \,
  \left(
    \hat{f}^{\dagger}_{\mathbf{k}m} c_{\mathbf{k}m} - 
    c_{\mathbf{k}m}^{\dagger} \hat{f}_{\mathbf{k}m}
  \right).\nonumber \\[-1ex]
  \label{PAM6} 
  &&
\end{eqnarray}

The parameter $A_{\mathbf{k}}(\lambda,\Delta\lambda)$ of the generator 
$ X_{\lambda,\Delta\lambda}$ needs to be chosen in
such a way that Eq.~\eqref{Bed4}, 
$
  \mathbf{Q}_{(\lambda-\Delta\lambda)} 
\mathcal{H}_{(\lambda-\Delta\lambda)} = 0
$,
is fulfilled. However, as already discussed before, 
this requirement only determines the part
$\mathbf{Q}_{(\lambda-\Delta\lambda)} X_{\lambda,\Delta\lambda}$
of the generator \eqref{PAM6} of the unitary transformation whereas 
$\mathbf{P}_{(\lambda-\Delta\lambda)} X_{\lambda,\Delta\lambda}$ can be chosen
arbitrarily. Thus, 
$\mathbf{P}_{(\lambda-\Delta\lambda)} X_{\lambda,\Delta\lambda} = 0$ is
usually chosen to perform the minimal transformation to match the requirement
\eqref{Bed4}. In this way, the impact of approximations necessary for every
renormalization step can be minimized.  

On the other hand, the approach of "minimal" approximations can also lead to
some problems if a steplike renormalization behavior for the parameter of the
Hamiltonian is found. This is the case for the PRM approach of
Refs.~\onlinecite{HB_2005,HB_2006} where a constant renormalized
$f$ energy $\tilde{\varepsilon}_{f}$ have been used for all cutoff values
$\lambda$ to ensure a continuous behavior of the one-particle energies as
required for physical reasons. Therefore, in the following
$\mathbf{P}_{(\lambda-\Delta\lambda)} X_{\lambda,\Delta\lambda}$ shall 
again be chosen non-zero in order to ensure a more continuous 
renormalization of all parameters of the Hamiltonian. 
In close analogy to subsection \ref{FA_rev}, we choose a proper
generator $A''_{\mathbf{k}}(\lambda,\Delta\lambda)\sim \Delta\lambda$, not yet
specified, which almost completely integrates out
interactions \textit{before} the cutoff energy $\lambda$ approaches their
corresponding transition energies. 
In the limit of small $\Delta\lambda$,
we again expect an exponential decay for the
hybridization $V_{\mathbf{k},\lambda}$ in this way.

\subsection{Renormalization equations}

In comparison to the approach of Refs.~\onlinecite{HB_2005,HB_2006}, the
derivation of the renormalization equation is simplified: 
Having in mind  $A''_{\mathbf{k}}(\lambda,\Delta\lambda)\sim \Delta \lambda$, 
where $\Delta \lambda$ is a small quantity, 
we can restrict ourselves to first order renormalization contributions
and neglect the   $A'_{\mathbf{k}}(\lambda,\Delta\lambda)$
part of $X_{\lambda, \Delta \lambda}$ altogether. 
Thus, eliminating excitations
within the energy shell between $\lambda-\Delta\lambda$ and $\lambda$, the
renormalized Hamiltonian $\mathcal{H}_{(\lambda-\Delta\lambda)}$ can be
calculated based on Eq.~\eqref{B17a}. 

To derive the renormalization equations for the parameters of the Hamiltonian,
we compare the coefficients of the different operator terms in the
renormalization ansatz \eqref{PAM2} at cutoff $\lambda-\Delta\lambda$ and in
the explicitly evaluated Eq.~\eqref{B17a}. Thus, based on similar
approximations as the approach of Refs.~\onlinecite{HB_2005} and
\onlinecite{HB_2006}, we obtain the following equations:
\begin{eqnarray}
  \label{PAM8}
     \varepsilon_{\mathbf{k},\lambda-\Delta\lambda} -
    \varepsilon_{\mathbf{k},\lambda}
    & = &
  - \, 2 D \, A''_{\mathbf{k}}(\lambda,\Delta\lambda) \,
    V_{\mathbf{k},\lambda} \\
&& \nonumber \\
  \label{PAM9}
  \Delta_{\mathbf{k},\lambda-\Delta\lambda} - \Delta_{\mathbf{k},\lambda}
  & = &
  - \, \frac{1}{D}
  \left[
    \varepsilon_{\mathbf{k},\lambda-\Delta\lambda} -
    \varepsilon_{\mathbf{k},\lambda}
  \right],
\end{eqnarray}
\begin{eqnarray}
  \label{PAM10}
  \lefteqn{ 
    e_{f,\lambda-\Delta\lambda} - e_{f,\lambda}
    \, = \,
  }
  && \\
  &=& -\,
  \frac{1}{D}  \frac{1}{N} \sum_{\mathbf{k}}
  \left[
    \varepsilon_{\mathbf{k},\lambda-\Delta\lambda} -
    \varepsilon_{\mathbf{k},\lambda}
  \right]
  \nonumber \\
  && \qquad \times \,
  \left\{
    1 + \left( \nu_{f} - 1 \right) 
    \left\langle c_{\mathbf{k}m}^{\dagger} c_{\mathbf{k}m} \right\rangle
  \right\}  
  \nonumber \\
  && + \,
  \frac{\nu_{f} - 1}{N} \sum_{k}
  \Theta\left( \mathbf{k}, \lambda - \Delta\lambda \right) \, 
  A''_{\mathbf{k}}(\lambda,\Delta\lambda) \,
  \left( \Delta_{\mathbf{k},\lambda} - \bar\Delta_{\lambda} \right)
  \nonumber \\
  && \qquad \times \,
  \left\langle 
    \hat{f}_{\mathbf{k}m}^{\dagger} c_{\mathbf{k}m} + \mathrm{h.c.}
  \right\rangle ,
  \nonumber
\end{eqnarray}
\begin{eqnarray}
  \label{PAM11}
  \lefteqn{
    V_{\mathbf{k},\lambda-\Delta\lambda} - V_{\mathbf{k},\lambda}
    \, = \,
  } && \\
  &=& - \,
  A''_{\mathbf{k}}(\lambda,\Delta\lambda) \,
  \left[
    e_{f,\lambda} + 
    D\left( \Delta_{\mathbf{k},\lambda} - \bar{\Delta}_{\lambda} \right) -
    \varepsilon_{\mathbf{k},\lambda}
  \right]
  \nonumber
\end{eqnarray}
\begin{eqnarray}
  \label{PAM12}
  E_{(\lambda - \Delta\lambda)} - E_{\lambda} &=&
  -\, N \langle \hat{n}_{i}^{f} \rangle
  \left[
    e_{f,\lambda-\Delta\lambda} - e_{f,\lambda}
  \right] \\
  && - \,
  \frac{\langle \hat{n}_{i}^{f} \rangle}{D}
  \sum_{\mathbf{k}}
  \left[
    \varepsilon_{\mathbf{k},\lambda-\Delta\lambda} -
    \varepsilon_{\mathbf{k},\lambda}
  \right] .
  \nonumber
\end{eqnarray}
Here, the condition 
$
  V_{\mathbf{k},\lambda-\Delta\lambda} = 
  \Theta(\mathbf{k},\lambda-\Delta\lambda)
  V_{\mathbf{k},\lambda-\Delta\lambda}
$ has
to be fulfilled. Note that higher order terms in these equations have been 
evaluated in Refs.~\onlinecite{HB_2005} and \onlinecite{HB_2006} for the case 
that the generator $X_{\lambda, \Delta \lambda}$ was fixed by  
${\bf Q}_{(\lambda -\Delta \lambda)}X_{\lambda, \Delta \lambda}$.

In deriving the renormalization equations \eqref{PAM8} - \eqref{PAM12} a
factorization approximation has been employed in order to trace back all terms
to operators appearing in the renormalization ansatz \eqref{PAM2}. Thus, the
renormalization equations still depend on expectation values which have to be
determined simultaneously. Following the approach of
Ref.~\onlinecite{HB_2006}, we neglect the $\lambda$ dependency of all
expectation values and calculate them with respect to the full Hamiltonian
$\mathcal{H}$. As discussed in subsection \ref{GenUT}, there are two 
strategies to obtain
such expectation values: The first one is based on the free energy which we
will use later for the analytical solution in \ref{PAM_analytical}. However,
the evaluation of the free energy is complicated as long as the renormalized
Hamiltonian contains Hubbard operators $\hat{f}_{\mathbf{k}m}$. Thus, here it
would be more convenient to use the second strategy to calculate expectation
values and to derive renormalization equations for additional operator
expressions (see Refs.~\onlinecite{HB_2005} and \onlinecite{HB_2006} for more
details). However, such involved approach is only needed in case of a
numerical treatment of the renormalization equations which will be discussed
below. 

The further calculations can be simplified by considering the limit
$\Delta\lambda\rightarrow 0$ and to transform the difference equations
\eqref{PAM8} - \eqref{PAM12} into differential equations. For this purpose we
define 
\begin{eqnarray}
\label{PAM12a}
  \alpha_{\mathbf{k}}(\lambda) & = & 
  \lim_{\Delta\lambda\rightarrow 0}
  \frac{A''_{\mathbf{k}}(\lambda,\Delta\lambda)}{\Delta\lambda}
\end{eqnarray}
so that we obtain
\begin{eqnarray}
  \label{PAM13}
  \frac{\mathrm{d}\varepsilon_{\mathbf{k},\lambda}}{\mathrm{d}\lambda} &=& 
  2 D \, \alpha_{\mathbf{k}}(\lambda) \, V_{\mathbf{k},\lambda} , \\[1ex]
  \label{PAM14}
  \frac{\mathrm{d}\Delta_{\mathbf{k},\lambda}}{\mathrm{d}\lambda} &=&
  - \, \frac{1}{D} 
  \frac{\mathrm{d}\varepsilon_{\mathbf{k},\lambda}}{\mathrm{d}\lambda} \\[1ex]
  \frac{\mathrm{d}e_{f,\lambda}}{\mathrm{d}\lambda} &=&
  -\, \frac{1}{D}\frac{1}{N}\sum_{\mathbf{k}}
  \left\{
    1 + \left( \nu_{f} - 1 \right) 
    \left\langle c_{\mathbf{k}m}^{\dagger} c_{\mathbf{k}m} \right\rangle
  \right\}
  \frac{\mathrm{d}\varepsilon_{\mathbf{k},\lambda}}{\mathrm{d}\lambda} ,
  \nonumber \\
  && -\,
  \frac{\nu_{f} - 1}{N} \sum_{k}
  \Theta( \mathbf{k}, \lambda ) \, 
  \alpha_{\mathbf{k}}(\lambda) \,
  \left( \Delta_{\mathbf{k},\lambda} - \bar\Delta_{\lambda} \right)
  \nonumber \\
  \label{PAM15}
  &&\qquad \times \,
  \left\langle 
    \hat{f}_{\mathbf{k}m}^{\dagger} c_{\mathbf{k}m} + \mathrm{h.c.}
  \right\rangle , \\[1ex]
  \label{PAM16}
  \frac{\mathrm{d}V_{\mathbf{k},\lambda}}{\mathrm{d}\lambda} &=&
  \left[
    e_{f,\lambda} + 
    D\left( \Delta_{\mathbf{k},\lambda} - \bar{\Delta}_{\lambda} \right) -
    \varepsilon_{\mathbf{k},\lambda}
  \right] \,
  \alpha_{\mathbf{k}}(\lambda) , \\[1ex]
  \label{PAM17}
  \frac{\mathrm{d}E_{\lambda}}{\mathrm{d}\lambda} &=& 
  -\, N \langle \hat{n}_{i}^{f} \rangle
  \frac{\mathrm{d}e_{f,\lambda}}{\mathrm{d}\lambda} - 
  \frac{\langle \hat{n}_{i}^{f} \rangle}{D}
  \sum_{\mathbf{k}}
  \frac{\mathrm{d}\varepsilon_{\mathbf{k},\lambda}}{\mathrm{d}\lambda} .
\end{eqnarray}

\subsection{Analytical solution}
\label{PAM_analytical}

In the following, we concentrate on an analytical solution of the 
renormalization equations \eqref{PAM13}-\eqref{PAM17}
by assuming a $\lambda$ independent energy of the $f$ electrons. 
The aim is to demonstrate that
the analytical solution of Ref.~\onlinecite{HB_2005} can also be derived from
the renormalization equations \eqref{PAM13}-\eqref{PAM17}
or likewise \eqref{PAM8}-\eqref{PAM12} obtained here.
In particular, we want to derive an analytical
solution that describes HF behavior. As in Ref.~\onlinecite{HB_2005}, we use
the following approximations:
\begin{enumerate}
  \item[(i)]
  All expectation values (which appear due to the employed factorization
  approximation) are considered as independent from the renormalization
  parameter $\lambda$ and are calculated with respect to the full Hamiltonian
  $\mathcal{H}$. 

  \item[(ii)]
  As mentioned, the $\lambda$ dependence of the renormalized $f$ level is neglected and we
  approximate 
  $e_{f,\lambda} - D\bar{\Delta}_{\lambda} \approx \tilde{\varepsilon}_{f}$
  to decouple the renormalization of the different $\mathbf{k}$ values. Note
  that such a renormalized $f$ energy is also used from the very beginning in
  the SB theory.

  \item[(iii)]
  To obtain the analytical solution of Ref.~\onlinecite{HB_2005} we set
  $\frac{1}{N}\sum_{\mathbf{k}}\tilde{\Delta}_{\mathbf{k}} = 0$ for further
  simplification. 

  \item[(iv)]
  The Hubbard operators are replaced by usual fermionic operators where we
  employ
  \begin{eqnarray*}
    \sum_{\mathbf{k}} \hat{f}_{\mathbf{k}m}^{\dagger}\hat{f}_{\mathbf{k}m} &=&
    \sum_{\mathbf{k}} f_{\mathbf{k}m}^{\dagger} f_{\mathbf{k}m}
    \quad\mbox{and} \\
    \left( 
      \hat{f}_{\mathbf{k}m}^{\dagger}\hat{f}_{\mathbf{k}m} 
    \right)_{\mathrm{NL}} &=&
    D \left( f_{\mathbf{k}m}^{\dagger} f_{\mathbf{k}m} \right)_{\mathrm{NL}}.
  \end{eqnarray*}
  Thus, on a mean-field level, the system is prevented from generating
  unphysical states but a multiple occupation of $f$ sites is \textit{not}
  completely suppressed by this approximation. Therefore, we can only obtain
  useful results as long as only very few $f$ type states below the Fermi
  level are occupied.
\end{enumerate}
It turns out that the analytical solution of Ref.~\onlinecite{HB_2005} is
obtained if the approximations (i)-(iii) are applied to the renormalization
equations \eqref{PAM13}-\eqref{PAM17}. 

Employing approximation (iv), the desired renormalized Hamiltonian 
$\tilde{\cal H}=\lim_{\lambda\rightarrow 0}\mathcal{H}_{\lambda}$ 
is a free system consisting of two non-interacting fermionic quasi-particle
bands, 
\begin{eqnarray}
  \label{PAM18}
  \tilde{\cal H} &=& 
  \sum_{\mathbf{k},m} \tilde{\varepsilon}_{\mathbf{k}} \,
  c_{\mathbf{k}m}^{\dagger} c_{\mathbf{k}m} \\
  && + \,
  \sum_{\mathbf{k},m} \left(
    \tilde{\varepsilon}_{f} + D \tilde{\Delta}_{\mathbf{k}}
  \right) \,
  f_{\mathbf{k}m}^{\dagger} f_{\mathbf{k}m} + 
  \tilde{E}.
  \nonumber
\end{eqnarray}
Eqs.~\eqref{PAM14} and \eqref{PAM12} can be easily integrated between
$\lambda=0$ and the cutoff $\Lambda$ of the original model,
\begin{eqnarray}
  \label{PAM19}
  \tilde{\Delta}_{\mathbf{k}} &=&
  - \, \frac{1}{D}
  \left[ \tilde{\varepsilon}_{\mathbf{k}} - \varepsilon_{\mathbf{k}} \right],
  \\[1ex]
  \tilde{E} &=&
  -\, N \langle \hat{n}_{i}^{f} \rangle
  \left[ \tilde{\varepsilon}_{f} - \varepsilon_{f} \right] +
  \frac{D-1}{D} \langle \hat{n}_{i}^{f} \rangle \sum_{\mathbf{k}}
  \left[ \tilde{\varepsilon}_{\mathbf{k}} - \varepsilon_{\mathbf{k}} \right]
  \nonumber\\[-1ex]
  \label{PAM20}
  &\approx&
  -\, N \langle \hat{n}_{i}^{f} \rangle
  \left[ \tilde{\varepsilon}_{f} - \varepsilon_{f} \right],
\end{eqnarray}
where approximation (iii) has been used. The equation \eqref{PAM13} can also
be solved if the renormalizations of the different ${\bf k}$ values are 
decoupled from each other by approximations (i) and (ii). Thus, 
Eq.~\eqref{PAM16} can be rewritten as 
\begin{eqnarray*}
  \alpha_{\mathbf{k}}(\lambda) &=&
  \frac{1}{
    \tilde{\varepsilon}_{f} + \varepsilon_{\mathbf{k}} -
    2 \varepsilon_{\mathbf{k},\lambda}
  } \,
  \frac{\mathrm{d}V_{\mathbf{k},\lambda}}{\mathrm{d}\lambda}
\end{eqnarray*}
and inserted into \eqref{PAM13} so that we obtain
\begin{eqnarray}
  \label{PAM21}
  0 &=& 
  \frac{\mathrm{d}}{\mathrm{d}\lambda} \left\{
    \varepsilon_{\mathbf{k},\lambda}^{2} - 
    \left( \tilde{\varepsilon}_{f} + \varepsilon_{\mathbf{k}} \right)
    \varepsilon_{\mathbf{k},\lambda} + 
    D V_{\mathbf{k},\lambda}^{2}
  \right\}.
\end{eqnarray}
Eq.~\eqref{PAM21} can easily be integrated and a quadratic equation for 
$
  \tilde{\varepsilon}_{\mathbf{k}} = 
  \lim_{\lambda\rightarrow 0}\varepsilon_{\mathbf{k},\lambda}
$
is obtained. Our recent work on the PAM \cite{HB_2005,HB_2006} has shown that
the quasi-particles in the final Hamiltonian $\tilde{\mathcal{H}}$
\eqref{PAM18} do not change their ($c$ or $f$) character as function of the
wave vector $\mathbf{k}$. Therefore, $\tilde{\varepsilon}_{\mathbf{k}}$ jumps
between the two solutions of the obtained quadratic equation in order to
minimize its deviations from the original $\varepsilon_{\mathbf{k}}$,
\begin{eqnarray}
  \label{PAM22}
  \tilde{\varepsilon}_{\mathbf{k}} &=& 
  \frac{\tilde{\varepsilon}_{f} + \varepsilon_{\mathbf{k}}}{2} - 
  \frac{\mathrm{sgn}(\tilde{\varepsilon}_{f} - \varepsilon_{\mathbf{k}})}{2} 
  W_{\mathbf{k}} ,
  \\
  \label{PAM23}
  W_{\mathbf{k}} &=& 
  \sqrt{
    \left( \varepsilon_{\mathbf{k}} - \tilde{\varepsilon}_{f} \right)^{2} + 
    4D \left| V_{\mathbf{k}} \right|^{2}
  }.
\end{eqnarray}
The second quasi-particle band is given by
\begin{eqnarray}
  \tilde{\omega}_{\mathbf{k}} &:=&
  \tilde{\varepsilon}_{f} + D \tilde{\Delta}_{\mathbf{k}} \,=\,
  \frac{\tilde{\varepsilon}_{f} + \varepsilon_{\mathbf{k}}}{2} +
  \frac{\mathrm{sgn}(\tilde{\varepsilon}_{f} - \varepsilon_{\mathbf{k}})}{2} 
  W_{\mathbf{k}} .
  \nonumber\\[-1ex]
  \label{PAM24} &&
\end{eqnarray}
Thus, we have obtained the same effective Hamiltonian \eqref{PAM18} and the
same quasi-particle energies \eqref{PAM22} and \eqref{PAM24} as found in
Ref.~\onlinecite{HB_2005}. 

Finally, we need to determine the renormalized $f$ energy
$\tilde{\varepsilon}_{f}$ and the expectation values. Because the final
renormalized Hamiltonian \eqref{PAM18} consists of non-interacting fermionic
quasi-particles, it is straightforward to calculate all desired quantities
from the free energy as it was done in Ref.~\onlinecite{HB_2005}. Because 
the effective model $\tilde{\mathcal{H}}$ is connected with the original
Hamiltonian $\mathcal{H}$ by an unitary transformation the free energy can
also be calculated from $\tilde{\mathcal{H}}$,
\begin{eqnarray*}
  F &=& -\, \frac{1}{\beta} \ln \mathrm{Tr} \, e^{-\beta\tilde{\mathcal{H}}}.
\end{eqnarray*}
The expectation value of the $f$ occupation is found from the free energy by
functional derivative,
\begin{eqnarray}
  \label{PAM25}
  \langle \hat{n}_{i}^{f} \rangle &=&
  \frac{1}{N} \frac{\partial F}{\partial \varepsilon_{f}} \,=\,
  \frac{1}{N} \left\langle
    \frac{\partial \tilde{\mathcal{H}}}{\partial \varepsilon_{f}}
  \right\rangle_{\tilde{\mathcal{H}}} .
\end{eqnarray}
Thus, we finally obtain a relation of the following structure
\begin{eqnarray}
  \label{PAM26}
  0 &=&
  \{ \dots \} \left( 
    \frac{\partial \tilde{\varepsilon}_{f}}{\partial \varepsilon_{f}} 
  \right) + 
  \{ \dots \} \left(
    \frac{\partial \langle \hat{n}_{i}^{f} \rangle}{\partial \varepsilon_{f}} 
  \right).
\end{eqnarray}
In the cases of mixed valence and heavy Fermion behavior the derivatives in
Eq.~\eqref{PAM26} are non-zero so that both brace expressions can be set equal
to zero to find equations of self-consistency for the renormalized $f$ level
and the averaged $f$ occupation number,
\begin{eqnarray}
  \label{PAM27}
  \langle \hat{n}_{i}^{f} \rangle &=&
  \frac{\nu_{f}}{N}
  \sum_{\mathrm{k}}
  f(\tilde{\varepsilon}_{\mathbf{k}})
  \left\{
    \frac{1}{2} +
    \mathrm{sgn}(\tilde{\varepsilon}_{f} - \varepsilon_{\mathbf{k}})
    \frac{\varepsilon_{\mathbf{k}}-\tilde{\varepsilon}_{f}}{2W_{\mathbf{k}}} \,
  \right\} \\[1ex]
  && +\,
  \frac{\nu_{f}}{N} \sum_{\mathbf{k}}
  f(\tilde{\omega}_{\mathbf{k}})
  \left\{
    \frac{1}{2} +
    \mathrm{sgn}(\varepsilon_{\mathbf{k}} - \tilde{\varepsilon}_{f} )
    \frac{\varepsilon_{\mathbf{k}}-\tilde{\varepsilon}_{f}}{2W_{\mathbf{k}}} \,
  \right\} ,
  \,\phantom{a}
  \nonumber
\end{eqnarray}
\begin{eqnarray}
  \label{PAM28}
  \tilde{\varepsilon}_{f} - \varepsilon_{f} &=&
  \frac{\nu_{f}-1}{N}
  \sum_{\mathbf{k}}
  \mathrm{sgn}(\tilde{\varepsilon}_{f} - \varepsilon_{\mathbf{k}}) \,
  f(\tilde{\varepsilon}_{\mathbf{k}})
  \frac{|V_{\mathrm{k}}|^{2}}{W_{\mathbf{k}}} \\[1ex]
  && + \,
  \frac{\nu_{f}-1}{N}
  \sum_{\mathbf{k}}
  \mathrm{sgn}(\varepsilon_{\mathbf{k}} - \tilde{\varepsilon}_{f} ) \,
  f(\tilde{\omega}_{\mathbf{k}})
  \frac{|V_{\mathrm{k}}|^{2}}{W_{\mathbf{k}}}
  \nonumber.
\end{eqnarray}
These equations are quite similar to the results of the SB theory
\cite{FKZ_1988}. In particular, the limit $\nu_{f}\rightarrow\infty$ of
Eqs.~\eqref{PAM27} and \eqref{PAM28} leads to the SB equations. Note that
expectation values $\langle c_{\mathbf{k}m}^{\dagger} c_{\mathbf{k}m} \rangle$
and 
$
  \langle 
    \hat{f}_{\mathbf{k}m}^{\dagger} c_{\mathbf{k}m} + \mathrm{h.c.}
  \rangle
$
can be calculated similar to Eq.~\eqref{PAM25}, see Ref.~\onlinecite{HB_2005}
for details.

\subsection{Numerical solution} 

Note that for the analytical solution in the preceeding subsection 
an explicit expression for the 
generator $A'' (\lambda, \Delta \lambda)$, was not needed. 
The reason was that a   $\lambda$ independent
$f$ electron energy $\varepsilon_{f, \lambda}$ was
assumed in close analogy to what is done in the well known slave boson 
mean field approach for the periodic Anderson model. 
For an improved treatment  an explicit expression 
for $A''_{\bf k}(\lambda,\Delta \lambda)$ should be used. 
Following the discussion in subsection \ref{FA_rev}
we make the following  ansatz for 
$A''_{\mathbf{k}}(\lambda,\Delta\lambda)$ 
\begin{eqnarray}
  \label{PAM7}
  \lefteqn{A''_{\mathbf{k}}(\lambda,\Delta\lambda) \,=\,} && \\
  &=&
  \frac{
    \left(
        e_{f,\lambda} + 
        D\left( \Delta_{\mathbf{k},\lambda} - \bar{\Delta}_{\lambda} \right) - 
        \varepsilon_{\mathbf{k},\lambda}
    \right)
    V_{\mathbf{k},\lambda}
  }{
    \kappa
    \left[
      \lambda - \left|
        e_{f,\lambda} + 
        D\left( \Delta_{\mathbf{k},\lambda} - \bar{\Delta}_{\lambda} \right) - 
        \varepsilon_{\mathbf{k},\lambda}
      \right|
    \right]^{2}
  }
  \, \Delta\lambda  \nonumber .
\end{eqnarray}


\begin{figure}
  \begin{center}
   \scalebox{0.65}{
\includegraphics*{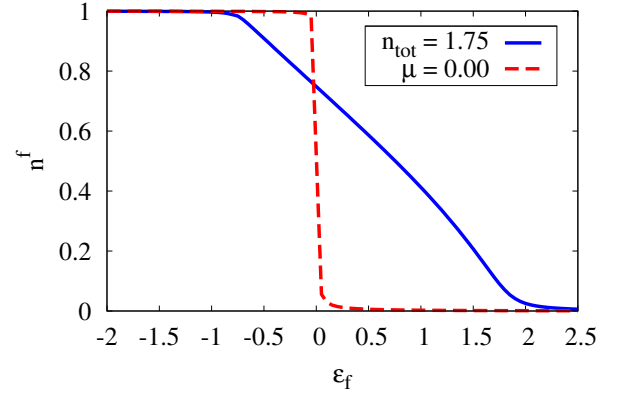}
    }
  \end{center}
  \caption{(Color online) $f$-electron occupation number 
$n^f= \langle \hat{n}_i^f \rangle$ as function 
of the bare energy $\varepsilon_f$
for an one-dimensional lattice with $10 000$ sites for two cases:  
i) the total particle occupation  $n_{\mbox{tot}}= n^f + n^c
=1.75$ is fixed (in red)  and (ii) the chemical potential $\mu$ (in green)
is fixed. Moreover, $\nu_f = 2$, $ V = 0.10 (4t) $ and the 
temperature $T = 0$}
  \label{Fig_n_f}
\end{figure}

In the limit of
small $\Delta\lambda$, we again expect an exponential decay for the
hybridization $V_{\mathbf{k},\lambda}$ in this way. In Eq.~\eqref{PAM7},
$\kappa$ denotes an energy constant to ensure a dimensionless 
$A''_{\mathbf{k}}(\lambda,\Delta\lambda)$. Note that
$A''_{\mathbf{k}}(\lambda,\Delta\lambda)$ is chosen proportional to
$\Delta\lambda$ to reduce the impact of the actual value of $\Delta\lambda$ on 
the final results of the renormalization. 
Using \eqref{PAM12a} and \eqref{PAM7} the basic renormalization equations 
\eqref{PAM13} - \eqref{PAM17} was solved 
numerically in Ref.~\onlinecite{Mai_2007}. 

FIG.~\ref{Fig_n_f} shows the $f$ 
occupation $n^f =\langle \hat{n}_i^f \rangle $ as 
function of the bare $f$ energy $\varepsilon_f$ at degeneracy $\nu_f=2$ for two
cases, (i) for fixed total particle occupation  $n_{\mbox{tot}}= n^f + n^c
=1.75$ (in red) and (ii) for fixed chemical potential $\mu$ (in green). 
Here, $n^c = (1/N)\sum_{{\bf k},\sigma} \langle c_{{\bf k},
  \sigma}^\dagger  c_{{\bf k}, \sigma} \rangle$ is the 
conduction electron occupation. For the first case the result from the PRM
approach shows a rather smooth decay from the integer valence region with
$n^f= 1$, when $\varepsilon_f$ is located far below the  Fermi level, to an 
empty state with no $f$ electrons $n^f=0$, when $\varepsilon_f$ is far above 
the Fermi level (black line). Note that this analytical PRM result almost
completely agrees with the result from recent DMRG calculations 
from Ref.~\onlinecite{Myake_2006} for the
same parameter values. For comparison, the figure also
contains a curve obtained from the PRM approach when 
the chemical potential $\mu$ instead of $n_{\mbox{tot}}$
was fixed in the calculation (red curve). Note that in
this case $n^f$ as function of $\varepsilon_f$
shows an abrupt change from an completely filled to an empty $f$ state. 
Obviously the latter behavior can easily be understood as change of the $f$
charge when $\varepsilon_f$ crosses the fixed chemical potential. In contrast,
for fixed total occupation $n_{\mbox{tot}}$ the Fermi level is shifted
upwards, when the $f$ level is partially depleted when    $\varepsilon_f$
comes closer to the Fermi level. 
For details we refer to Ref.\onlinecite{Mai_2007}.


\section{
 Crossover behavior in the metallic one-dimensional Holstein model 
}
\label{Holstein_CROSS}

In this section we discuss the one-dimensional Holstein model. 
As is well known, this model shows a 
quantum phase transition between a metallic and a charge ordered state
as function of the electron-phonon coupling. 
In the present section we restrict ourselves to the metallic state.

Let us start with the Hamiltonian of the one-dimensional Holstein model 
of spinless fermions (HM) which reads,
\begin{eqnarray}
  \label{HM1}
  {\cal H} &=&
  - t \sum_{\langle i,j\rangle} ( c_{i}^\dagger c_{j} + \mathrm{h.c.} )
  + \omega_0 \sum_i  \; b_i^\dagger b_i \\
  &&
  + \, g \sum_i \; (b_i^\dagger + b_i)n_i .
  \nonumber
\end{eqnarray}
This model is perhaps the simplest realization of an electron-phonon (EP)
system and describes the interaction between the local electron
density $n_{i} = c^{\dagger}_{i}c_{i}$ and dispersion-less phonons with 
frequency $\omega_0$. Here, the $c^{\dagger}_{i}$ ($b^{\dagger}_{i}$) denote 
creation operators of electrons (phonons), and the summation 
$\langle i,j\rangle$ runs over all pairs of neighboring lattice sites. With 
increasing EP coupling $g$, the HM undergoes the quantum-phase transition from 
a metallic to a charge-ordered insulating state. At half-filling, the 
insulating state of the HM is a dimerized Peierls phase.

Because the HM is not exactly solvable, a number of different analytical and
numerical methods have been applied: strong coupling expansions
\cite{HF_1983}, Monte Carlo simulations \cite{HF_1983, MHM_1996}, variational
\cite{ZFA_1989} and renormalization group \cite{HM_RG} approaches, exact
diagonalization (ED) techniques \cite{HM_ED}, density matrix renormalization
group \cite{BMH_1998,JZW_1999,FWH_2005} and dynamical mean-field theory
(DMFT) \cite{MHB_2002}. However, most of these approaches are restricted in
their application, and the infinite phononic Hilbert space (even for finite
systems) demands the application of truncation schemes 
in numerical methods or involved reduction procedures.

The PRM represents an alternative analytical approach. In the following the
PRM is applied to the HM where we mainly follow
Refs.~\onlinecite{SHB_2006_1}, and \onlinecite{SHB_2006_2}. Here we focus on the  
investigation of the change of physical properties by passing from the 
adiabatic to the anti-adiabatic limit.   
Furthermore, we discuss electronic and phononic
quasi-particle energies as well as the impact of the system filling.

\subsection{Metallic solutions}
\label{HM_metal}

For the metallic phase of the HM a very simple renormalization scheme 
is sufficient where only the electronic and phononic one-particle 
energies are renormalized.

Following Refs.~\onlinecite{SHBWF_2005} and \onlinecite{SHB_2006_1}, we make
the following ansatz for the renormalized Hamiltonian
\begin{eqnarray}
  \label{HM2}
  \mathcal{H}_{\lambda} &=& \mathcal{H}_{0,\lambda} + \mathcal{H}_{1,\lambda}, 
  \\[1ex]
  \mathcal{H}_{0,\lambda} &=&
  \sum_{k} \varepsilon_{k,\lambda} c^{\dagger}_{k}c_{k} +
  \sum_{q} \omega_{q,\lambda} b^{\dagger}_{q} b_{q} + E_{\lambda}, 
  \nonumber\\[1ex]
  \mathcal{H}_{1,\lambda} &=&
  \frac{g}{\sqrt{N}}\sum_{k,q}
  \Theta_{k,q,\lambda} \,
  \left(
    b^{\dagger}_{q} c^{\dagger}_{k}c_{k+q} +
    b_{q} c^{\dagger}_{k+q}c_{k}
  \right)
  \nonumber
\end{eqnarray} 
Here, all excitations with energies larger than a given cutoff
$\lambda$ are thought to be integrated out. Moreover, we have 
defined $\Theta_{k,q,\lambda} =
\Theta( \lambda - |\omega_{q,\lambda} + \varepsilon_{k,\lambda} -
\varepsilon_{k+q,\lambda}|)$. Note that Fourier-transformed
one-particle operators have been used for convenience. 
Next, all transitions 
within the energy shell between $\lambda-\Delta\lambda$ and $\lambda$ 
will be removed by use of a unitary transformation (Eq.~\eqref{B16}),
\begin{eqnarray}
  \label{HM3}
  \mathcal{H}_{(\lambda-\Delta\lambda)} &=& 
  e^{X_{\lambda, \Delta\lambda}}\; \mathcal{H}_{\lambda}\;  
  e^{-X_{\lambda, \Delta\lambda}} \;,
\end{eqnarray}
where the following ansatz is made for the generator
$X_{\lambda,\Delta\lambda}$ of the transformation
\begin{eqnarray}
  X_{\lambda,\Delta\lambda} &=&
  \frac{1}{\sqrt{N}}\sum_{k,q} A_{k,q}(\lambda,\Delta\lambda)
  \left(
    b^{\dagger}_{q} c^{\dagger}_{k}c_{k+q} -
    b_{q} c^{\dagger}_{k+q}c_{k}
  \right).
  \nonumber \\[-1ex]
  \label{HM4} &&
\end{eqnarray}
The part
$\mathbf{P}_{(\lambda-\Delta\lambda)}X_{\lambda,\Delta\lambda}$ has been
set equal to zero. Therefore $A_{k,q}(\lambda,\Delta\lambda)$ reads
\begin{eqnarray*}
  A_{k,q}(\lambda,\Delta\lambda) &=& 
  A'_{k,q}(\lambda,\Delta \lambda) \, 
  \Theta_{k,q,\lambda}\,[1 - \Theta_{k,q,\lambda - \Delta \lambda}].
\end{eqnarray*}
As before, the ansatz \eqref{HM4} is suggested by the form of 
the first order expression \eqref{B17b} of the generator 
$X_{\lambda,\Delta\lambda}$. Later, 
the coefficients $A'_{k,q}(\lambda,\Delta \lambda)$ will be fixed in a way that
$\mathbf{Q}_{(\lambda-\Delta\lambda)}\mathcal{H}_{(\lambda-\Delta\lambda)}=0$
is fulfilled, so that $\mathcal{H}_{(\lambda-\Delta\lambda)}$ contains no
transitions larger than the new cutoff $\lambda-\Delta\lambda$. 

By evaluating \eqref{HM3}, terms with four
fermionic and bosonic one-particle operators and higher order terms 
are generated. In order to restrict the renormalization scheme to the terms 
included in the ansatz \eqref{HM2}, a factorization approximation has to be
employed, 
\begin{eqnarray*}
  c^{\dagger}_{k}c_{k} c^{\dagger}_{k-q}c_{k-q} &\approx&
  c^{\dagger}_{k}c_{k} \langle c^{\dagger}_{k-q}c_{k-q} \rangle +
  \langle c^{\dagger}_{k}c_{k} \rangle c^{\dagger}_{k-q}c_{k-q} \\
  && -
  \langle c^{\dagger}_{k}c_{k} \rangle
  \langle c^{\dagger}_{k-q}c_{k-q} \rangle, \\[1ex]
  b^{\dagger}_{q} b_{q} c^{\dagger}_{k}c_{k} &\approx&
  b^{\dagger}_{q} b_{q} \langle c^{\dagger}_{k}c_{k} \rangle +
  \langle b^{\dagger}_{q} b_{q} \rangle c^{\dagger}_{k}c_{k}  -
  \langle b^{\dagger}_{q} b_{q} \rangle
  \langle c^{\dagger}_{k}c_{k} \rangle .
\end{eqnarray*}
In this way, it is possible to sum up the series expansion from
transformation \eqref{HM3}. 

The parameters $A'_{k,q}(\lambda,\Delta \lambda)$ as well as 
the renormalization equations for
$\varepsilon_{k,\lambda}$, $\omega_{q,\lambda}$, $g_{k,q,\lambda}$, and
$E_{\lambda}$ can be found by comparing the final result obtained 
from the explicit evaluation of the unitary transformation \eqref{HM3} with 
the renormalization ansatz \eqref{HM2}, where  $\lambda$ is 
replaced by $\lambda - \Delta \lambda$. 
The result is given in Ref.~\onlinecite{SHBWF_2005}. It can be further 
simplified in the thermodynamic limit $N \rightarrow \infty$. By expanding  
the renormalization equations from Ref.~\onlinecite{SHBWF_2005} in powers of
$g$, one finds that only terms of quadratic or linear order in $g$ survive. 
The final equations read
\begin{eqnarray}
  \label{TLe}
  \lefteqn{ 
    \varepsilon_{k,(\lambda - \Delta \lambda)} - \varepsilon_{k,\lambda}
    \, = \,
  }
  && \\
  &=&
  \frac{1}{N} \sum_{q} \left( n_q^{\rm b} + n_{k+q}^{\rm c} \right) 
  \frac{g^2 \Theta_{k,q}(\lambda,\Delta \lambda)}
  {\omega_{q,\lambda} + \varepsilon_{k,\lambda} - \varepsilon_{k+q,\lambda}}
  \nonumber
  \\
  && - \frac{1}{N} 
  \sum_{q} \left( n_q^{\rm b} - n_{k-q}^{\rm c} + 1\right) 
  \frac{g^2 \Theta_{k-q,q}(\lambda,\Delta \lambda)}
  {\omega_{q,\lambda} + \varepsilon_{k-q,\lambda} - \varepsilon_{k,\lambda}},
  \nonumber
\end{eqnarray}
\begin{eqnarray}
  \label{TLw}
  \lefteqn{ 
    \omega_{q,(\lambda - \Delta \lambda)} - \omega_{q,\lambda}
    \, = \,
  }
  && \\
  &=&
  \frac{1}{N} \sum_{k} \left( n_k^{\rm c} 
   - n_{k+q}^{\rm c} \right) \frac{g^2 \Theta_{k,q}(\lambda,\Delta
  \lambda)}{\omega_{q,\lambda} + \varepsilon_{k,\lambda} 
  - \varepsilon_{k+q,\lambda}}
  \nonumber
\end{eqnarray}
where $n_k^{\rm c} = \langle c_k^\dag c_k
\rangle$, $n_q^{\rm b} = \langle b_q^\dag b_q
\rangle$, and $\Theta_{k,q}(\lambda,\Delta
  \lambda) = \Theta_{k,q,\lambda}\,[1 - \Theta_{k,q,\lambda
  - \Delta \lambda}]$.

Note that the renormalization equations still depend on unknown 
expectation values $\langle c_{k}^{\dagger} c_{k} \rangle$ and
$\langle b_{q}^{\dagger} b_{q} \rangle$ which follow from the 
factorization approximation. Following
Ref.~\onlinecite{SHB_2006_1}, they are best evaluated with
respect to the full Hamiltonian $\mathcal{H}$. 

Exploiting
$
 \langle\mathcal{A}\rangle = 
 \lim_{\lambda\rightarrow 0}\langle
   \mathcal{A}_{\lambda}
 \rangle_{\mathcal{H}_{\lambda}}
$,
we derive additional renormalization equations for
the fermionic and bosonic one-particle operators, $c_{k}^{\dagger}$ and
$b_{q}^{\dagger}$. They have the following form 
 according to Refs.~\onlinecite{SHBWF_2005} and \onlinecite{SHB_2006_2}, 
\begin{eqnarray}
  c_{k,\lambda}^{\dagger} &=& 
  \alpha_{k,\lambda} \, c_{k}^{\dagger} + 
  \sum_{q} \left(
    \beta_{k,q,\lambda} \, c_{k+q}^{\dagger} b_{q} + 
    \gamma_{k,q,\lambda} \, c_{k-q}^{\dagger} b_{q}^{\dagger}
  \right), 
  \nonumber \\[-1ex]
  \label{HM5} && \\
  \label{HM6}
  b_{q,\lambda}^{\dagger} &=&
  \phi_{q,\lambda} \, b_{q}^{\dagger} + \eta_{q,\lambda} \, b_{-q} +
  \sum_{k} \psi_{k,q,\lambda} \, c_{k+q}^{\dagger} c_{k} .
\end{eqnarray}

The set of renormalization equations has to be solved self-consistently:
One chooses some values for the expectation values. With these values,
the numerical
evaluation starts from the cutoff $\Lambda$ of the original model
$\mathcal{H}$ and proceeds step by step to $\lambda = 0$. For $\lambda = 0$, 
the Hamiltonian and the one-particle operators are fully 
renormalized. The case $\lambda = 0$ allows the re-calculation of all
expectation values, and the renormalization procedure starts again
with the improved expectation values 
by reducing again the cutoff from $\Lambda$ to $\lambda=0$. 
After a sufficient number of such cycles, the expectation values are
converged and the renormalization equations are solved
self-consistently. Thus, we finally obtain an effectively free model,
\begin{eqnarray}
  \label{HM7}
  \tilde{\mathcal{H}} &=&
  \sum_{k} \tilde{\varepsilon}_{k}
  c^{\dagger}_{k}c_{k} +
  \sum_{q} \tilde{\omega}_{q}
  b^{\dagger}_{q} b_{q} +
  \tilde{E},
\end{eqnarray}
where we have introduced the renormalized dispersion relations
$\tilde{\varepsilon}_{k}=\lim_{\lambda\rightarrow 0}\varepsilon_{k,\lambda}$
and 
$\tilde{\omega}_{q}=\lim_{\lambda\rightarrow 0}\omega_{q,\lambda}$, and the
energy shift $\tilde{E}=\lim_{\lambda\rightarrow 0}E_{\lambda}$. 

For the numerical evaluation of the renormalization equations we choose a
lattice size of $N=1000$ sites. The temperature is fixed to $T=0$.

\subsection{Adiabatic case}

\begin{figure}
  \begin{center}
    \scalebox{0.71}{
      \includegraphics*{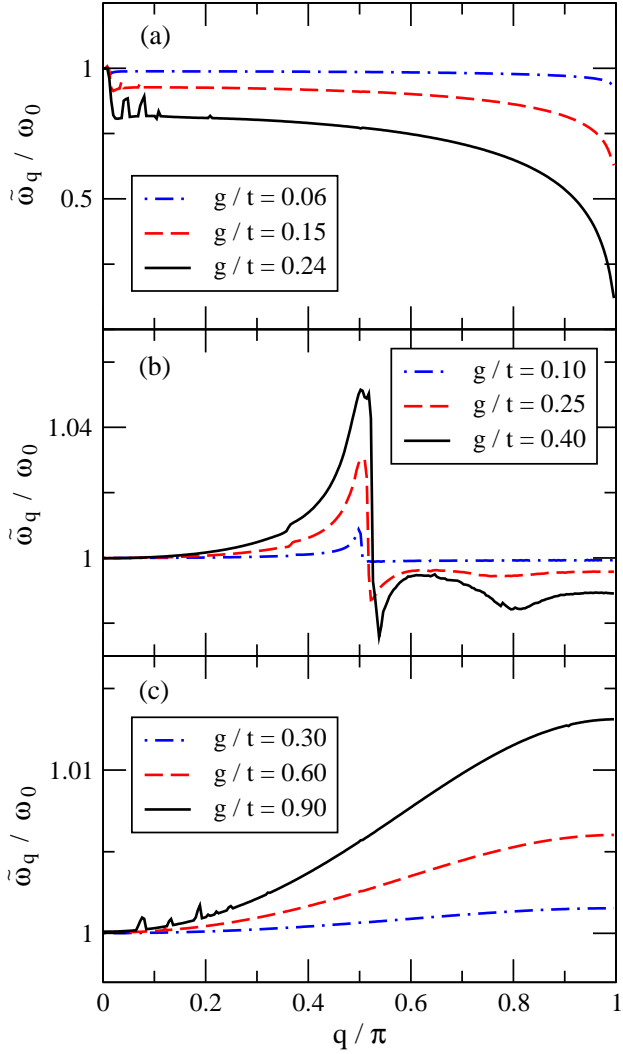}
    }
  \end{center}
  \caption{
    (Color online) Bosonic quasi-particle energies $\tilde{\omega}_{q} /
    \omega_0$ at half-filling as function of $q$ for 
    different values of the EP coupling $g$
    in the adiabatic case $\omega_{0} / t = 0.05$ (panel (a)), the
    intermediate case $\omega_{0} / t = 2.8$ (panel (b)),
    and the anti-adiabatic case $\omega_{0} / t = 6.0$ (panel (c)). 
  }
  \label{Fig_phonon_dispersion_2}
\end{figure}

\begin{figure}
  \begin{center}
    \scalebox{0.71}{
      \includegraphics*{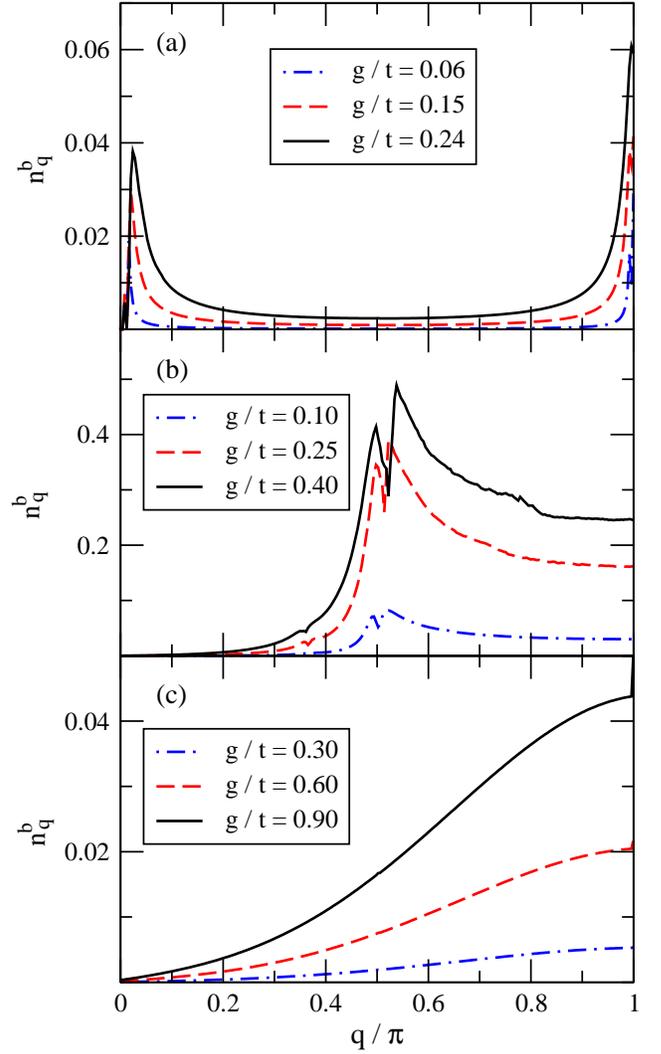}
    }
  \end{center}
  \caption{
    (Color online) Phonon distribution $n_q^{\rm b} =
    \langle b_q^\dag b_q \rangle$ as function of $q$ for the same parameters
    as in Fig.~\ref{Fig_phonon_dispersion_2}.
  }
  \label{Fig_phonon_expect}
\end{figure}

\begin{figure}
  \begin{center}
    \scalebox{0.71}{
      \includegraphics*{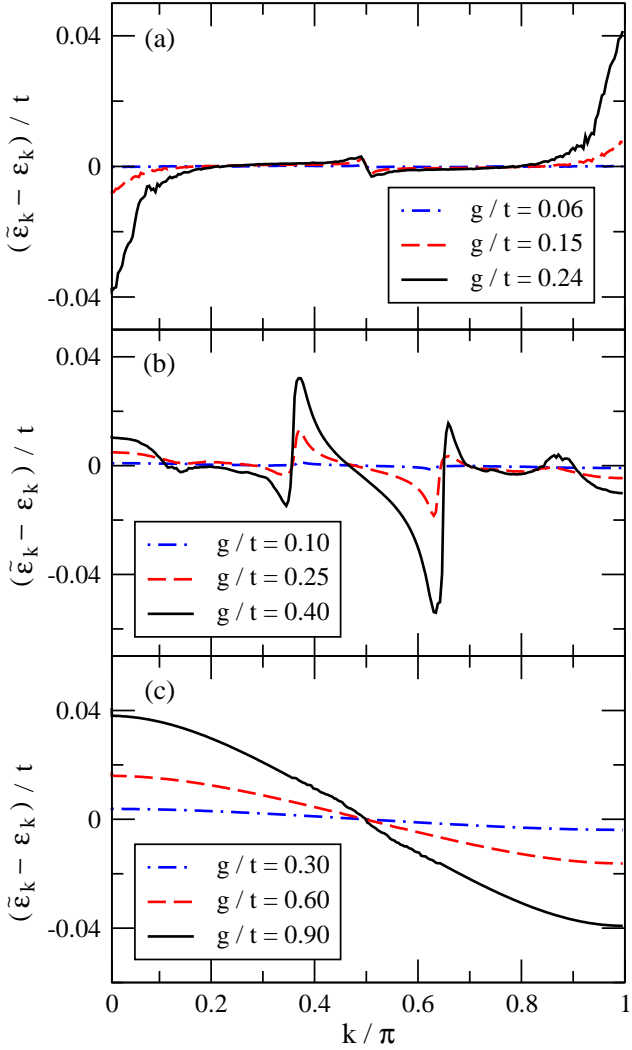}
    }
  \end{center}
  \caption{
    (Color online) Fermionic quasi-particle energies $(\tilde{\varepsilon}_{k}
    - \varepsilon_{k}) / t$ as function of $k$ for the same parameters
    as in Fig.~\ref{Fig_phonon_dispersion_2}. Here $\varepsilon_{k}$ is the
    original electronic dispersion.
  }
  \label{Fig_electron_dispersion}
\end{figure}

\begin{figure}
  \begin{center}
    \scalebox{0.65}{
      \includegraphics*{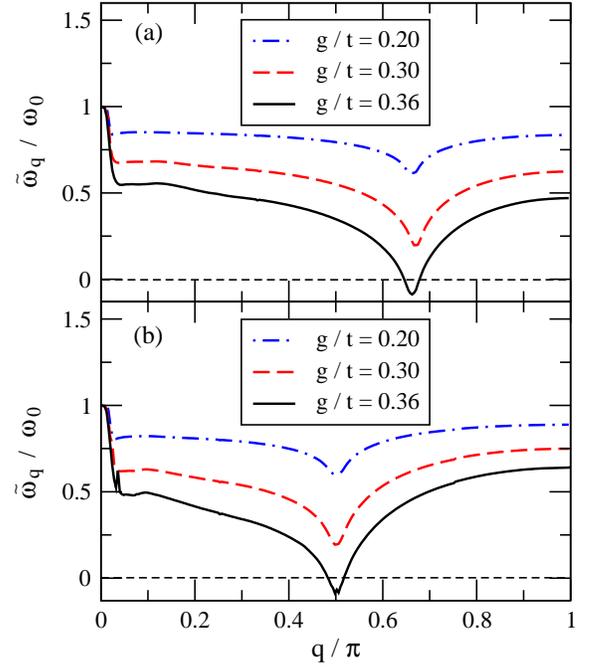}
    }
  \end{center}
  \caption{
    (Color online) (a) Phononic quasi-particle energy $\tilde{\omega}_{q}$ in 
    unit of $\omega_0$ of the one-dimensional HM
    with 500 lattice sites for filling 1/3 and
    different values of the EP coupling $g$. $\omega_{0}/t=0.05$. (b)
    Same quantity $\tilde{\omega}_{q} / \omega_0$ for filling 1/4.
  }
  \label{Fig_phonon_dispersion}
\end{figure}

At first, let us discuss our results for the so-called 
adiabatic case $\omega_{0} \ll
t$. They are shown in panel (a) of Figs.~\ref{Fig_phonon_dispersion_2},
\ref{Fig_phonon_expect}, \ref{Fig_electron_dispersion}, 
and in panels (a) and (b) of Fig.~\ref{Fig_phonon_dispersion}. First, 
according to Fig.~\ref{Fig_phonon_dispersion_2}a the phononic
quasi-particle energies $\tilde{\omega}_{q}$ (half-filling) are found 
to gain dispersion due to the coupling between electronic and 
phononic degrees of freedom in particular around $q=\pi$. Furthermore, 
if the 
coupling exceeds a critical value $g_{c}$ non-physical negative
energies at $q=\pi$ occur. This feature signals the break-down of the
present description for the metallic phase at the quantum-phase transition to
the insulating Peierls state. 


Whereas at half-filling the phonon softening occurs at the Brillouin-zone
boundary, soft phonon modes are found at $2k_{F}=2\pi/3$ 
and at $2k_{F}=\pi/2$ for filling $1/3$ and $1/4$, respectively. 
This can be seen in Fig.~\ref{Fig_phonon_dispersion}.
Since the phonon softening can be considered as a precursor effect of the
metal-insulator transition, the type of the broken symmetry in the
insulating phase \textit{strongly} depends on the filling of the
electronic band. Note that  
the critical EP coupling $g_{c}$ of the phase transition may be 
determined from the vanishing of the phonon mode 
(see Ref.~\onlinecite{SHBWF_2005}). At half-filling and for $\omega_0 = 0.1t$, 
a value of $g_{c}=0.31t$ is found, which is somewhat
larger than the DMRG result of $g_{c}=0.28t$ of
Refs.~\onlinecite{BMH_1998} and \onlinecite{FWH_2005}. In
subsection~\ref{HM_uniform} the determination of the critical coupling $g_c$
within our PRM approach will be discussed in more detail.

Fig.~\ref{Fig_phonon_expect}a shows the phonon distribution
$n_q^{\rm b} = \langle b_q^\dag b_q \rangle$ for the same parameter values
as in Fig.~\ref{Fig_phonon_dispersion_2}a. There are two pronounced
maxima found at wave numbers $q = \pi$ and $q \approx 0$. The peak at $q
= \pi$ is directly connected to the softening of $\tilde{\omega}_q$ at 
the zone boundary and can therefore be considered as
a precursor  of the transition to a dimerized state. For 
the critical EP coupling $g= g_c$ a divergency of
$n_q^{\rm b}$ should appear at $q = \pi$. 
The second peak around $q\approx 0$ follows from renormalization
contributions which become strong for small $q$ for the adiabatic 
case $\omega_0 \ll t$. This will be explained in more detail 
in the discussion part below.   
    
Finally, in Fig.~\ref{Fig_electron_dispersion}a the renormalized fermionic
one-particle energy $\tilde{\varepsilon}_k$ is shown in relation to  
the original dispersion $\varepsilon_k = -2t \cos ka$ for the same parameter
values as in Fig.~\ref{Fig_phonon_dispersion_2}a. 
Though the absolute changes are quite small, the difference between
$\tilde{\varepsilon}_k$ and $\varepsilon_k$ is strongest in the
vicinity of $k=0$ and $k=\pi$. In particular, we find $\tilde{\varepsilon}_k <
\varepsilon_k$ for $k=0$ and $\tilde{\varepsilon}_k >
\varepsilon_k$ for $k=\pi$, so that the renormalized
bandwidth becomes larger than 4t, i.e.~larger than the original bandwidth.

\subsection{Intermediate case}

Next, let us discuss the results for phonon frequencies $\omega_0$ 
of the  order of the hopping matrix element $t$
(intermediate case). The results are found in the panels (b) of 
Figs.~\ref{Fig_phonon_dispersion_2},
\ref{Fig_phonon_expect}, \ref{Fig_electron_dispersion}. In contrast to the
adiabatic case, the renormalized phonon energy   $\tilde{\omega}_q$
(Fig.~\ref{Fig_phonon_dispersion_2}b) now 
shows a noticeable 'kink' at an intermediate wave vector 
(for $\omega_0 / t = 2.8$).  This particular $q$ value, which will be
called $q_k$ in the following strongly
depends on the initial phonon energy $\omega_0$.
The appearance of such a 'kink' at $q_k < \pi$ is a specific feature of the
intermediate case. The wave number $q_k$ is characterized by a strong
renormalization of the phonon energy in a small $q$-range around $q_k$, where
$\tilde{\omega}_q / \omega_0 > 1$ for $q < q_k$ and $\tilde{\omega}_q /
\omega_0 < 1$ for $q > q_k$ holds. The origin of these
features will be discussed in more detail below.    

Similar to $\tilde{\omega}_q$, also the phonon distribution $n_q^{\rm b}$ in 
Fig.~\ref{Fig_phonon_expect}b shows a pronounced structure of considerable
weight around $q_k$. Finally, in 
Fig.~\ref{Fig_electron_dispersion}b the difference of the  fermionic
one-particle energies $(\tilde{\varepsilon}_k - \varepsilon_k)$ is shown.
Again a remarkable structure is found, though the absolute changes are  
small for the present $g$-values.

\subsection{Anti-adiabatic case}

Finally, let us discuss the results for the anti-adiabatic case $\omega_0
\gg t$. In panels (c) of 
Figs.~\ref{Fig_phonon_dispersion_2},
\ref{Fig_phonon_expect}, \ref{Fig_electron_dispersion} a value of
$\omega_0 / t = 6.0$ was used. As most important feature 
a stiffening of the renormalized
phonon frequency  $\tilde{\omega}_q$ (Fig.~\ref{Fig_phonon_dispersion_2}c)
is found instead of a softening as in the adiabatic case.
In particular, for large values of the EP coupling no softening
of the phonon modes is found at $q=\pi$.
Moreover, no large renormalization contributions occur 
in any limited $q$-space regime
which would lead to peak-like structures.  
Instead an overall smooth behavior is found in the entire
Brillouin zone.  

Also the phonon distribution $n_q^{\rm b}$ (Fig.~\ref{Fig_phonon_expect}c)
shows a smooth behavior with a maximum at $q=\pi$. The lack of strong 
peak-like structures in $q$ space indicates that there is 
no phonon mode that gives a dominant contribution 
to the renormalization processes.  

If one compares the renormalized electronic bandwidth for the anti-adiabatic
case (Fig.~\ref{Fig_electron_dispersion}c)
with that of the adiabatic case (Fig.~\ref{Fig_electron_dispersion}a), 
one observes a relatively strong reduction of the bandwidth. 
This indicates the
tendency to localization in the anti-adiabatic case. It also indicates that the
metal-insulator transition in the anti-adiabatic limit can be understood as
the formation of small immobile polarons with electrons surrounded by
clouds of phonon excitations. In the present PRM approach, a renormalized 
one-particle excitation like $\tilde{\varepsilon}_k$
corresponds to a quasiparticle of the coupled many-particle system. 
Therefore, a completely flat $k$ dependence of 
$\tilde{\varepsilon}_k$ would be expected to be found 
in the insulating regime.

\subsection{Discussion}

It may be worthwhile to demonstrate that the PRM approach has the 
advantage that all features of the results for 
$\tilde{\omega}_q$ and $n_q^{\rm b}$ or $\tilde{\varepsilon}_k$
can easily be understood
on the basis of the former renormalization equations. For simplicity, 
we shall restrict ourselves to the case of half-filling and to the 
renormalization of the phonon energies $\tilde{\omega}_{q}$. 

The basic equation is the 
renormalization equation \eqref{TLw}. Due to the 
$\Theta$-functions $\Theta_{k,q}(\lambda,\Delta
 \lambda)$ in all equations a renormalization approximately
  occurs when the energy difference $|\omega_{q,\lambda} 
+ \varepsilon_{k,\lambda} -
\varepsilon_{k+q,\lambda}|$ lies within a small energy shell between $\lambda$
and $\lambda - \Delta \lambda$. As one can see from \eqref{TLw} 
the most dominant renormalization
processes take place for small values of the cutoff $\lambda$. 
Therefore, the largest renormalization contributions come 
from $k$ and $q$ values that fulfill the condition 
\begin{equation}
\label{cond1}
\varepsilon_{k+q,\lambda} - \varepsilon_{k,\lambda} \approx
\omega_{q,\lambda} .
\end{equation}
From \eqref{TLw} directly follows a second condition for the renormalization
contributions to $\omega_{q,\lambda}$. Due to the expectation values
$(n_k^{\rm c} - n_{k+q}^{\rm c})$ in \eqref{TLw} 
the renormalization of $\omega_{q,\lambda}$ is caused from the coupling to
particle-hole excitations. Therefore, 
the energies $\varepsilon_{k,\lambda}$ and
$\varepsilon_{k+q,\lambda}$
have to be either below or above the Fermi level,
i.e. $|k| < k_F$ and $|k+q| > k_F$ or $|k| > k_F$ and $|k+q| < k_F$.

Let us first discuss the adiabatic case $\omega_0 \ll t$. 
The most dominant contributions to the renormalization are expected 
when both conditions are simultaneously fulfilled. 
This is the case for $q \approx \pm \pi$ or partially also 
for $ q \approx 0$. Note that for $q=\pi$ practically all 
$k$-values can contribute 
to the renormalization of \eqref{TLw}, which 
is not the case for $q$-values different from $\pi$. 
For instance, for $q \approx 0$ only few $k$ points from the sum in
\eqref{TLw} can contribute which are located in a small region around the
Fermi momentum $k_F$. On the other hand, for $q\approx 0$, 
the energy denominator is almost zero so that still some noticeable 
renormalization structures are found in Fig.~\ref{Fig_phonon_dispersion_2}a.
Moreover, for the adiabatic case, where $\omega_{q,\lambda}$ is small,
the energy denominator of  \eqref{TLw}  can be replaced by  
$(\varepsilon_{k,\lambda} - \varepsilon_{k+q,\lambda})$.
Therefore, almost all particle-hole
contributions to $\omega_{q,\lambda}$ are negative 
because $(n_k^{\rm c} - n_{k+q}^{\rm c})$
and $(\varepsilon_{k,\lambda} - \varepsilon_{k+q,\lambda})$ 
have always different signs. One concludes that in the
adiabatic case $\omega_{q,\lambda}$ will be renormalized 
to smaller values where 
the renormalization at $q = \pi$ should be dominant.

The behavior of $\tilde{\omega}_q$ for the case 
of intermediate phonon frequencies ($\omega_0 / t = 2.8$ in 
Fig.~\ref{Fig_phonon_dispersion_2}b
and Fig.~\ref{Fig_phonon_expect}b) can again be understood on the basis
of the renormalization equations \eqref{TLw} and condition \eqref{cond1}. 
As was already discussed, particle-hole excitations 
lead to the renormalization of $\omega_{q,\lambda}$. Therefore, 
from the sum over $k$ in Eq.~\eqref{TLw} only $k$ terms
contribute where either  $|k| < k_F$ and $|k+q| > k_F$ 
or $|k| > k_F$ and $|k+q| < k_F$. For the latter case 
always $(\varepsilon_{k,\lambda} - \varepsilon_{k+q,\lambda}) > 0$ is
valid so that \eqref{cond1} can not be fulfilled. Therefore, we can restrict
ourselves to contributions $|k| <
k_F$ und $|k+q| > k_F$, for which  always $(\varepsilon_{k,\lambda} 
- \varepsilon_{k+q,\lambda}) < 0$ and 
$(n_k^{\rm c} - n_{k+q}^{\rm c}) > 0$ holds. 
The largest renormalization should result from 
a small $q$ region around some $q$ vector $q_k$ for which 
$\varepsilon_{k+q_k} -\varepsilon_k = \omega_0$ is 
approximately fulfilled. Since $\omega_0$ is of the order of $t$,
$q_k$ is located somewhere in the middle of the Brillouin zone and depends
strongly on  $\omega_0$. From Eq.~ \eqref{TLw} also follows that 
renormalization contributions to  $\tilde{\omega}_q$ change their sign 
at $q_k$ due to the sign change in the energy denominator. 

Finally, from equation \eqref{TLw} one may point out also the stiffening of the
phonon modes in the anti-adiabatic case  $\omega_0 / t = 6.0$. In this case 
the phonon energy $\omega_0$ is much larger than the electronic 
bandwidth. Therefore, for all $\lambda$ a positive energy
denominator  $(\omega_{q,\lambda} + \varepsilon_{k,\lambda} -
\varepsilon_{k+q,\lambda})$ is obtained. Nevertheless, for half-filling 
in the $k$ sum 
on the right hand side of \eqref{TLw} 
there are as many negative as positive
terms due to the factor  $(n_k^{\rm c} - n_{k+q}^{\rm c})$.
Since from 
$(n_k^{\rm c} - n_{k+q}^{\rm c}) < 0$ always follows 
$(\varepsilon_{k,\lambda} -\varepsilon_{k+q,\lambda}) > 0$, the negative 
terms have larger energy denominators and are always smaller
than the positive terms. The resulting renormalization of $\omega_{q,\lambda}$
is therefore positive for all $q$ values and largest for $q=\pi$ due to the
smallest energy denominator.


\section{Quantum Phase transition in the 
one-dimensional Holstein model} 
\label{Holstein_QP}

In this section we want to demonstrate the ability of the PRM approach to 
describe also quantum phase transitions. In particular, we shall
investigate the transition from the metallic to the insulating charge ordered
phase when the electron-phonon coupling $g$ exceeds a critical value.  

\subsection{
  Uniform description of metallic and insulating phases at half-filling
}
\label{HM_uniform}

In the following we present a uniform description that covers the
metallic as well as the insulating phase of the HM in the adiabatic 
case. We mainly follow the
approach of Ref.~\onlinecite{SHB_2006_1} where we have discussed methodological
aspects in more detail.
As already mentioned above, the simple approach of subsection \ref{HM_metal}
breaks down for EP couplings $g$ larger than some critical 
value $g_c$ where a long-range charge density wave
occurs and the ions are shifted away from their symmetric positions. An
adequate theoretical description needs to take into account a
broken symmetry field. For this purpose, the underlying 
idea of subsection \ref{EP_BCS} to take 
such a term into account in the renormalization ansatz will
be transferred to the present case. As 
one can see from Fig.~\ref{Fig_phonon_dispersion}, the
order parameter of the insulating phase strongly depends on the filling of the
electronic band. Therefore, in the following we restrict ourselves to 
the case of half-filling. Here, the unit cell is doubled and a
dimerization occurs in the insulating phase. 

Following Ref.~\onlinecite{SHB_2006_1}, the Hamiltonian in the
reduced Brillouin zone including symmetry breaking fields reads
\begin{eqnarray}
  \label{HM8}
  \mathcal{H}_{\lambda} &=& 
  \mathcal{H}_{0,\lambda} + \mathcal{H}_{1,\lambda},  \\[1ex]
  \mathcal{H}_{0,\lambda} &=&  
  \sum_{k>0,\alpha}
  \varepsilon_{\alpha,k,\lambda}
  c_{\alpha,k}^{\dag} c_{\alpha,k} +
  \sum_{q>0, \gamma}
  \omega_{\gamma,q,\lambda}
  b_{\gamma,q}^{\dag} b_{\gamma,q} 
  \nonumber \\
  && + \,
  E_{\lambda} +
  \sum_{k}
  \Delta_{k,\lambda}^{\mathrm{c}}
  \left(
    c_{0,k}^{\dag} c_{1,k} + \mathrm{h.c.}
  \right)
  \nonumber \\
  && + \,
  \sqrt{N}
  \Delta_{\lambda}^{b}
  \left( b_{1,Q}^{\dag} + \mathrm{h.c.} \right), 
  \nonumber \\[1ex]
  \mathcal{H}_{1,\lambda} & = &
  \frac{1}{\sqrt{N}}
  \sum_{
    \genfrac{}{}{0pt}{1}{
      \genfrac{}{}{0pt}{1}{k, q>0}{
        \alpha,\beta,\gamma
      }
    }{}
  }
  g_{k,q,\lambda}^{\alpha,\beta,\gamma}
  \left\{ \delta(b_{\gamma,q}^{\dag})
  \delta(c_{\alpha,k}^{\dag}
  c_{\beta,k+q}) + \mathrm{h.c.} \right\} .
  \nonumber
\end{eqnarray}
where $\Delta^c_{k,\lambda}$ and $\Delta^b_\lambda$ are the appropriate 
order parameters for the electronic and the phononic symmetry breaking fields. 
Note that the reduced Brillouin zone leads to additional band indices
$\alpha,\beta,\gamma=0,1$ of both electronic and phononic one-particle
operators. Furthermore, we defined 
$\delta\mathcal{A} = \mathcal{A} - \langle\mathcal{A}\rangle$ and
$Q=\pi/a$. The ansatz \eqref{HM8}
is restricted to the one-dimensional case at half-filling. To extend the
approach to higher dimensions one would need to take into account all
$\mathbf{Q}$ wave vectors of the Brillouin zone boundary. 

Before we can proceed we need to diagonalize $\mathcal{H}_{0,\lambda}$. For
this purpose a rotation in the fermionic subspace and a translation to new
ionic equilibrium positions are performed in order
to diagonalize  $\mathcal{H}_{0,\lambda}$
\begin{eqnarray}
  \label{HM9}
  \mathcal{H}_{0,\lambda} &=&
  \sum_{k>0} \sum_{\alpha}
  \varepsilon_{\alpha,k,\lambda}^{C}
  C_{\alpha,k,\lambda}^{\dag} C_{\alpha,k,\lambda} \\
  &&
  + \sum_{q>0}
  \sum_{\gamma} \omega_{\gamma,q,\lambda}^{B}
  B_{\gamma,q,\lambda}^{\dag}
  B_{\gamma,q,\lambda} - E_{\lambda}
  \nonumber
\end{eqnarray}
with new fermionic and bosonic creation an annihilation
operators, $C_{\alpha,k,\lambda}^{(\dag)}$ and
$B_{\gamma,q,\lambda}^{(\dag)}$, and we 
rewrite $\mathcal{H}_{1,\lambda}$ in terms of the new 
operators, $C_{\alpha,k,\lambda}^{(\dag)}$ and
$B_{\gamma,q,\lambda}^{(\dag)}$. 

Finally, we have to transform 
${\cal H}_\lambda$ to  ${\cal H}_{(\lambda - \Delta \lambda)}$ 
according to \eqref{B16} 
to derive the renormalization equations for the parameters of
$\mathcal{H}_{\lambda}$.  Here the ansatz 
\begin{eqnarray*}
  X_{\lambda, \Delta \lambda}
  &=&
  \frac{1}{\sqrt N} \sum_{k,q} \sum_{\alpha,\beta,\gamma}
  A_{k,q,\lambda,\Delta \lambda}^{\alpha,\beta,\gamma} \\
  &&
  \quad \times
  \left\{
    \delta B_{\gamma,q}^\dagger 
    \delta(
      C_{k,\lambda}^\dagger C_{\beta,k+q,\lambda}
    ) 
    - \mathrm{h.c.}
  \right\}
\end{eqnarray*}
is used. The
coefficients  $A_{k,q,\lambda,\Delta \lambda}^{\alpha,\beta,\gamma}$ have to
be fixed in such a way so that only excitations with energies smaller than
$(\lambda-\Delta\lambda)$  contribute to 
$\mathcal{H}_{1,(\lambda- \Delta \lambda)}$. The renormalization equations 
for the parameters $\varepsilon_{\alpha,k,\lambda}, \Delta^c_{k,\lambda},
\omega_{\gamma, q, \lambda}, \Delta^b_\lambda$, and $g^{\alpha, \beta,
  \gamma}_{k, q, \lambda}$ are finally obtained by comparison with \eqref{HM8}
 after the creation and annihilation operators
$C^{(\dagger)}_{\alpha,k,\lambda},B^{(\dagger)}_{\gamma, q,\lambda}$ 
have been transformed back to the original operators
$c^{(\dagger)}_{\alpha,k}, b^{(\dagger)}_{\gamma, q}$. The actual calculations
are done in close analogy to subsection~\ref{HM_metal}. 
Note that again a factorization 
approximation was used and only operators of the same 
structure as in \eqref{HM8} are kept. Therefore, the final
renormalization equations still depend on unknown expectation values,
which are evaluated with the full Hamiltonian $\mathcal{H}$. Note that  
in order to evaluate the expectation values  
$\langle{\cal A}\rangle = \langle {\cal A}_\lambda \rangle_{{\cal H}_\lambda} 
$
additional renormalization equations have also to be found for 
the fermionic
and bosonic one-particle operators, $c_{\alpha,k}^{\dag}$ and
$b_{\gamma,q}^{\dag}$. By using the same approximations as for the
Hamiltonian a resulting set of renormalization equations 
is derived. It is solved numerically
where the equations for the expectation values are taken into account 
in a self-consistency loop.

By eliminating all excitations in steps $\Delta \lambda$ we finally 
arrive at cutoff $\lambda=0$ which again provides an effectively free model  
$
  \tilde{\mathcal{H}} = \lim_{\lambda\rightarrow 0} \mathcal{H}_{\lambda}
  = \lim_{\lambda\rightarrow 0} \mathcal{H}_{0,\lambda}
$. It reads 
\begin{eqnarray}
\label{G6}
  \tilde{ \mathcal{H}} &=&  \sum_{k>0,\alpha} 
\tilde{\varepsilon}_{\alpha,k}
   c_{\alpha,k}^{\dag} c_{\alpha,k} + 
   \sum_{k>0}
   \tilde{\Delta}_{k}^{\mathrm{c}} 
   \left( 
     c_{0,k}^{\dag} c_{1,k} + \mathrm{h.c.} 
   \right) \\
   &+& \sum_{q>0, \gamma}
\tilde{\omega}_{\gamma,q}
   b_{\gamma,q}^{\dag} b_{\gamma,q} + \sqrt{N}
   \tilde{\Delta}^{b} 
   \left( b_{1,Q}^{\dag} + b_{1,Q} \right)
  - \tilde{E} \nonumber
\end{eqnarray}
where  it was defined
$
\tilde{\varepsilon}_{\alpha, k} = 
\lim_{\lambda \rightarrow 0}{\varepsilon}_{\alpha, k,\lambda}
$, 
$
  \tilde{\Delta}_{k}^{c} = 
  \lim_{\lambda\rightarrow 0}\Delta_{k,\lambda}^{c}
$,
$
\tilde{\omega}_{\gamma, q}=
\lim_{\lambda \rightarrow 0} \omega_{\gamma, q, \lambda}
$,
and
$
  \tilde{\Delta}^{b} = 
  \lim_{\lambda\rightarrow 0}\Delta_{\lambda}^{b}
$.
Note that all excitations from ${\cal H}_{1,\lambda}$ were used up to
renormalize the parameters of $\tilde{\cal H}_0$. 
The expectation values are also calculated in 
the limit $\lambda\rightarrow 0$. Because $\tilde{\mathcal{H}}$ 
is a free model they can easily be determined from  
$
  \langle A \rangle_{\mathcal{H}} 
  = \langle A_{\lambda} \rangle_{\mathcal{H}_{\lambda}}
  = \langle 
      (\lim_{\lambda\rightarrow 0} A_{\lambda}) 
    \rangle_{\tilde{\mathcal{H}}} \, . 
$

\subsection{Results}
\label{results}

In the following, we first demonstrate that the PRM can be used to
investigate the Peierls transition of the one-dimensional spinless Holstein
model \eqref{HM1} at half-filling. The phonon energy is fixed to
$\omega_0 = 0.1t$. In particular, our analytical approach
provides a simultaneous theoretical description for both 
the metallic and the
insulating phase.  Finally, we compare our results with recent 
DMRG calculations \onlinecite{BMH_1998,FWH_2005}.

\begin{figure}
  \begin{center}
    \scalebox{0.62}{
      \includegraphics*{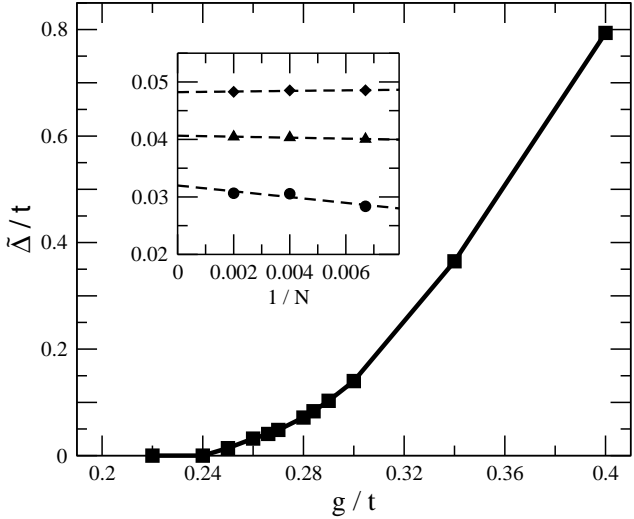}
    }
  \end{center}
  \caption{
    Electronic excitation gap of the one-dimensional HM at half-filling where
    the data are extrapolated to an infinite chain. The inset shows the
    finite-size scaling for $g$ values of the EP coupling of $0.26t$
    (circles), $0.266t$ (triangles), and $0.27t$ (diamonds).
  }
  \label{Fig_HM_delta}
\end{figure}

First, let us consider the critical electron-phonon coupling  $g_{c}$.
For that purpose, in Fig.~\ref{Fig_HM_delta} a characteristic 
electronic excitation gap $\tilde{\Delta}$ for infinite system size
is plotted as function of the EP coupling $g$, where
$\tilde{\Delta}$
was determined from the opening of a gap in the 
quasi-particle energy $\tilde{\varepsilon}_k$ (see text below). A closer 
inspection of the data shows
that an insulating phase with a finite excitation gap is obtained for $g$
values larger than the critical EP coupling $g_{c}\approx 0.24t$. 
A comparison with the critical value $g_{c}\approx 0.28t$ obtained from DMRG
calculations \onlinecite{BMH_1998,FWH_2005} shows that the 
critical values from the PRM
approach might be somewhat too small. However, this difference can be
attributed to the exploited factorization approximation in the PRM which
suppresses fluctuations so that the ordered insulating phase is
stabilized. 
Note that in order to determine $g_{c}$ a careful finite-size scaling 
was performed as
shown for some $g$ values in the inset of Fig.~\ref{Fig_HM_delta}. 
A linear regression was applied to extrapolate our results 
to infinite system size. Note that the finite size scaling 
may be affected by two
different effects: Suppression of long-range fluctuations by the finite
cluster size and by the used factorization approximation so that a
rather unusual dependence on the system size is found. 

In contrast to other methods, the PRM directly provides the quasi-particle
energies: After the renormalization equations were
solved self-consistently the electronic and phononic quasi-particle energies
of the system, $\tilde{\varepsilon}_{k}$ and 
$\tilde{\omega}_{q}$, respectively, are given by the limit 
$\lambda\rightarrow 0$ of the parameters 
$\varepsilon_{\alpha,k,\lambda}^{C}$ and 
$\omega_{\gamma,q,\lambda}^{B}$ of the diagonal Hamiltonian 
$\mathcal{H}_{0,(\lambda \rightarrow 0)}$ of
\eqref{HM9}. 
\begin{figure}
  \begin{center}
    \scalebox{0.65}{
      \includegraphics*{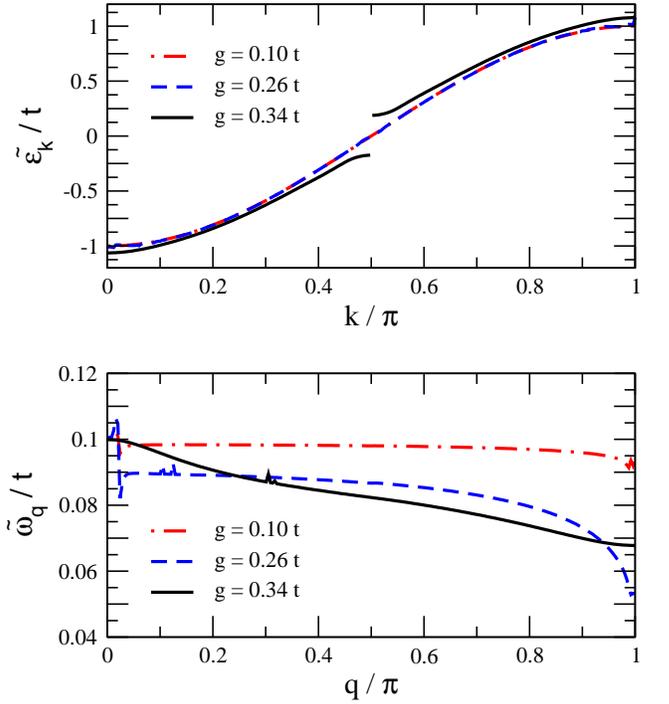}
    }
  \end{center}
  \caption{
    (Color online) Fermionic quasi-particle energy $\tilde{\varepsilon}_k=
     \varepsilon^C_{\alpha=0, k, \lambda=0}$ (upper panel) 
     and bosonic quasi-particle energy 
     $\tilde{\omega}_q= \omega^B_{\alpha=0, q, \lambda=0}$ (lower panel)
     of a chain with $500$ lattice sites for different
     EP couplings $g$.
  }
  \label{Fig_4}
\end{figure}
In Fig.~\ref{Fig_4} the renormalized one-particle energies 
$\tilde{\varepsilon}_k= \varepsilon^C_{\alpha=0, k, \lambda=0}$
and $\tilde{\omega}_q =\omega^B_{\gamma=0, q, \lambda=0}$ 
as  quasi-particle of the full system 
are shown for different values of
the EP coupling $g$. The upper panel shows that the electronic
one-particle energies depend only slightly on $g$ as long as $g$ is smaller
than the critical value $g_{c}\approx 0.24t$. If the EP coupling $g$ is
further increased a gap $\tilde{\Delta}$ opens at the Fermi energy so that the
system becomes an insulator. Remember 
that the gap $\tilde{\Delta}$ has been used
as order parameter to determine the critical EP coupling $g_{c}$ of the
metal-insulator transition (see Fig.~\ref{Fig_HM_delta}). The lower panel of
Fig.~\ref{Fig_4} shows the results for the phononic one-particle energy 
$\tilde{\omega}_{q}$. 
One can see that $\tilde{\omega}_{q}$ gains dispersion 
due to the coupling $g$ between the electronic and phononic
degrees of freedom. In particular, the phonon mode at momentum $2k_F$, i.e.~at
the Brillouin-zone boundary becomes soft if the EP coupling is increased up to
$g_{c}\approx 0.24t$. However, in contrast to the metallic solution of
subsection~\ref{HM_metal} $\tilde{\omega}_q$ at $2k_F$ always remains positive 
though it is very small. Note that for $g$ values larger than $g_c$
the energy $\tilde{\omega}_q$ increases again.
This phonon softening at the phase transition has to be
interpreted as a lattice instability which leads to the formation of the
insulating Peierls state for $g>g_{c}$. The phase transition is
associated with a shift of the ionic equilibrium positions. A lattice
stiffening occurs if $g$ is further increased to values 
much larger than the critical value  $g_{c}\approx 0.24t$.

Note also that the critical coupling $g_{c}\approx 0.24t$
obtained from the opening of the gap in $\tilde{\varepsilon}_k$ 
is significantly smaller than the 
$g_{c}$ value of $\approx 0.31t$ which was found 
from the vanishing of the phonon mode 
at the Brillouin zone boundary in the metallic solution of 
subsection~\ref{HM_metal}. 
Instead, one would expect that both the gap in  
$\tilde{\varepsilon}_{k}$ and the vanishing of $\tilde{\omega}_{q}$
should occur at the same $g_c$ value. This inconsistency 
can again be understood from the factorization 
approximation in the PRM: As discussed above, 
the inclusion of additional fluctuations
leads to a less stable insulating phase so that a $g_c$ value 
larger than $0.24 t$ would follow. 
On the other hand, the dispersion of 
$\tilde{\omega}_q$ due to renormalization 
processes would be enhanced by taking 
additional fluctuations into account. Thus,  
a $g_c$ value smaller than $\approx 0.31 t$ would follow. 
In this way, both ways to determine $g_c$ would be consistent with each other 
and could lead to a common result for $g_c$ 
in between $0.24 t$ and $0.31 t$. This would
be in agreement with the DMRG value of $g_c \approx 0.28 t$       
(\onlinecite{BMH_1998,FWH_2005}).


\section{Charge ordering and superconductivity in the 
two-dimensional Holstein model}
\label{Holstein_SC}
As a second example for a quantum phase transition, we now study 
the competition of charge-density waves (CDW) and superconductivity
(SC) for the two-dimensional half-filled Holstein model 
by use of the projector-based renormalization method.  
In one dimension the coupling of electrons to
phonons gives rise to a metal-insulator transition. 
In two dimensions the electron-phonon 
interaction may also be responsible  for the formation of Cooper pairs. 
In the following, the competing influence of superconductivity and 
charge order will be discussed for two dimensions. The PRM not 
only allows to study SC and CDW correlation functions 
but gives direct access to the order parameters. The discussion closely
follows the approach of Ref.~\onlinecite{SHB_2007} \\

The relationship between a possible superconducting and an insulating 
Peierls-CDW phase in the 2d-Holstein model has been subject to a 
number of studies in the literature (for details we refer to 
Ref.~\onlinecite{SHB_2007}). In general, it is believed that the onset of strong 
SC correlations suppresses the development
of CDW correlations and vice versa. Thus close to the
phase transition, both types of correlations must be taken into account.

\subsection{Unified description of SC and CDW phases at half-filling}
\label{SP}

To find a uniform description of both the superconducting (SC) 
and the insulating CDW phase, 
two fields, which break the translation 
and the gauge symmetry should be added to the Hamiltonian.
Thus, the model on a square Lattice is given by 
\begin{eqnarray}
\label{1}
\mathcal{H} &=&  {\cal H}_0 + {\cal H}_1 \\
&& \nonumber \\
{\cal H}_0 &=& \sum_{{\bf k},\sigma} \varepsilon_{\bf k} c_{{\bf
 k},\sigma}^{\dag} c_{{\bf k},\sigma} +
\omega_0  \sum_{\bf q} b_{\bf q}^{\dag} b_{\bf q} \\
&& \nonumber \\ 
\label{2}
&+&  
   \sum_{\bf k}   \left(       \Delta^{\mathrm{s}}_{\bf k} 
c_{{\bf k},\uparrow}^{\dag}  c_{-{\bf k},\downarrow}^{\dag} + 
 {\Delta^{\mathrm{s}}_{\mathbf k}}^* 
c_{-{\bf  k},\downarrow} c_{{\bf k},\uparrow} \right) \nonumber  \\
\label{3}
&+&
\frac{1}{2} \sum_{{\bf k},\sigma} \left( \Delta^{\mathrm{p}}_{\bf k} \,
c_{{\bf k},\sigma}^{\dag} c_{{\bf k}-{\bf Q},\sigma} +{\rm h.c} \right) 
+ \sqrt{N}\Delta^b (b_{\bf Q}^\dagger+ b_{\bf Q}) \nonumber \\
&& \nonumber \\
{\cal H}_1 &=&
\frac{1}{\sqrt{N}}  g  \sum_{{\bf k},{\bf q},\sigma}  
\left\{ b_{\bf q}^{\dag} c_{{\bf k},\sigma}^{\dag}
 c_{{\bf k} + {\bf q},\sigma} + b_{\bf q} c_{{\bf k} + {\bf q},\sigma}^{\dag}
 c_{{\bf k},\sigma} \right\}. 
\label{4}
\end{eqnarray}
where ${\bf k}$ is the wave vector on the reciprocal lattice and  
${\bf Q}$ is the characteristic wave vector of the CDW phase 
 ${\bf Q}= (\pi/a,\pi/a) $.
Assuming an electron hopping between nearest-neighbor sites, 
the electronic dispersion is given by 
$\varepsilon_{\mathbf k}= -2t (\cos k_xa + \cos k_ya) - \mu$, where $\mu$ 
is the chemical potential. Moreover, $\omega_0$ is 
the dispersionless phonon energy, and
$g$ denotes the coupling strength between the electrons and phonons.  
At the beginning of the renormalization the two symmetry breaking fields 
$\Delta_{\bf k}^{\mathrm{s}}$
and $\Delta_{\bf k}^{\mathrm{p}}$, as well as $\Delta^b$,
are assumed to be 
infinitesimally small ($\Delta^{\mathrm s}_{\bf k}\rightarrow 0$,
$\Delta^{\mathrm{p}}_{\bf k} \rightarrow 0, \Delta^b  \rightarrow 0$).

The unperturbed Hamiltonian ${\cal H}_0$ 
can be diagonalized, since its electronic part 
is quadratic in the fermionic operators.  Note that  
due to the doubling of the unit cell in the insulating
phase, in ${\cal H}_0$ the creation operator $c_{{\bf k}, \sigma}^\dagger$
is coupled to  $c_{{\bf k}- {\bf Q}, \sigma}$. In addition 
the coupling of $c_{{\bf k}, \uparrow}^\dagger$ to  
$c_{{\bf -k}, \downarrow}^\dagger $ is caused by  superconductivity. 
Therefore, the eigenmodes of ${\cal H}_0$ can be represented 
as a linear combination of the following four operators 
\begin{eqnarray}
\label{5}
\left( 
\begin{array}{llll}
c_{-{\bf k}-{\bf Q},\downarrow} & c_{{\bf k},\uparrow}^\dag & 
c_{-{\bf k},\downarrow} & c_{{\bf k}+{\bf Q},\uparrow}^\dag
\end{array}
\right)
\end{eqnarray}

In the renormalization procedure, all transitions 
with energies larger than $\lambda$ will be integrated out. As can be seen, 
the renormalized Hamiltonian can again be
divided into ${\cal H}_\lambda =  {\cal H}_{0,\lambda}+ {\cal H}_{1,\lambda}$.
If one denotes by $a_{\alpha,{\bf k},\lambda}^\dagger$ ($\alpha= 1 \cdots 4$) 
the $\lambda$ dependent eigenmodes of  ${\cal H}_{0,\lambda}$ 
the electronic part of the renormalized Hamiltonian 
${\cal H}_{0,\lambda}$can be written as 
\begin{eqnarray}
\label{10}
\mathcal{H}_{0,\lambda}^{\mathrm{el}} &=& \sum_{{\bf k} \in {\rm r.BZ}} 
  \left\{ 
E_{1,{\bf k},\lambda} \left( a_{1,{\bf k},\lambda}^{\dag} a_{1,{\bf
  k},\lambda} + a_{2,{\bf k},\lambda}^{\dag} a_{2,{\bf
  k},\lambda} \right)  \right. \nonumber \\
&& 
\left.
\, + E_{2,{\bf k},\lambda} \left( a_{3,{\bf k},\lambda}^{\dag} 
a_{3,{\bf  k},\lambda} + a_{4,{\bf k},\lambda}^{\dag} a_{4,{\bf k},
\lambda} \right) 
 \right\} 
\end{eqnarray}
where the eigenenergies are given by 
\begin{eqnarray}
\label{11}
&& E_{1/2,{\bf k},\lambda} = 
 \frac{\varepsilon_{{\bf k},\lambda} +
  \varepsilon_{{\bf k}-{\bf Q},\lambda}}{2} \pm W_{{\bf k},\lambda} \\
  && W_{{\bf k},\lambda} = 
\sqrt{ \left(
  \frac{\varepsilon_{{\bf k},\lambda} -
  \varepsilon_{{\bf k}-{\bf Q},\lambda}}{2} \right)^{2} + 
  |\Delta_{{\bf k},\lambda}^{\mathrm{p}}|^{2} + 
  |\Delta_{{\bf k},\lambda}^{\mathrm{s}}|^{2}} \nonumber
\end{eqnarray}
for $\varepsilon_{{\bf k},\lambda} 
+ \varepsilon_{{\bf k}-{\bf Q},\lambda} > 0$, 
whereas for $\varepsilon_{{\bf k},\lambda} + 
\varepsilon_{{\bf k}-{\bf Q},\lambda} \leq 0$ the $\pm$-signs have to be 
reversed. Note  that in \eqref{11} the sum of 
the two order parameters squared enter the energies 
$E_{\alpha,{\bf k},\lambda}$ of \eqref{11}.   


In order to derive the renormalization equations,
the unitary transformation \eqref{B16} has to be evaluated explicitly. 
Thereby, also the interaction ${\cal H}_{1,\lambda}$ has to be 
expressed in terms of the eigenmodes 
$a_{\lambda, {\bf k}, \lambda}$ of ${\cal H}_{0,\lambda}$. Moreover, an 
ansatz for   $X_{\lambda, \Delta \lambda}$ has to be made in analogy to 
what was done in the previous sections. The explicit calculation is found in
Ref.~\onlinecite{SHB_2007}.
 
\subsection{Results and Discussion}

\begin{figure}
  \begin{center}
    \scalebox{0.33}{
      \includegraphics*{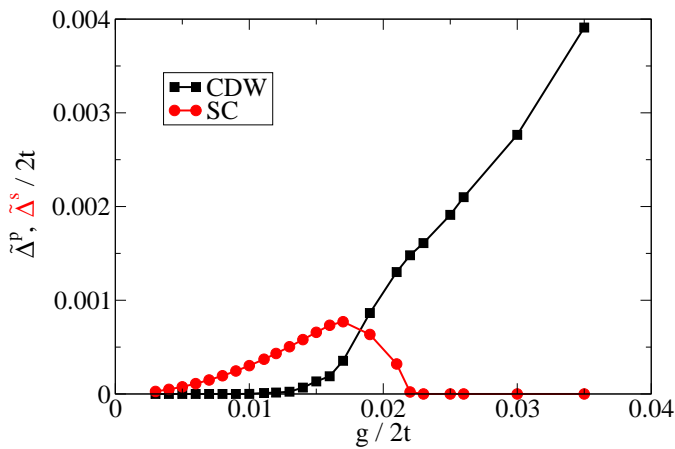}
    }
  \end{center}
  \caption{
    (Color online)
    Renormalized values of the Peierls gap
  $\tilde{\Delta}_{\bf k}^{\rm p}$ (black line) and of the superconducting gap
  $\tilde{\Delta}_{\bf k}^{\rm s}$ (red line) at wave vector 
  ${\bf k} = (\pi / 2 , \pi / 2)$. A square
  lattice with 144 lattice sites at half-filling was taken,  
  $\omega_0 / t = 0.1$ and $T=0$. 
  }
  \label{Fig_1/2}
\end{figure}

%

For the numerical evaluation of the renormalization 
equations, we consider a square lattice with $N=144$ sites.
The temperature is set equal to $T=0$, and
a small value of $\omega_{0} = 0.1t$ is chosen. For simplicity, we also restrict 
ourselves to $s$-wave-like superconducting solutions.

The results are shown in Fig.~\ref{Fig_1/2}, where the 
${\bf k}$-dependent symmetry
breaking fields $\tilde{\Delta}_{\bf k}^{\rm p}$ (black) and 
$\tilde{\Delta}_{\bf k}^{\rm s}$ (red) for 
${\bf k} = (\pi / 2 , \pi / 2)$ are plotted as function of 
the electron-phonon coupling $g$. The
coupling  $g$ is restricted to small values  $g/2t \leq 0.04$.
As can be seen from Fig.~\ref{Fig_1/2},  
for small values of $g/2t < 0.010$ the system is in a pure
superconducting state, i.e.~no charge order
is present. For small $g$, the superconducting 
gap increases roughly proportional to $g^2$.
In the intermediate  $g$ range,  $0.010 < g/2t < 0.023$, a coexistence
of both phases is found. The system 
is in a combined superconducting-charge ordered phase. Here, the 
$g$ dependence of   
$\tilde{\Delta}^{\mathrm{s}}_{\bf k}$ is no longer quadratic 
as in the small $g$ regime. 
Instead, $\tilde{\Delta}^{\mathrm{s}}_{\bf k}$
reaches a maximum value and drops down to zero with increasing 
$g$. Finally, for  $g/2t > 0.023$ the superconducting phase is
completely suppressed and the system is in a pure 
charge ordered state.

\section{Summary}
\label{summary}

The aim of this contribution was to discuss the basic ideas of 
a new theoretical approach for many-particle systems 
which is called
projector-based renormalization method (PRM) and its application to a number
of nontrivial physical problems. Instead of 
eliminating high-energy states as in usual renormalization group 
methods in the PRM high-energy transitions
are successively eliminated. Thereby, a unitary
transformation is used where all states of the 
unitary space of the interacting system are kept. 
In that respect, the PRM is closely
related to the  similarity transformation introduced by Wilson and Glazek 
and to Wegner's flow equation method though both approaches start from a 
continuous formulation of the unitary transformation.   
The PRM starts from a Hamiltonian
which can be decomposed into a solvable unperturbed part 
and a perturbation, ${\cal H} = {\cal H}_0 +
{\cal H}_1$, where the latter part induces transitions 
between the eigenstates of ${\cal H}_0$. 

Suppose 
a renormalized Hamiltonian ${\cal H}_\lambda$ has been constructed
which only contains transitions with transition 
energies smaller than some given 
cutoff energy $\lambda$.  
The Hamiltonian ${\cal H}_\lambda$  
can be further renormalized by eliminating all transitions from, 
roughly speaking, the energy shell between the cutoff $\lambda$ and a reduced 
cutoff $(\lambda - \Delta \lambda)$, and so on.  This is done by   
a unitary transformation $H_{(\lambda - \Delta \lambda)} =
e^{X_{\lambda,\Delta \lambda}}\, {\cal H}_\lambda\, 
e^{-X_{\lambda,\Delta \lambda}}$ which guarantees that the eigenspectrum is
not changed. The generator of the unitary transformation
$X_{\lambda, \Delta \lambda}$ is specified by the condition 
${\bf Q}_{\lambda - \Delta \lambda}{\cal H}_{(\lambda - \Delta \lambda)}=0$
where  ${\bf Q}_{\lambda - \Delta \lambda}$ is the projector on all 
transitions with energy differences larger than 
$(\lambda - \Delta \lambda)$. The latter condition implies that 
all transitions from the 'shell' between $\lambda$ and $\lambda - 
\Delta \lambda$ are eliminated and lead to a renormalization of  
${\cal H}_{(\lambda - \Delta \lambda)}$. Note that only 
the equivalent part ${\bf Q}_{\lambda - \Delta \lambda}
X_{\lambda, \Delta \lambda}$
of $X_{\lambda, \Delta \lambda}$ is fixed whereas 
the orthogonal part  ${\bf P}_{\lambda - \Delta \lambda}
X_{\lambda, \Delta \lambda}$ can be chosen arbitrarily. Note that this
additional freedom can be used in a different way. Whereas in the original 
version of the PRM the remaining part ${\bf P}_{\lambda - \Delta \lambda}
X_{\lambda, \Delta \lambda}$ of $X_{\lambda, \Delta \lambda}$ was set equal to
zero for simplicity this part was used in Wegner's flow equation method as 
the only relevant part when the transformation was performed 
continuously. In this case, the interaction parameters were chosen 
to decay exponentially. By proceeding the renormalization up 
to the final cutoff $\lambda =0$ all transitions 
induced by ${\cal H}_{1,\lambda}$
are eliminated. The final renormalized Hamiltonian 
$\tilde{\cal H}={\cal H}_{0,\lambda=0}$ is diagonal and allows 
to evaluate in principle any correlation function of physical interest. 
In particular the one-particle excitations of $\tilde{\cal H}$ can be 
considered as quasi-particles of the coupled many-particle system since the 
eigenspectrum of the original interacting Hamiltonian ${\cal H}$ 
and of  $\tilde{\cal H}$ are in principle the same since both are connected by 
a unitary transformation.  

Note that the present approach has the advantage 
of formulating the renormalization quite universally.
By specifying the unitary transformation of the many-particle system 
both the PRM and Wegner's flow equation method can be derived from the same
basic ideas. However, the stepwise transformation of the PRM 
has its own merits. Firstly, as was shown in Sec.~\ref{EP_BCS},
Sec.~\ref{Holstein_QP}, and Sec.~\ref{Holstein_SC}
the physical behavior
on both sides of a quantum critical point can be 
described within the same PRM scheme. 
This seems not the case for the flow equation approach. 
In particular, by allowing symmetry breaking terms in the 'unperturbed' part
${\cal H}_{0,\lambda}$,  the transformation of 
eigenmodes of the Liouville operator 
${\bf L}_{0,\lambda}$ can be followed in each renormalization step.
This makes the description of quantum critical points possible. 
Secondly, in Sec.~\ref{B_perturbation} 
a perturbation theory for ${\cal H}_{\lambda}$ was given. This allows to
evaluate physical properties in perturbation theory. In contrast to a recent 
perturbation approach on the basis of the flow equation method, in the 
PRM no equidistant spectrum of ${\cal H}_0$ is required.

\section*{Acknowledgments}

We would like to acknowledge stimulating and enlightening discussions with
A. Mai and J. Sch\"{o}ne. This work was supported by the DFG through the
research program SFB 463.

\begin{appendix}

\section{Example: dimerized and frustrated spin chain}
\label{spin_chain}

In this appendix we are going to investigate ground-state properties 
of a dimerized and frustrated spin chain. We apply the projector-based 
perturbation theory and use expression \eqref{B11} 
for ${\cal H}_\lambda$ and chose $\lambda \rightarrow 0$ 
right from the beginning. In this case, the interaction 
${\cal H}_1$ is completely integrated out in one step. 
The starting Hamiltonian reads 
\begin{eqnarray}
  \label{B12}
  \mathcal{H} &=& 
  \mathcal{H}_{0} + \mathcal{H}_{1}, \\[1ex]
  \mathcal{H}_{0} &=&
  J \sum_{i} \mathbf{S}_{2i} \mathbf{S}_{2i+1}, \nonumber \\
  \mathcal{H}_{1} &=& 
  J \sum_{i} [\alpha \mathbf{S}_{2i} \mathbf{S}_{2i-1} + 
  \beta\left(
    \mathbf{S}_{2i} \mathbf{S}_{2i-2} + 
    \mathbf{S}_{2i-1} \mathbf{S}_{2i+1}
  \right)],
  \nonumber
\end{eqnarray}
The model itself is of some physical interest because 
it can be used to describe some
spin-Peierls compounds like CuGeO$_{3}$ or TTFCuBDT 
\cite{RD_1995, CCE_1995, BIJB_1983}.  

\bigskip
In the following we are interested in the limit of strong 
dimerization of the
model so that we start from isolated dimers as described by
$\mathcal{H}_{0}$. Every dimer can be in the singlet state or in
one of the three degenerated triplet states. Here, the dimer states are
energetically separated by the singlet-triplet splitting,
$\Delta = \varepsilon_{t} - \varepsilon_{s} = J$. Thus, triplets can be
considered as the basic excitations of the system.

Following the ideas of Refs. \onlinecite{KU_2000} and \onlinecite{SHB_2004},
the contributions to the perturbation $\mathcal{H}_{1}$ can be
classified according to the number of created or annihilated local triplets,
\begin{eqnarray}
  \label{B13}
  \lefteqn{\mathcal{H}_{1} \,=\,} &&\\ 
  &=& 
  \sum_{j} \left[
    \mathcal{T}_{-2}(j) + \mathcal{T}_{-1}(j) + \mathcal{T}_{0}(j) +
    \mathcal{T}_{1}(j) + \mathcal{T}_{2}(j)
  \right]\nonumber
\end{eqnarray}
The introduced excitation operators $\mathcal{T}_{m}(j)$ only act on the
local dimers with indices $j$ and $j-1$, and create $m$ local triplets. The
$\mathcal{T}_{m}(j)$ are eigenoperators of the Liouville operator
$\mathbf{L}_{0}$ and the corresponding eigenvalues are 
$\Delta_{m} = m\Delta$. The actual contributions to $\mathcal{T}_{0}(j)$, 
$\mathcal{T}_{1}(j)$, and $\mathcal{T}_{2}(j)$ are summarized in Table
\ref{Tab_dimer}, and $\mathcal{T}_{-1}(j)$ and $\mathcal{T}_{-2}(j)$ are given
by the relation $\mathcal{T}_{-m}(j) = [\mathcal{T}_{m}(j)]^{\dagger}$.

\begin{table}
  \caption{
    Action of the ${\cal T}_{m}(j)$ as used in the calculations. For
    convenience, the dimer indices of the states are suppressed.
  }
  \label{Tab_dimer}
 \begin{center},
    \scalebox{0.685}{\includegraphics{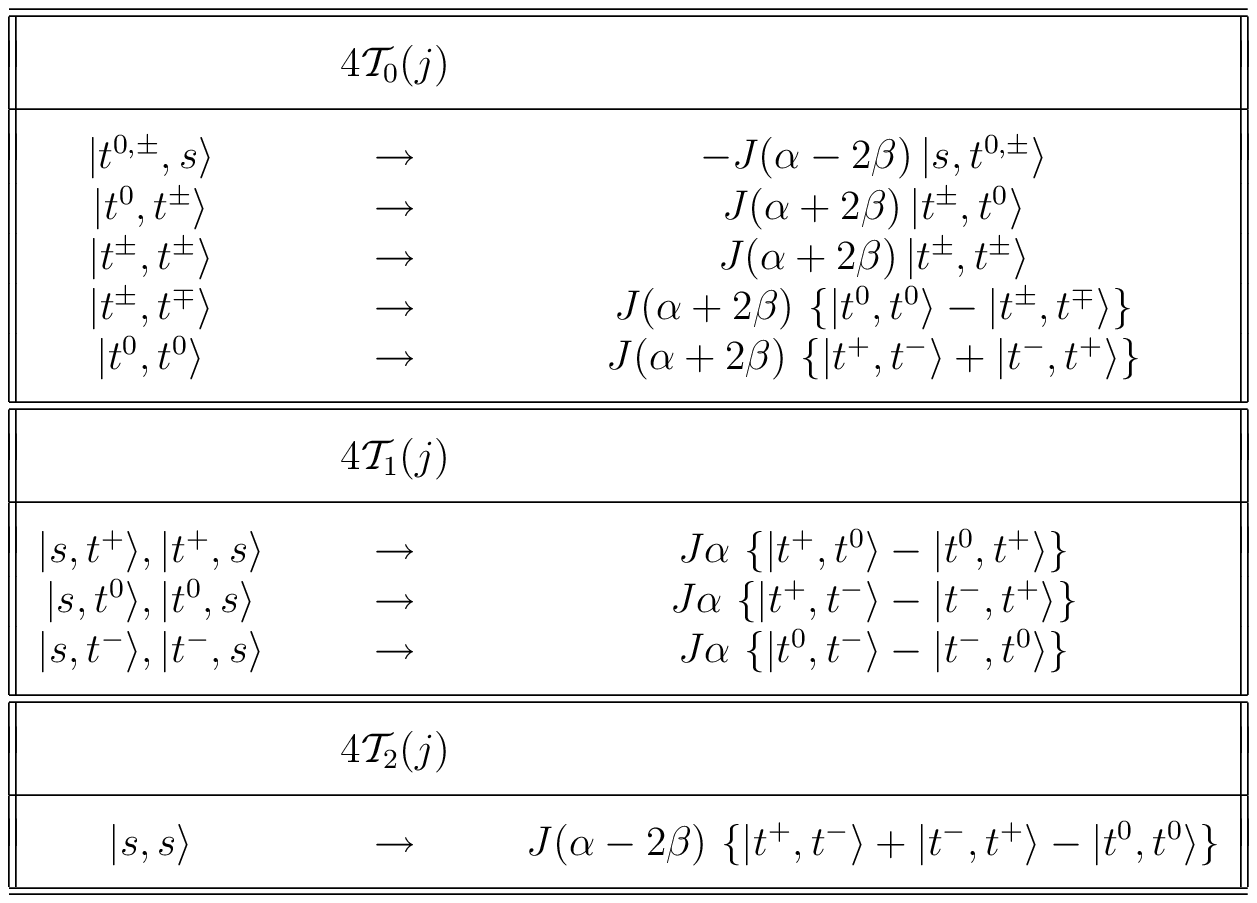}}
 \end{center}
\end{table}

In the limit of strong dimerization, Hilbert space sectors with different
numbers of triplets in the system are energetically separated because the
unperturbed part $\mathcal{H}_{0}$ of the Hamiltonian \eqref{B12}
does not change the number of triplets in the system, and the interaction 
$\mathcal{H}_{1}$ only leads to modest corrections. Consequently,
the evaluation of the effective Hamiltonian \eqref{B11} can be simplified if
one concentrates on a Hilbert space sector with a given fixed number of
triplets. In the following, actual calculations are presented for the two
energetically lowest sectors where the system contains no or only one triplet.

\bigskip
The subspace without triplets consists of a single state, i.e. the singlet
product state,
$
  \left|\Phi_{\mathrm{GS}}\right\rangle =
  \left| s_{1} \right\rangle \left| s_{2} \right\rangle \dots 
  \left| s_{N} \right\rangle
$.
Because the effective Hamiltonian $\mathcal{H}_{(\lambda\rightarrow0)}$ is
obtained from the original Hamiltonian $\mathcal{H}$ by means of a unitary
transformation, the ground-state energy can be calculated from 
\begin{eqnarray*}
  E_{\mathrm{GS}} &=&
  \lim_{\beta\rightarrow\infty} \langle \mathcal{H} \rangle \,=\,
  \lim_{\beta\rightarrow\infty} 
  \frac{
    \mathrm{Tr}\, \mathcal{H}_{(\lambda\rightarrow 0)} \,
    e^{-\beta \mathcal{H}_{(\lambda\rightarrow 0)}}
  }{\mathrm{Tr} \, e^{-\beta \mathcal{H}_{(\lambda\rightarrow 0)}}} ,\\
  &=&
  \left\langle \Phi_{\mathrm{GS}} \right| \mathcal{H}_{(\lambda\rightarrow 0)}
  \left|\Phi_{\mathrm{GS}}\right\rangle .
\end{eqnarray*}
Here ${\cal H}_{(\lambda \rightarrow 0)}$ is given by \eqref{B11} where
$
\mathbf{\bar{P}}_{(\lambda \rightarrow 0)}^{}
{\cal H}_{1} = 
\sum_{j} {\cal T}_0(j)
$
and 
$\mathbf{\bar{Q}}_{(\lambda \rightarrow 0)}^{}
{\cal H}_{1}$
is the remaining part of \eqref{B13}.
Using the notation of Ref.~\onlinecite{SHB_2004}, one easily finds  
\begin{eqnarray}
  \label{B14}
  E_{\mathrm{GS}} &=&
  NJ\left\{
    -\frac{3}{4} - \frac{3}{32}(\alpha - 2\beta)^{2}
  \right\} 
  + \mathcal{O}(\mathcal{H}_{1}^{3}).
\end{eqnarray}
This result agrees with findings of Refs. \onlinecite{KU_2000} and
\onlinecite{SHB_2004}. Note that higher order terms can easily be calculated
by implementing a computer based evaluation algorithm as discussed in
Ref.~\onlinecite{SHB_2004} where a cumulant method \cite{HVB_1999} was applied
to the same model.

\bigskip
The case of a single triplet in the system is more complex because a triplet
can easily move along the chain. Consequently, it is advantageous to introduce
momentum dependent states,
\begin{eqnarray*}
  \left|\Phi_{k}^{\nu}\right\rangle &=&
  \frac{1}{\sqrt{N}} \sum_{j} e^{ikR_{j}} 
  \left| s_{1} \right\rangle \left| s_{2} \right\rangle \dots
  \left| t_{j}^{\nu} \right\rangle \dots \left| s_{N} \right\rangle,
\end{eqnarray*}
and the eigenvalues of this Hilbert space sector can be calculated by
$
  E_{k}^{\nu} = 
  \lim_{\lambda\rightarrow0}
  \left\langle \Phi_{k}^{\nu} \right| \mathcal{H}_{\lambda}
  \left|\Phi_{k}^{\nu}\right\rangle
$.
We again employ the useful notation of Ref.~\onlinecite{SHB_2004} and obtain
\begin{eqnarray}
  \label{B15}
  E_{k}^{\nu} &=& 
  E_{\mathrm{GS}} + 
  J \left\{ 
    1 + \frac{3}{16} \left( \alpha - 2 \beta \right)^{2} - 
    \frac{1}{4} \alpha^{2}
  \right\} \\
  && - \,
  J\left\{
    \frac{1}{2} \left( \alpha - 2\beta \right) + \frac{1}{4} \alpha^{2}
  \right\} \cos(ka) 
  \nonumber\\
  && - \,
  \frac{1}{16}J \left( \alpha - 2\beta \right)^{2} \cos(2ka) + 
  \mathcal{O}(\mathcal{H}_{1}^{3}).
  \nonumber
\end{eqnarray}
Note that the energy gap of the system can easily be determined from 
Eq.~\eqref{B15} by considering the case $k=0$. The $k$ dependence of
$E_{k}^{\nu}$ describes the triplet dispersion relation. Furthermore, the
calculation can easily be extended to higher orders. 

\bigskip
The same model was also studied \cite{KU_2000} based on Wegner's flow
equation method \cite{W_1994} where both ground-state energy and triplet
dispersion relation were calculated in high orders. However, for this purpose 
a set of coupled differential equations had to be integrated so that this
approach is restricted to systems with an equidistant eigenvalue spectrum of
the unperturbed part $\mathcal{H}_{0}$ of the Hamiltonian.

\end{appendix}



\begin{thebibliography}{}

\bibitem{ED} see, for example, H.Q. Lin and J.E. Gubernatis,
  Comput. Phys. {\bf 7}, 400 (1993), and references therein.
\bibitem{NRG} K.G. Wilson, Rev. Mod. Phys. {\bf 47}, 773 (1975); for a recent
  review see R. Bulla, T. Costi, and T. Pruschke, cond-mat/0701105.
\bibitem{MC} W. von der Linden, Physics Rep. {\bf 220}, 53 (1992).
\bibitem{DMRG} S. White, Phys. Rev. Lett. {\bf 69}, 2863 (1992); for a recent
  review see P. Schollw\"{o}ck, Rev. Mod. Phys. {\bf 77}, 259 (2005).
\bibitem{DMFT} W. Metzner and D. Vollhardt, Phys. Rev. Lett. {\bf 62}, 324
  (1989); G. Kotliar and D. Vollhardt, Physics Today, March 2004, p. 53; for a
  review see A. Georges, G. Kotliar, W. Krauth, and M.J. Rozenberg,
  Rev. Mod. phys, {\bf 68}, 13 (1996). 
\bibitem{GW_1993} S.D.~G{\l}azek and K.G.~Wilson, Phys. Rev. D {\bf 48}, 5863
  (1993).
\bibitem{GW_1994} S.D.~G{\l}azek and K.G.~Wilson, Phys. Rev. D {\bf 49}, 4214
  (1994).
\bibitem{W_1994} F.~Wegner, Ann.~Phys. (Leipzig) {\bf 3}, 77 (1994);
see also S.~Kehrein, {\it The Flow Equation Approach to Many-Particle Systems}, 
Springer Tracts in Modern Physics, Springer-Verlag GmbH, 2006. 
\bibitem{RG} see, for example, J.~Zinn-Justin, {\it Quantum field theory and critical phenomena},
Oxford, Clarendon Press 2002.
\bibitem{BHS_2002} K.W.~Becker, A.~H\"ubsch, and T.~Sommer, Phys. Rev. B
  {\bf 66}, 235115 (2002).
\bibitem{C_1984} P. Coleman, Phys. Rev. B {\bf 29}, 3035 (1984).
\bibitem{FKZ_1988} For a review see, for example, P. Fulde, J. Keller, and
  G. Zwicknagl, in \textit{Solid State Physics}, edited by H. Ehrenreich and
  D. Turnbull (Academic, San Diego, 1988), Vol. 41, p. 1.
\bibitem{S_1997} J. Stein, J. Stat. Phys. {\bf 88}, 487 (1997).
\bibitem{KU_2000} C. Knetter and G.S. Uhrig, Eur. Phys. J. B {\bf 13}, 209
  (2000).
\bibitem{HVB_1999} A. H\"{u}bsch, M. Vojta, and K.W. Becker, J. Phys.:
  Condens. Matter {\bf 11}, 8523 (1999).
\bibitem{RD_1995} J. Riera and A. Dobry, Phys. Rev. B {\bf 51}, 16098 (1995).
\bibitem{CCE_1995} G. Castilla, S. Chakravarty, and V. J. Emery, Phys. Rev.
  Lett. {\bf 75}, 1823 (1995).
\bibitem{BIJB_1983} J.W. Bray, L.V. Interante, I.C. Jacobs, J.C. Bonner, in
  \textit{Extended Linear Chain Compounds}, edited by J.S. Miller (Plenum
  Press, New York, 1983), Vol. 3, p. 353.
\bibitem{SHB_2004} S. Sykora, A. H\"{u}bsch, and K.W. Becker, Phys. Rev. B
  {\bf 70}, 054408 (2004).
\bibitem{HB_2003} A. H\"{u}bsch and K.W. Becker, Eur. Phys. J. B {\bf 33}, 391
  (2003).
\bibitem{HB_2005} A. H\"{u}bsch and K.W. Becker, Phys. Rev. B {\bf 71}, 155116
  (2005). 
\bibitem{HB_2006} A. H\"{u}bsch and K.W. Becker, Eur. Phys. J. B {\bf 52}, 345
  (2006).
\bibitem{SHBWF_2005} S. Sykora, A. H\"{u}bsch, K.W. Becker, G. Wellein, and
  H. Fehske, Phys. Rev. B {\bf 71}, 045112 (2005).
\bibitem{A_1961} P.W. Anderson, Phys. Rev. {\bf 124}, 41 (1961).
\bibitem{F_1961} U. Fano, Phys. Rev. {\bf 124}, 1866 (1961).
\bibitem{BCS} J. Bardeen, L.N. Cooper, and J.R. Schrieffer, Phys. Rev. 
  {\bf 108}, 1175 (1957).
\bibitem{C_1956} L.N. Cooper, Phys. Rev. {\bf 104}, 1189 (1956).
\bibitem{F_1952} H. Fr\"{o}hlich, Proc. R. Soc. London A {\bf 215}, 291
  (1952). 
\bibitem{LW_1996} P. Lenz and F. Wegner, Nucl. Phys. B {\bf 482}, 693 (1996). 
\bibitem{M_1997} A. Mielke, Ann. Physik (Leipzig) {\bf 6}, 215 (1997).
\bibitem{B_1958} N.N. Bogoliubov, Nuovo Cim. {\bf 7}, 794 (1958).
\bibitem{SHB_2006_1} S. Sykora, A. H\"{u}bsch, and K.W. Becker,
  Eur. Phys. J. B {\bf 51}, 181 (2006).
\bibitem{SHB_2006_2} S. Sykora, A. H\"{u}bsch, and K.W. Becker,
  Europhys. Lett. {\bf 76}, 644 (2006).
\bibitem{LRSSW_1986} P.A. Lee, T.M. Rice, J.W. Serene, L.J. Sham, and
  J.W. Wilkins, Comments Condens. Matter Phys. {\bf 12}, 99 (1986).
\bibitem{FFF_2002} R. Franco, M.S. Figueira, M.E. Foglio, 
Phys. Rev. B {\bf 66}, 045112 (2002).
\bibitem{Mai_2007} A. Mai, P.V. Nham, A. H\"{u}bsch, 
and K.W. Becker, unpublished.
\bibitem{Myake_2006} Myake
\bibitem{HF_1983} J.E.~Hirsch and E.~Fradkin, Phys.~Rev.~B {\bf 27}, 4302
  (1983). 
\bibitem{MHM_1996} R.H.~McKenzie, C.J.~Hamer, and D.W.~Murray, Phys. Rev. B
  {\bf 53}, 9676 (1996).
\bibitem{ZFA_1989} H.~Zheng, D.~Feinberg, and M.~Avignon, Phys.~Rev.~B 
  {\bf 39}, 9405 (1989).
\bibitem{HM_RG} L.G.~Caron and C.~Bourbonnais, Phys.~Rev.~B {\bf 29}, 4230
  (1984); G.~Benfatto, G.~Gallovotti, and J.L.~Lebowitz, Helv. Phys. Acta 
  {\bf 68}, 312 (1995).
\bibitem{HM_ED} A.~Wei{\ss}e and H.~Fehske, Phys. Rev. B {\bf 58}, 13526
  (1998); H.~Fehske, M.~Holicki, and A.~Wei{\ss}e, Advances in Solid State
  Physics {\bf 40}, 235 (2000).
\bibitem{BMH_1998} R.J.~Bursill, R.H.~McKenzie, and C.J.~Hamer,
  Phys. Rev. Lett. {\bf 80}, 5607 (1998).
\bibitem{JZW_1999} E.~Jeckelmann, C.~Zhang, and S.R.~White, Phys. Rev. B 60,
  7950-7955 (1999).
\bibitem{FWH_2005} H.~Fehske, G.~Wellein, G.~Hager, A.~Wei\ss e, K.W.~Becker,
  and A.R.~Bishop, Physica B {\bf 359-361}, 699 (2005).
\bibitem{MHB_2002} D.~Meyer, A.C.~Hewson, and R.~Bulla, Phys. Rev. Lett.
  {\bf 89}, 196401 (2002).

\bibitem{Scalettar} R.T.~Scalettar, N.E.~Bickers, and D.J.~Scalapino,
Phys.~Rev.~B {\bf 40}, 197 (1989).
\bibitem{MarsiglioI} F.~Marsiglio, Phys.~Rev.~B {\bf 42}, 2416 (1990).
\bibitem{Vekic} M.~Vekic, R.M.~Noack, and S.R.~White, 
      Phys.~Rev.~B {\bf 46}, 271 (1992).
\bibitem{MarsiglioII} F.~Marsiglio,J.E.~Hirsch, 
Phys.~Rev.~B {\bf 49}, 1366 (1994).
\bibitem{Berger} E.~Berger, P.~Valasek, and W.~von der Linden, 
Phys.~Rev.~B {\bf 52}, 4806 (1995).
\bibitem{SHB_2007} S.~Sykora, A.~H\"ubsch, and K.W.~Becker, to be published.


\end{thebibliography}
\end{document}